\documentclass[sigconf,balance=false]{acmart}
\usepackage{popets}

\usepackage{multirow}
\usepackage{array}
\usepackage{makecell} 

\usepackage{pdfpages} 

\usepackage{float} 

\usepackage{enumitem}
\setitemize[1]{leftmargin=4mm}

\setcopyright{popets}
\copyrightyear{2025}

\acmYear{2025}
\acmVolume{2025}
\acmNumber{1}
\acmDOI{10.56553/popets-2025-0038}
\acmISBN{}
\acmConference{Proceedings on Privacy Enhancing Technologies}
\begin{document}

\title[The Role of Demographics in Browser Fingerprinting]{How Unique is Whose Web Browser? The Role of Demographics in Browser Fingerprinting among US Users
}


\author{Alex Berke}
\orcid{https://orcid.org/0000-0001-5996-0557}
\affiliation{%
  \institution{MIT Media Lab}
  \city{}
  \state{}
  \country{}}
\email{aberke@mit.edu}

\author{Enrico Bacis}
\orcid{}
\affiliation{%
  \institution{Google}
  \city{}
  \state{}
  \country{}}
\email{enricobacis@google.com}

\author{Badih Ghazi}
\orcid{}
\affiliation{%
  \institution{Google}
  \city{}
  \state{}
  \country{}}
\email{badihghazi@gmail.com}

\author{Pritish Kamath}
\orcid{}
\affiliation{%
  \institution{Google}
  \city{}
  \state{}
  \country{}}
\email{pritish@alum.mit.edu}

\author{Ravi Kumar}
\orcid{}
\affiliation{%
  \institution{Google}
  \city{}
  \state{}
  \country{}}
\email{ravi.k53@gmail.com}

\author{Robin Lassonde}
\orcid{}
\affiliation{%
  \institution{Google}
  \city{}
  \state{}
  \country{}}
\email{lassonde@google.com}

\author{Pasin Manurangsi}
\orcid{}
\affiliation{%
  \institution{Google}
  \city{}
  \state{}
  \country{}}
\email{pasin@google.com}

\author{Umar Syed}
\orcid{}
\affiliation{%
  \institution{Google}
  \city{}
  \state{}
  \country{}}
\email{usyed@google.com}


\renewcommand{\shortauthors}{Berke et al.}

\begin{abstract}
Browser fingerprinting can be used to identify and track users across the Web, even without cookies, by collecting attributes from users' devices to create unique "fingerprints". This technique and resulting privacy risks have been studied for over a decade. Yet further research is limited because prior studies used data not publicly available. Additionally, data in prior studies lacked user demographics. 
Here we provide a first-of-its-kind dataset to enable further research. It includes browser attributes with users' demographics and survey responses, collected with informed consent from 8,400 US study participants. 
We use this dataset to demonstrate how fingerprinting risks differ across demographic groups.  For example, we find lower income users are more at risk, and find that as users' age increases, they are both more likely to be concerned about fingerprinting and at real risk of fingerprinting. Furthermore, we demonstrate an overlooked risk: user demographics, such as gender, age, income level and race, can be inferred from browser attributes commonly used for fingerprinting, and we identify which browser attributes most contribute to this risk.
Our data collection process also conducted an experiment to study what impacts users' likelihood to share browser data for open research, in order to inform future data collection efforts, with responses from 12,461 total participants. Female participants were significantly less likely to share their browser data, as were participants who were shown the browser data we asked to collect.
Overall, we show the important role of user demographics in the ongoing work that intends to assess fingerprinting risks and improve user privacy, with findings to inform future privacy enhancing browser developments. The dataset and data collection tool we provide can be used to further study research questions not addressed in this work.

\end{abstract}


\maketitle

\section{Introduction}

Third-party cookies (3P cookies) have enabled tracking users across websites, which is highly valuable to companies serving targeted information, such as ads. This cross-site user tracking is also detrimental to user privacy and in recent years major browsers have made changes to remove 3P cookies, to better protect user privacy \cite{microsoftEdge3PC, firefox3PC2023, webkit3PC2020, chrome3PC2023}. However, third parties can still use browser "fingerprinting", in similar ways to 3P cookies, to identify and track users across websites~\cite{cookiegraph2023}, and even across browsers~\cite{crossbrowserFP2017}. While it is possible for users to block or delete cookies, these privacy protective measures are not available to prevent fingerprinting, and browser fingerprinting is now commonplace \cite{iqbal2021}, used on more than a third of the top 500 US websites~\cite{fowler2019}.

Browser fingerprinting is achieved by collecting attributes from users' browsers which are related to their installed software, device hardware, or configuration. Each attribute alone may be shared by many users, but when combined they can create a unique "fingerprint", identifying a user's device~\cite{rfc6973}.

In 2010, Eckersley and the Electronic Frontier Foundation published the first large browser fingerprinting study: "How Unique is Your Web Browser?" \cite{eckersley2010}. Eckersley collected data by creating the \href{http://panopticlick.eff.org}{Panopticlick} website, which collected browser attributes from site visitors, many of whom learned about the study via social media or popular tech-oriented websites. Similar studies followed. In 2014, Laperdrix et al. launched their fingerprinting study website, AmIUnique.org, to collect more state-of-the-art browser fingerprinting data. Like the Panopticlick study, they solicited volunteer participation via social media and tech-oriented websites \cite{Laperdrix2016}. A 2018 study by Gómez-Boix et al. \cite{GomezBoix2018} deployed fingerprinting scripts on a French news site to collect even more attributes. Each of these studies quantified how many site visitors had unique fingerprints, and quantified how each browser attribute contributed to uniqueness by computing entropy for each attribute, where higher entropy indicated higher risk.

These studies had real impacts on privacy enhancing developments.  For example, the Panopticlick study showed the Plugins list had highest entropy, and changes were since made in Chromium browsers to return Plugins as a hardcoded list~\cite{chromiumPlugins}. Other browser attributes that had relatively high entropy, such as User agent (related to installed browser version) and Platform (indicates operating system), have similarly undergone changes to reduce the amount of information they can pass to fingerprinters~\cite{senol2023, chromiumUserAgentReduction}.

Despite their impact on privacy enhancing browser developments, these previous studies have limitations. Firstly, the user-level data analyzed in these studies are not publicly available\footnote{Aggregated statistics from AmIUnique.org are available.}. This lack of available data limits opportunities to further research fingerprinting risks, as noted by other researchers \cite{pugliese2020, pau2023}. This also limits comparing results from the various studies, or evaluating how effectively browser developments have curtailed fingerprinting. Secondly, data in these previous studies lacked demographics associated with the browser users~\cite{pugliese2020}, and there is reason to believe the data had bias. For example, the Panopticlick and AmIUnique study authors described their own datasets as biased due to how they were collected from participants interested in privacy, and they noted how this limited generalizing their results to broader populations~\cite{eckersley2010, Laperdrix2016}. A related fingerprinting study, which similarly collected data from volunteer participants interested in online privacy, also collected participants' demographics \cite{pugliese2020}. The participants were overwhelmingly (76.5\%) male, raising further questions about gender biases in other works. Did demographic biases impact previous results, with implications for how browser developers should prioritize privacy related changes? Without demographic data, potential demographic biases in prior works are unknown, and without access to datasets that include both browser data and user demographics, the potential impacts cannot be tested or controlled for.

In this work we address these previous limitations by collecting a dataset that includes browser attributes with users' demographics, collected with users' informed consent, which we provide for the research community. In addition, we ask new research questions. \\

\noindent \textbf{RQ1: What impacts users’ likelihood to share browser data for open research?} While collecting our dataset we conducted an experiment to help study this question and inform future data collection efforts. \\

\noindent \textbf{RQ2: Are some demographic groups more at risk than others?} In other words we ask, privacy for whom?  \\

\noindent \textbf{RQ3: Can sensitive demographic information be inferred from device fingerprinting signals?} This potential risk goes beyond user tracking, presenting risks for both users and advertisers.  \\

Our main contributions can be summarized as follows:

\begin{itemize}
    \item 
We develop a tool to collect browser attributes along with users' demographics and survey question responses, while prioritizing informed consent. 
    \item 
We provide a first-of-its-kind dataset with browser attributes and demographics collected from 8400 users, enabling this research and future research by others.
    \item 
We use survey data to quantify how demographics impact users' likelihood to share their browser data for open research, as well as their perceived understanding and concerns regarding fingerprinting.
    \item 
We show how fingerprinting risks differ across demographics.
    \item 
We identify and demonstrate an overlooked risk: user demographics can be inferred from browser data commonly used as fingerprinting signals.
\end{itemize}

Overall, we show the important role of user demographics in the ongoing work that intends to assess fingerprinting risks and improve user privacy. We provide an initial dataset to support such work, and provide a method to collect datasets for further analyses.


\section{Related Work}
\label{section:related work}

In 2009 Mayer described how Web 2.0 browser features could be used to deanonymize web users, and conducted a small empirical study, collecting fingerprints from 1,328 browser instances~\cite{mayer2009}. 
The 2010 paper by Eckersley was the first large study published on the topic~\cite{eckersley2010} and became the canonical fingerprinting work.  Eckersley collected data from 470,161 browsers whose users visited the Panopticlick\footnote{\url{https://panopticlick.eff.org}} site, most of whom heard about Panopticlick through websites like Slashdot, BoingBoing, Lifehacker, Ars Technica, io9, and through social media channels. Eckersley reported 83.6\% to 94.2\% of the collected fingerprints were unique, depending on the availability of Flash or Java. The study also reported data for 8 specific browser features, calculating the anonymity set sizes and Shannon entropy for each feature. Similarly, the 2016 study by Laperdrix et al.~\cite{Laperdrix2016} collected 17 browser attributes from 118,934 browsers via the AmIUnique\footnote{\url{https://amiunique.org}} site, reporting 90\% and 81\% of the fingerprints were unique for desktop and mobile devices, respectively. The 2018 study by Gómez-Boix et al.~\cite{GomezBoix2018} collected over 2 million fingerprints by deploying a script on two pages of a French news site (weather forecast and political news pages), and reported, in contrast to prior work, only 33.6\% of the fingerprints were unique when using the same 17 attributes as the AmIUnique study.  They also reported metrics on these 17 attributes to further compare to the Panopticlick and AmIUnique studies. Entropy for specific attributes notably differed, potentially due to the different populations they sampled from. For example, the Timezone attribute had substantially lower entropy, which may be expected when collecting data from users on a French news site versus a more geographically distributed sample, as with Panopticlick and AmIUnique. Additional data were collected via the French news site and then used in later analyses on the durability of fingerprints and how fingerprinting can be used for web authentication~\cite{Andriamilanto2021a, Andriamilanto2021b}. The AmIUnique site also recruited participants for additional data collection by asking them to download browser extensions for Chrome and Firefox. Researchers then used this data from 1,905 browsers to study how browser attributes change over time and how fingerprinting algorithms can track users despite these changes~\cite{vastel2018}, as well as how IP addresses can be used to track users~\cite{mishra2020}.

For the most part, related studies have not collected user demographics. An exception is the 2020 study by Pugliese et al.~\cite{pugliese2020}, which collected fingerprinting data from 1,304 participants. Rather than studying fingerprint uniqueness, they studied the long term trackability of users via fingerprinting. They used surveys to collect participants’ demographics as well as responses to questions about users’ perceptions of fingerprinting.
They recruited participants via university mailing lists, press releases, private contacts, and announcements at computer science and security conferences and noted their resulting sample was biased towards German tech-savvy, well educated males (76.5\% male). They found older participants were easier to track over time, yet did not study relationships between demographics and responses to survey questions about fingerprinting due to low survey response rates (n=243). In our study we repurpose two of their survey questions and do study relationships between responses and participant demographics (n=12,461). In contrast to their study, we recruit participants via a crowdworker platform. Our recruitment approach was also taken in a 2022 study~\cite{Chalise2022} which paid 2,064 Amazon MTurk crowdworkers to participate in a study on the effectiveness of web audio fingerprints.
Our specific data collection methods borrow from a recent survey experiment that crowdsourced Amazon users' purchases data~\cite{amazonCSCW2024}. This consent-driven study design gave participants the option to decline to share their sensitive data while still receiving payment for participation, and was designed as an experiment to study which factors impacted participants' likelihood to share the data. They found significant differences in share rates when comparing genders and race groups.

The present work also studies how fingerprinting related privacy concerns differ by demographics, building on studies dating back to 2001 which have shown online privacy concerns differ by demographics~\cite{oneil2001}, particularly by gender, with females expressing higher levels of concern~\cite{gerber2018, Potoglou2015, coopamootoo2022}. There are also reported gender gaps in how users learn about privacy risks and privacy enhancing technologies~\cite{coopamootoo2023}.

\section{Data collection and experiment design}
\label{section:data collection}

We developed a survey tool and designed the data collection process to achieve multiple objectives:
\begin{itemize}
    \item 
Collect users’ web browser data along with their demographics, and only with users’ informed consent.

\item 
Collect responses to survey questions regarding users’ privacy concerns and perceptions, as informed by prior work.
\item 
Conduct a randomized controlled experiment, where a random subset of participants were shown the browser data we asked to collect, while others were simply asked to share the data without the data shown to them. This was to help us study: What impacts users’ likelihood to share browser data for open research? We studied this question in order to help inform future data collection efforts.
\end{itemize}

\begin{figure}
    \centering
    \includegraphics[width=0.35\textwidth]{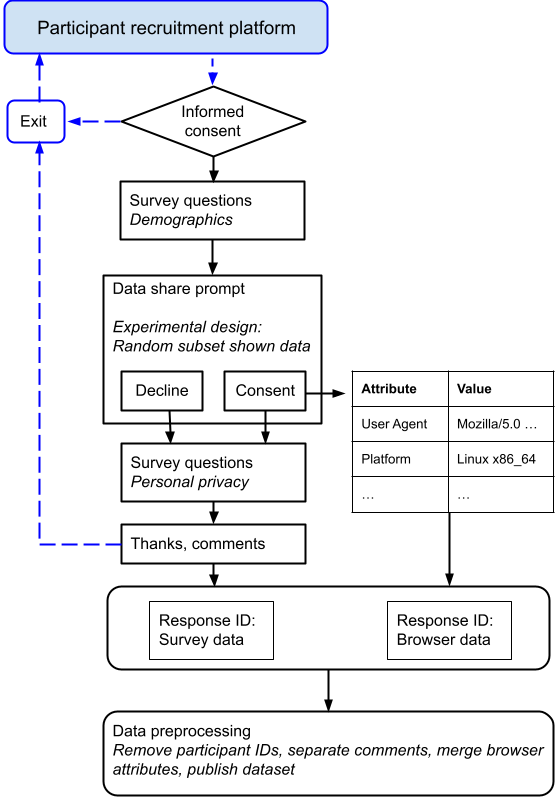}
    \caption{Flowchart summarizing data collection.}
    \label{fig:data_collection_flowchart}
\end{figure}

The data collection described in this section is summarized in Figure~\ref{fig:data_collection_flowchart}. The survey tool is available in our open repository, with more information about it in Appendix~\ref{SI:data collection tool}.

Data were collected in December of 2023. 

\subsection{Ethics and informed consent}
The data were collected and published by the first author, who received approval for the study from their institutional review board (IRB). 
The other authors work at a technology company. They only contributed analysis on the published dataset after approval from their company.

Before beginning the survey, participants were informed they would be asked about their demographics and that they would be asked to share information collected from their browser – the same kind of information used to track users across the web. They were also informed they would have the option to decline to share the browser data and their payment for participation would not depend on that choice. We designed the survey software so that the requested browser data was only collected from participants' devices with their explicit consent. Participants were also informed that any data they shared would be published in an open dataset.  Given this information, they could choose to exit without payment or continue to the survey with informed consent.

Even with participants' informed consent to publish the data, publishing this data may present potential risks to participants. We note Vastel et al. (2018) find browser attributes change frequently, where 50\% of browser instances in their sample changed their fingerprints in less than 5 days, 80\% in less than 10 days~\cite{vastel2018}. In order to lower the potential risk of re-identification, we avoid publishing a precise date of collection, only naming the month and year. 

\subsection{Participants}
Participants were limited to English speaking US residents who were 18 years or older. Participants were recruited via the online platform Prolific\footnote{https://www.prolific.com} with a male/female gender balance and were offered \$0.60 for an estimated 2 minute survey. According to Prolific, the actual median survey time was 1:13 (average \$29.59/hr). Recent studies of users' online behavior and privacy perceptions have similarly recruited US participants from crowdworker platforms and research shows their data generalize to the larger US population~\cite{redmiles2019, tang2022}. A comparison of crowdworker platforms found Prolific provides higher quality data~\cite{douglas2023}. The Appendix (\ref{SI:participants and data collection details}) provides further details.

\subsection{Experiment design}
\label{section: experiment}

\begin{figure}
    \centering
    \includegraphics[width=0.49\textwidth]{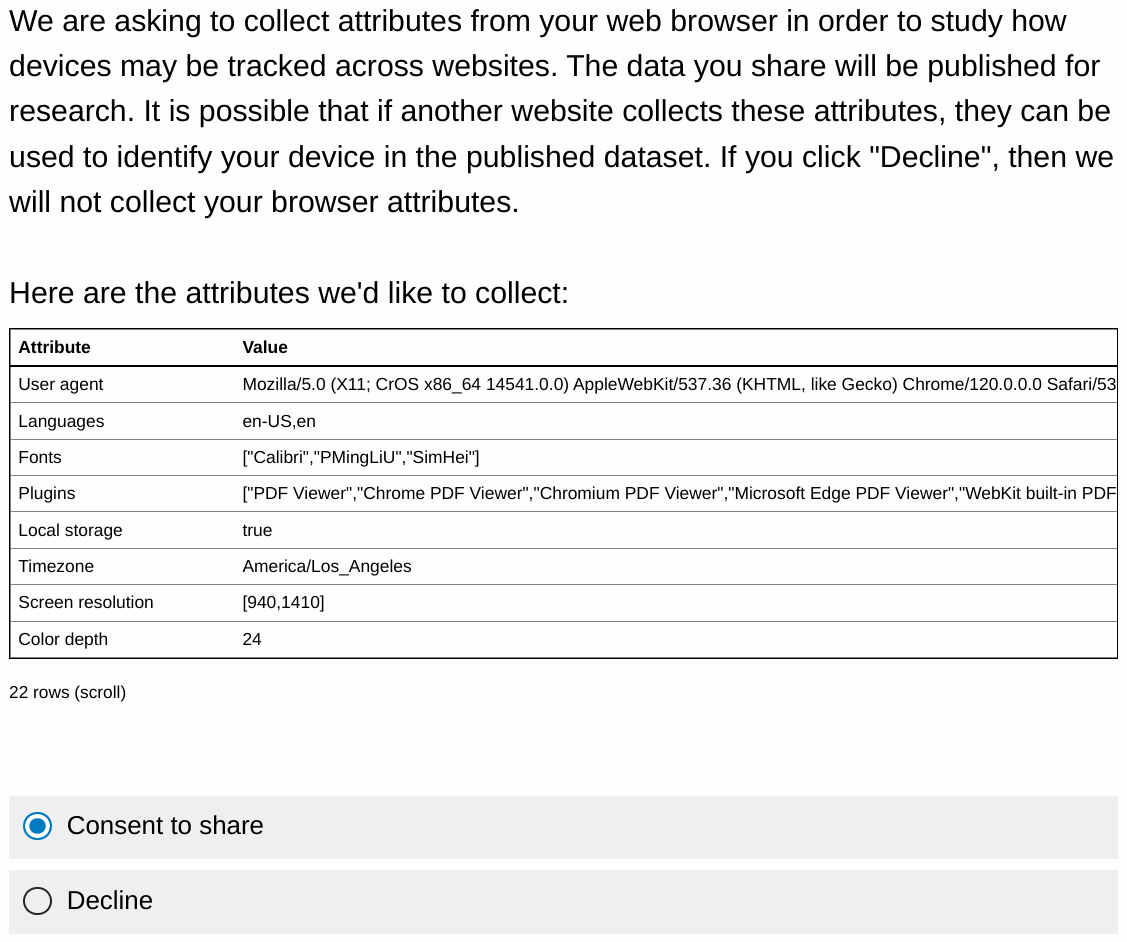}
    \caption{Screenshot from survey displaying experimental treatment that showed participants their browser attributes.}
    \label{fig:survey_screenshot_showdata}
\end{figure}

The survey tool included an experiment to allow studying whether showing participants the browser attributes they were asked to share impacted their likelihood of sharing the requested data. Upon entering the survey, participants had a random 50\% chance of being assigned the ``showdata'' experiment arm.  At the section of the survey where participants were prompted to share their browser data, the random half of participants in the ``showdata'' experiment arm were also shown the browser data in a scrollable interface, as shown in Figure~\ref{fig:survey_screenshot_showdata}. This interface was a table with a row for each browser attribute, including the attribute name and the value queried from the participant's device, which would be collected by the survey software given their consent. The other half of participants were not shown the data. Aside from this section of the survey, the survey was identical for participants regardless of the experiment arm.

\subsection{Survey questions}
After providing informed consent, participants continued to a section of the survey that asked for their demographic information. This included age, household income for the past year (with a "Prefer not to say" option), gender (male/female/other), ethnicity (whether or not of Hispanic origin), and race (multiple choice). The ethnicity and race questions were designed to be both inclusive and provide better comparison to data from the US Census, which collects these as separate demographic attributes. This section also included an attention check in order to improve data quality. Participants who failed this attention check were automatically directed out of the survey and their data were discarded.

Participants then continued to the data share prompt. All participants saw the following text with the options to "Consent to share" or "Decline". 
A random 50\% were also shown the browser attributes data (see Section~\ref{section: experiment}). 

``We are asking to collect attributes from your web browser in order to study how devices may be tracked across websites. The data you share will be published for research. It is possible that if another website collects these attributes, they can be used to identify your device in the published dataset. If you click ``Decline'', then we will not collect your browser attributes."

Participants then continued to a survey section that asked them about their level of perceived understanding of browser fingerprinting as well as concern.  In particular, it presented two statements and asked for participants' level of agreement (Strongly disagree/Somewhat disagree/Neutral/Somewhat agree/Strongly agree).  
Statement 1: “I think that I understand how browser fingerprinting works.” 
Statement 2: “I am concerned that websites and companies try to fingerprint/track my browser.” 
This section was designed to be comparable to the prior browser fingerprinting study by Pugliese et al. (2020) \cite{pugliese2020}, where survey participants were asked for their agreement level to these same statements, with the same agreement level options.
Participants then continued to a section where they were thanked for their time, with a text field to provide optional comments, and were then redirected to Prolific's platform.
The text for all questions and answer options, as well as the survey tool interface, are available via the Appendix (\ref{SI:data collection tool}).

\begin{table*}[h!]
  \caption{Browser attributes used in analysis. }
  \label{tab:browser_attributes_studied}
\small
\begin{tabular}{lllll}
\toprule
Attributes & \makecell[l]{Distinct\\values} & \makecell[l]{\% \\Unique} & Entropy & \makecell[l]{Example (most frequent) value} \\
\hline
User agent & 434 & 2.8 & 4.613 & \makecell[l]{Mozilla/5.0 (Windows NT 10.0; Win64; x64) AppleWebKit/537.36 (KHTML, like Gecko)...}\\
\hline
Languages & 264 & 2.4 & 1.73 & en-US,en \\
\hline
Timezone & 49 & 0.2 & 2.064 & America/New\_York \\
\hline
Screen resolution & 572 & 4.5 & 5.51 & [1920,1080] \\
\hline
Color depth & 3 & 0 & 0.616 & 24 \\
\hline
Platform & 12 & 0 & 2.114 & Win32 \\
\hline
Touch points & 11 & 0 & 1.463 & 0 \\
\hline
\makecell[l]{Hardware concurrency} & 24 & 0.1 & 2.34 & 4 \\
\hline
Device memory & 7 & 0 & 1.611 & 8 \\
\hline
WebGL Vendor & 3 & 0 & 0.465 & WebKit \\
\hline
\makecell[l]{WebGL Unmasked Vendor} & 36 & 0.1 & 3.313 & Google Inc. (Intel) \\
\hline
WebGL Renderer & 36 & 0.1 & 0.782 & WebKit WebGL \\
\hline
\makecell[l]{WebGL Unmasked Renderer} & 654 & 3.2 & 6.833 & Apple GPU \\
\hline
Fingerprint & 5973 & 60.2 & 12.101 & -1872714840963660000 \\
\bottomrule
\end{tabular}
\end{table*}

\subsection{Data collection software and processing}
The survey was implemented using Qualtrics\footnote{Copyright (c) 2023 Qualtrics. Qualtrics and all other Qualtrics product or service names are registered trademarks or trademarks of Qualtrics, Provo, UT, USA.} survey software. We developed a custom integration to query the participants’ browser attributes, display them for participants in the  ``showdata" experiment arm, and send the browser attributes data to Qualtrics only when participants did consent to share them.

The browser attributes collected are described in Section~\ref{section:browser attributes}.

We collected the browser attributes via JavaScript. To collect many of the attributes, we used the FingerprintJS library~\cite{fingerprintjsv3}, which is developed by the company FingerprintJS, Inc. and is the most widely used JavaScript fingerprinting library~\cite{webalmanac2021}. 
We used the open license version (v3) of the library, last updated mid 2023, rather than the more recent version, which does not have an open license. This may present limitations for the dataset, as it may not match what more modern fingerprinting scripts can collect due to more recent optimizations. We wrote additional JavaScript to query for attributes not handled by this library. The Appendix~\ref{SI:browser attributes collected} includes details on the code used to query for each attribute. 

Upon entering the survey, each participant was assigned a random ResponseID to which their responses were then attached. If they consented to share their browser attributes data, a file containing the attribute name and value pairs was then uploaded to the Qualtrics survey, with a filename corresponding to the ResponseID.  Our data preprocessing pipeline then used the ResponseID to attach participants' survey response data to their browser attributes. In addition, the preprocessing dropped any responses that were incomplete, had failed attention checks (less than 1\%), or were from participants living outside the US.

\section{Dataset}

The data contains survey responses from N=12,461 participants, of whom 8,400 shared their browser data, and is provided as two data tables. Both tables contain the survey question responses and are keyed by the same ResponseID.

The first table includes data from all 12,461 participants, and is used for the survey and experiment analysis. It includes a flag indicating whether participants were in the experiment arm where they were shown their browser data (showdata=true). It does not include browser attributes.

The second table contains data for the 8,400 participants who shared their browser attributes data, with a column for each attribute collected.

\subsection{Browser attributes}
\label{section:browser attributes}

The browser attributes used in this analysis are listed in Table~\ref{tab:browser_attributes_studied}. This is a subset of the attributes we collected. The full set is described in the Appendix (\ref{SI:browser attributes collected}).

We selected which attributes to collect based on those in prior related work by Eckersley (2010), Laperdrix (2016), Gómez-Boix et al (2018) and what is used in practice (by FingerprintJS
~\cite{fingerprintjs2024}). Including attributes from prior work enables comparisons across studies. Yet some of the attributes from prior work are outdated or less useful for fingerprinting. In order for browser attributes to be useful for device fingerprinting, they should be both stable, meaning they do not frequently change, and contribute to a fingerprint's uniqueness. 
To better assess real risk, our analyses use a subset of attributes chosen for their relative stability and uniqueness. For example, Plugins were collected in the 2010 study by Eckersley and yielded the highest entropy of the attributes studied. However, browsers have since implemented privacy protective changes to either return random values in the Plugins list (e.g. Brave~\cite{BraveRandomPlugins}) or a hardcoded list (e.g. Chrome~\cite{chromiumPlugins}). This is similarly the case for Fonts~\cite{brave2020}. For this reason, we collect Plugins and Fonts but do not use them in the analyses. 

For each attribute used in this analysis, Table~\ref{tab:browser_attributes_studied} shows the number of distinct values, the percent of users with a unique value, and the Shannon entropy, computed from our dataset. The most frequently occurring value is also provided as an example. Note these values are specific to our dataset; a larger dataset may yield lower uniqueness and higher entropy values. 
In addition to analyzing these 13 attributes separately, we combine and hash them to create an overall ``Fingerprint" for each user, similarly to the FingerprintJS library~\cite{fingerprintjs2024}. Approximately 60\% of users in our dataset have a unique overall Fingerprint, which matches the 60\% accuracy that FingerprintJS advertises~\cite{fingerprintJSAccuracy}.

The W3C, which is the interest group that develops standards and guidelines for the Web, distinguishes between "passive fingerprinting", which uses content in Web requests, versus "active fingerprinting", where JavaScript or other code runs on the local client~\cite{W3Cfingerprinting2019}. 
We collected all browser attributes via client-side JavaScript using methods common among fingerprinting scripts. However, some of these browser attributes are also sent via Web requests in HTTP headers, meaning they can be collected by servers, exposing users to "passive fingerprinting", without detection by browsers. These include the User agent~\cite{fielding2014} and Languages~\cite{alvestrand2002}.

In order to better describe the dataset, we use the Platform attribute to infer each user’s operating system (OS). Table~\ref{tab:os_market_share} compares the sample OS distribution to US market share estimates from StatCounter~\cite{statcounter2023} for December 2023 (the month data were collected). Note the table groups Android, Chrome OS, and Linux because recent Chromium browsers show Linux for all three~\cite{chromium2021}.

\begin{table}[]
\centering
\caption{OS market share distribution.}
\label{tab:os_market_share}
\small
\begin{tabular}{lrrrr}
\toprule
&\multicolumn{2}{r}{\textbf{Sample}} &\textbf{US \cite{statcounter2023}} \\
\cmidrule{2-4}
\textbf{OS} &\textbf{n} &\textbf{\%} &\textbf{\%} \\
\midrule
Windows &4214 &50.17 &35.29 \\
iOS &1568 &18.67 &26.77 \\
Android or Chrome OS or Linux &1526 &18.17 &22.38 \\
OS X &1092 &13 &14.03 \\
Other & & &1.53 \\
\bottomrule
\end{tabular}
\end{table}

\subsection{Demographics}

Table \ref{tab:participant_attributes} shows survey participants' demographic attributes, comparing all participants and those who shared browser data, with US Census data. Participants were required to be US residents and 18 years or older, so when possible we use census data limited to the 18+ population.

\begin{table}
  \caption{Survey participant demographic attributes.}
  \label{tab:participant_attributes}
  \small
\begin{tabular}{l|ll|ll|l}
\toprule
\multirow{3}{*}{ } & \multicolumn{4}{ c }{\textbf{Survey participants}} & \multirow{2}{*}{ \textbf{\makecell{US \\ Census}}} \\
& \multicolumn{2}{l}{\textbf{All}} & \multicolumn{2}{l}{\textbf{Shared data}} & \\
\hline
& N & \% & N & \% & \% \\
\textbf{Total} & 12461 & & 8400 & & \\
\hline
\textbf{Gender} & & & & & \\
Female & 6076 & 48.8 & 3990 & 47.5 & 50.9 \\
Male & 6134 & 49.2 & 4227 & 50.3 & 49.1 \\
Other & 251 & 2 & 183 & 2.2 & \\
\textbf{Age} & & & & & \\
18 - 24 years & 1727 & 13.9 & 1302 & 15.5 & 11.9 \\
25 - 34 years & 4136 & 33.2 & 2859 & 34 & 17.3 \\
35 - 44 years & 3039 & 24.4 & 2024 & 24.1 & 16.8 \\
45 - 54 years & 1803 & 14.5 & 1158 & 13.8 & 15.3 \\
55 - 64 years & 1164 & 9.3 & 712 & 8.5 & 15.7 \\
65 or older & 592 & 4.8 & 345 & 4.1 & 22.9 \\
\textbf{Household income} & & & & & \\
Less than \$25,000 & 1594 & 12.8 & 1097 & 13.1 & 15.5 \\
\$25,000 - \$49,999 & 2724 & 21.9 & 1842 & 21.9 & 17.9 \\
\$50,000 - \$74,999 & 2554 & 20.5 & 1692 & 20.1 & 18.7 \\
\$75,000 - \$99,999 & 1918 & 15.4 & 1267 & 15.1 & 12.1 \\
\$100,000 - \$149,999 & 2042 & 16.4 & 1411 & 16.8 & 15.1 \\
\$150,000 or more & 1363 & 10.9 & 951 & 11.3 & 20.7 \\
Prefer not to say & 266 & 2.1 & 140 & 1.7 & \\
\textbf{Hispanic origin} & & & & & \\
Yes & 1343 & 10.8 & 923 & 11 & 19.1 \\
\textbf{Race} & & & & & \\
White & 8832 & 70.9 & 5911 & 70.4 & 75.5 \\
Black & 1400 & 11.2 & 938 & 11.2 & 13.6 \\
Asian & 1179 & 9.5 & 842 & 10 & 6.3 \\
American Indian & 75 & 0.6 & 53 & 0.6 & 1.3 \\
Native Pacific Islander & 0 & 0 & 0 & 0 & 0.3 \\
Other or mixed & 975 & 7.8 & 656 & 7.8 & 3 \\
\bottomrule
\end{tabular}
\end{table}

We compare our dataset to US census data in order to highlight potential limitations. Our data are biased towards younger participants, and greatly underrepresents the 65+ population~\cite{NC-EST2022-ALLDATA}. Our data also underrepresents people in households making \$150,000 or more each year~\cite{USCensus-HINC-01}, as well as people of Hispanic origin \cite{USCensusQuickFacts}. 

Our survey allowed selecting multiple options for race, including "Other".
When comparing our data to Census data, we report on races "alone" and group all responses with either multiple race selections or "Other" to "Other or mixed" and compare this to census data for "Two or More Races"~\cite{USCensusQuickFacts}. 
Our dataset reports a higher number for "Other or mixed" compared to the census data. In the following analyses, we further group participants identifying as American Indian/Alaska Native or Native Hawaiian/Pacific Islander into the "Other or mixed" category, due to their small numbers. In addition, due to few participants selecting the gender "Other" or income "Prefer not to say" options, we leave these groups out of some analyses.

We also collected participants’ US state of residence, which we compare to 18+ population estimates from the US Census~\cite{census2022SCPRC-EST2022-18+POP}. Our participants' geographic distribution is overall representative of US state populations, with a Pearson correlation of r=0.988 (p<0.001). See the Appendix Table~\ref{tab:participant_states} for details. 

In the following analyses we refer to Gender, Age, Income, Hispanic, and Race as demographic "categories", which are the bold higher level categories in Table~\ref{tab:participant_attributes}, and we refer to the subgroups within these categories as demographic "groups".

\section{Survey and experiment analysis}

In this section of analysis, we use the survey data to address RQ1: What impacts users’ likelihood to share browser data for open research?
In particular, we assess the impact of users' demographics, their understanding and concerns regarding fingerprinting, and whether they were shown the browser data we asked to collect. We further analyze responses to the survey questions about participants' understanding and concerns regarding fingerprinting. We first present overall numbers, and then use multivariate regression analysis to assess the impacts of the various factors we study.

\subsection{Overall survey results}

Overall, 67.4\% of survey participants shared browser data, where the share rate was slightly lower for participants shown their data (65.8\% vs 69.1\%; p<0.05).

In addition to the main analyses, we ran a simple statistical test to probe whether participants inspected their data when shown to them. We find that, on average, participants shown their data had significantly longer survey duration times versus participants not shown their data, suggesting participants spent time inspecting their data. See Appendix~\ref{SI:survey_durations} for details.

Figure~\ref{fig:statement_responses} displays the responses to the survey questions about participants' understanding and concerns of fingerprinting. Overall, more than 70\% of participants expressed concerns about browser fingerprinting/tracking.  However, only 43\% of participants said they understand how browser fingerprinting works, with a small minority (6\%) strongly agreeing with that statement.  These results suggest an opportunity to educate concerned users about fingerprinting. The following analyses highlight how these results differ across demographic groups.

\begin{figure}
    \centering
    \includegraphics[width=0.48\textwidth]{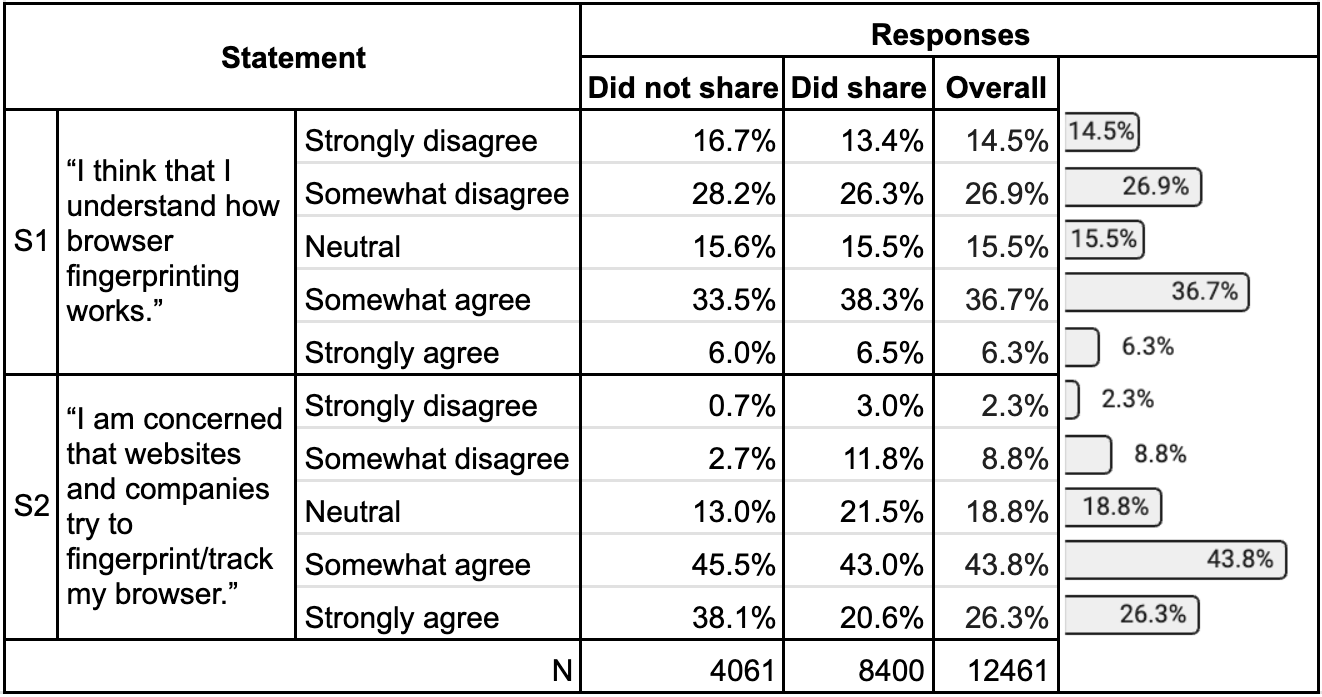}
    \caption{Participants' responses for statements S1 and S2. See Table~\ref{tab:regression_results_consolidated} for a consolidated statistical analysis of statement responses and sharing behavior.}
    \label{fig:statement_responses}
\end{figure}

\subsection{Multivariate regression analysis}
\label{section:survey regression analysis}

In this section we use logistic regression to conduct a multivariate analysis, assessing which factors impacted participants' likelihood of sharing their browser data, as well as their responses to survey questions.

We refer to the survey questions about participants' level of understanding and concern regarding fingerprinting as S1 and S2, respectively (see Figure~\ref{fig:statement_responses}).
Responses to S1 and S2 are coded as booleans in the regression models, where 1 indicates agreement ("Somewhat agree" or "Strongly agree"). Whether participants shared their browser data, and whether participants were in the experiment arm that showed them their browser data, are also coded as booleans, where share=1 indicates the participant shared their browser data and showdata=1 indicates they were shown their data.

We analyze three models, one for each survey question, where the dependent variable is S1 or S2, and one for studying participants' likelihood to share, where the dependent variable is whether they shared.
Table~\ref{tab:regression_results_consolidated} shows a consolidated view of the regression results, reporting the odds ratio (OR) for each explanatory variable. The OR is interpreted as the impact of each such explanatory variable on the dependent variable, in comparison to the reference variable, where an OR > 1 indicates the variable increased the odds.
The full set of model results with 95\% confidence intervals, along with robustness checks, are in Appendix~\ref{SI: survey regression analysis}.
While our results show that variables we tested for have statistical significance, the models are limited by the variables we collected; much of the variance in the data is unexplained by the models.
For this reason, the OR magnitudes should not be taken as exact values and we focus on whether variables had a positive versus negative impact.

\begin{table}
  \caption{Regression results analyzing participants' survey responses (S1, S2) and likelihood to share browser data (Share).}
  \label{tab:regression_results_consolidated}
  \small
  \begin{tabular}{r|ccc}
  \toprule
    & \multicolumn{3}{c}{\textbf{Odds Ratio (OR)}} \\
    \textbf{Predictor} & \textbf{S1 Agree} & \textbf{S2 Agree} & \textbf{Share} \\
\midrule
    Intercept & 1.129** & 3.626*** & 3.629*** \\
    \textbf{Gender (Ref: Male)} & & & \\
    Female & 0.515*** & 1.012 & 0.909* \\
    \textbf{Age (Ref: 35 - 44 years)} & & & \\
    18 - 24 years & 0.721*** & 0.803** & 1.517*** \\
    25 - 34 years & 0.858** & 1.004 & 1.139* \\
    45 - 54 years & 1.076 & 1.272** & 0.959 \\
    55 - 64 years & 1.038 & 1.582*** & 0.888 \\
    65 and older & 0.774** & 1.694*** & 0.795* \\
    \textbf{Race (Ref: White)} & & & \\
    Asian & 0.794** & 1.419*** & 1.152* \\
    Black & 1.324*** & 0.975 & 0.959 \\
    Other or mixed & 1.089 & 1.236** & 0.951 \\
    \hline
    share & & 0.331*** & \\
    showdata & & 0.983 & 0.871** \\
    S1 agree & & 1.871*** & 2.648*** \\
    S2 agree & & & 0.41*** \\
    S1 x S2 agree & & & 0.472*** \\
    \hline
    N & 12210 & 12210 & 12210 \\
    pseudo R-squared & 0.024 & 0.059 & 0.05 \\
    \bottomrule
    \end{tabular}
\end{table}

The results show statistically significant differences across gender, age, and race in participants' likelihood to share their browser data or agree with statements S1 and S2. In particular, females were significantly less likely 
to agree they understand web browser fingerprinting versus their male counterparts. They were also less likely to share their browser data, even when this lack of perceived understanding (variable S1) is controlled for in the model. When assessing which factors impacted participants' concern with fingerprinting/tracking (S2), there is a monotonic increase corresponding to age: the older participants are, the more likely they are to be concerned. Participants who agreed they understood fingerprinting were also more likely to express concern with fingerprinting/tracking.

When assessing what impacted participants' likelihood of sharing the browser data, we find showing the data had an overall negative impact. Younger participants were more likely to share their data, and as to be expected, participants who expressed concern about fingerprinting/tracking were less likely to share their data. Whether participants agreed they understood fingerprinting had a positive impact on sharing, yet when considering participants who both agreed they understood and were concerned regarding fingerprinting, there is an overall negative effect (seen in the interaction effect S1xS2).

\section{Fingerprinting risk and demographic differences}
\label{section: fingerprinting risk}

The analysis in this section addresses RQ2: Are some demographic groups more at risk than others?

\begin{figure*}
    \centering
    \includegraphics[width=0.7\textwidth]{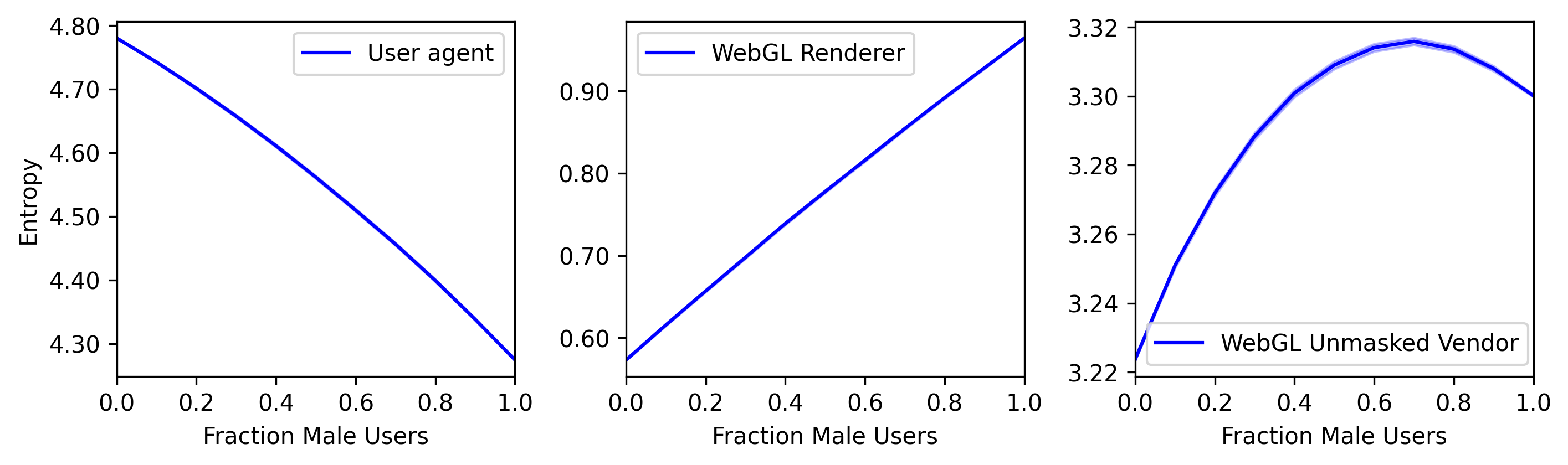}
    \caption{Entropy estimates for varying gender ratios.}
    \label{fig:entropy_changes_with_gender_3_attrs}
\end{figure*}

To motivate this analysis, we briefly demonstrate how data representativeness can impact analyses that intend to assess fingerprinting risk.  Prior studies assessed fingerprinting risk by computing Shannon entropy for the browser attributes they collected~\cite{eckersley2010, Laperdrix2016, GomezBoix2018, Laperdrix2020}; however, the demographics of their users were unknown. Authors of the large Panopticlick and AmIUnique studies noted likely demographic biases in their datasets as limitations~\cite{eckersley2010, Laperdrix2016}.  A later study by Pugliese et al.~\cite{pugliese2020}, which similarly reached participants interested in privacy, yet also collected demographics, showed a strong male bias (76.5\% male), raising the question of whether the prior studies suffered from gender bias as well. Here we show how such biases can potentially impact prior results by recomputing entropy metrics with our dataset, which includes browser attributes from the prior studies, as well as the demographic data they lacked.

We use our dataset to show how if our sample were skewed towards male participants, results could be different.
We simulate scenarios going from a sample of 0 to 100\% male users, and compute Shannon entropy for each attribute in our dataset over this range (n=4000, drawn with repeated random sampling). Figure~\ref{fig:entropy_changes_with_gender_3_attrs} shows the resulting changes in entropy estimates for the User agent, WebGL Render, and WebGL Unmasked Vendor attributes. These results estimate a more than 10\% decrease in entropy for User agent and 68\% increase in entropy for WebGL Renderer, when comparing 0 and 100\% male samples. The User agent and WebGL Renderer attributes are of interest because they have been featured in the prior fingerprinting studies which measured their entropy~\cite{eckersley2010, Laperdrix2016, GomezBoix2018, Laperdrix2020} and particular attention in anti-fingerprinting browser development has gone to reducing entropy for the User agent~\cite{senol2023, chromiumUserAgentReduction}. They also represent different types of attributes.  User agent is related to the browser installation and entropy is higher for female users in our data, while WebGL attributes are related to graphics hardware and tend to be more unique for males in our data. The varying entropy estimates for the WebGL Unmasked Vendor attribute, which peak between the extremes of 0 and 100\% male, demonstrate how entropy may sometimes be highest when the user sample is more balanced or diverse. See the Appendix (\ref{SI:entropy changes with gender}) for results for the other attributes. 
This analysis is not meant to communicate a statistical result, only to motivate further analysis.

\subsection{Metrics and sampling methods}

Previous work used entropy to assess fingerprinting risk across entire datasets. In this work we instead compare fingerprinting risk between groups of users in our dataset. In order to do this, we use the following metrics and sampling methods.

\subsubsection{Metrics}

\paragraph{Average Anonymity Set Size:}
An ``anonymity set" refers to the set of users sharing a distinct value for a given attribute, or overall fingerprint.  This metric was used by Eckersley (2010) to compare browser attributes~\cite{eckersley2010}. For greater intuition, see Appendix Figure~\ref{fig:anonymity_set_sizes}. For each attribute, and the overall fingerprint, we assign each user their ``anonymity set size", which is the number of users in their anonymity set (k).  For a group of users, we compute the group “average anonymity set size” as the mean over the anonymity set sizes (k) of the users in the group. When computing this metric for a specific group, an anonymity set can include any users in the sample, not just users in the group.

Smaller average anonymity set sizes indicate greater risk: Even if an attribute alone is not uniquely identifying, if it puts a user in a smaller anonymity set, then it may be easier to use this attribute in combination with others to uniquely identify the user.

\paragraph{\% Unique:}
For a group of users, we compute the percent that are uniquely identified by a given attribute, or overall fingerprint (i.e. anonymity set size is k=1). Whether users are counted as unique is determined by including users outside the specific group.
A higher percent unique indicates greater risk.

Note these metrics are not scale invariant. As sample size increases, average anonymity set size tends to increase linearly, while \% unique tends to decrease towards a limit (see Appendix~\ref{SI:fingerprinting metrics scaling}).

\subsubsection{Sampling strategy to make comparisons between groups}
This analysis compares fingerprinting risks for demographic groups by comparing the percent unique and average anonymity set size metrics across groups. To do this, we take an extra step to handle that our sample contains different numbers of users in each group. Consider age comparisons. If the 65 or older group has a higher \% unique metric, is that because they are truly easier to fingerprint or because they are a smaller group in the dataset? Given that browser attributes may be related to demographics, and our metrics are not scale invariant, we adopt a sampling strategy to make more robust comparisons. In particular, we use repeated random sampling to create subsamples that contain  an equal number of users from each demographic group. We then calculate the metrics from these subsamples.

We do the following process separately for each demographic category (Gender, Age, Income, Hispanic, Race).  We randomly sample an equal number of users from each group in the demographic category, adding up to n=1800. For example, for Age, we sample 300 users from each of the 6 age groups. We sample without replacement because we are estimating metrics sensitive to uniqueness. For each such subsample, we compute the \% unique and average anonymity set size metrics, and repeat this process 1000 times to compute means and 95\% confidence intervals for each metric. We then only consider the resulting metrics as different when their confidence intervals do not overlap.

We also compute the metrics without the sampling process. When we evaluate the results, we only make statements that are consistent across the results from the sampling process and metrics computed without the sampling process.

For an example of why a sampling method is necessary, consider Hispanic users.  The US Census estimates Hispanics represent 19\% of the population and they represent 11\% of our sample. Without resampling, Hispanics have a higher \% unique metric for the overall fingerprint versus non-Hispanics (63\% vs 60\%). Yet when computed from subsamples with equal portions Hispanic and non-Hispanic, the \% unique values are the same (see Appendix).

\subsection{Results}



\begin{figure*}
    \centering
    \includegraphics[width=0.75\textwidth]{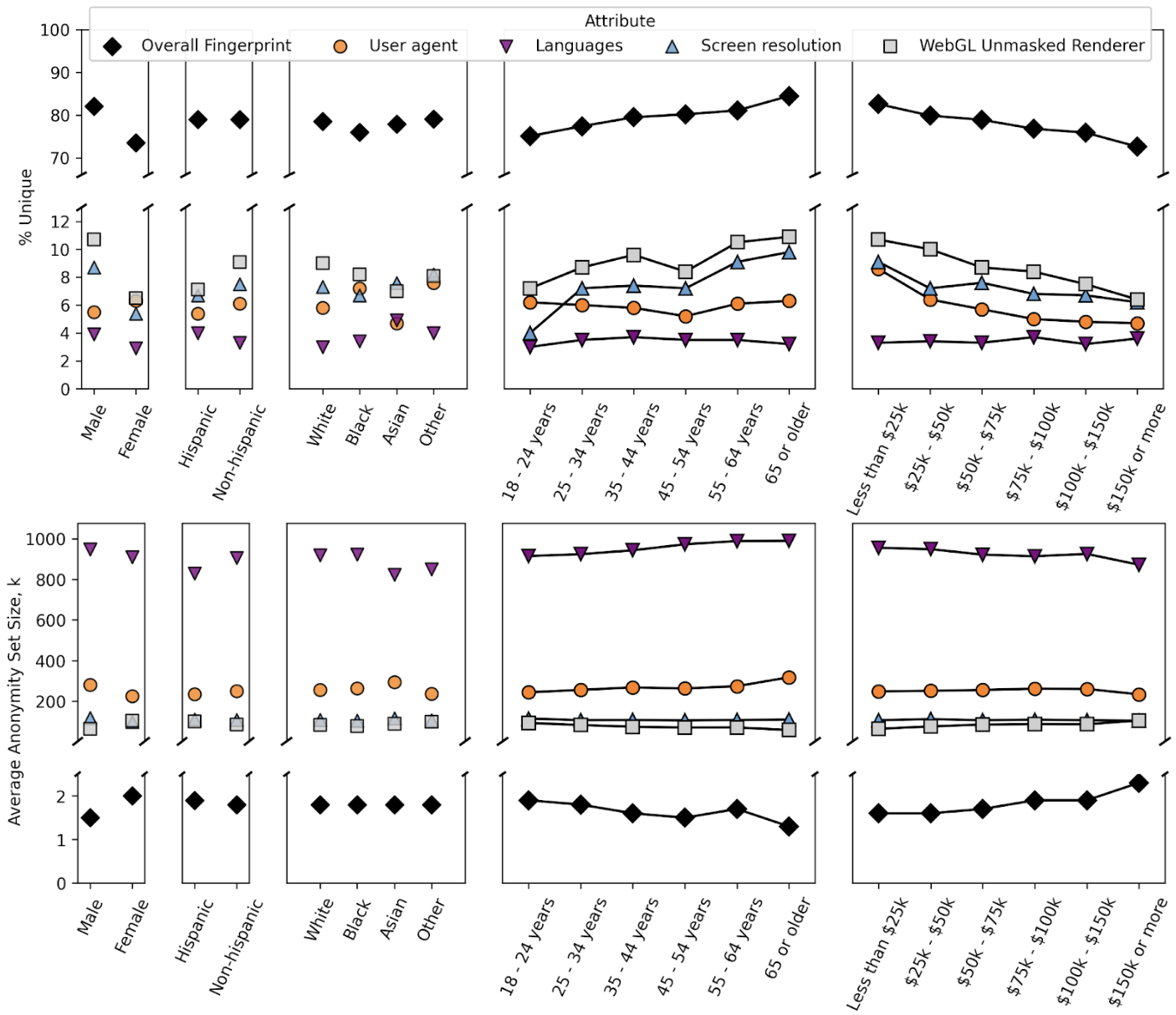}
    \caption{(Top) \% unique and (bottom) average anonymity set size by demographic group.}
    \label{fig:pct_unique_avg_k_by_demo_group}
\end{figure*}

Results comparing the \% unique and average anonymity set size values are shown in Figure~\ref{fig:pct_unique_avg_k_by_demo_group}. Only the 4 most uniquely identifying browser attributes, where \% unique is more than 1\%, are plotted (see Table~\ref{tab:browser_attributes_studied}), along with the overall fingerprint created by combining all 13 attributes in the study. Tables with the metrics for all browser attributes in this study, for each demographic category, both with and without the sampling process, are in the Appendix (\ref{SI:fingerprinting_metrics}).

The most uniquely identifying attributes in our study are User agent, Languages, Screen resolution, and WebGL Unmasked Renderer.  The User agent and Languages attributes are particularly sensitive because they are sent in Web request HTTP headers, enabling passive fingerprinting by servers without detection by browsers.  Also, the Languages attribute has higher \% unique and lower average anonymity set sizes for Hispanic (vs non-Hispanic) and non-White users, putting these minority groups at greater risk.

Male users in our data are more at risk of overall fingerprinting, which may be partly due to their more unique Screen resolution and WebGL Unmasked Renderer attributes. Yet female users have more unique User agent values, potentially putting them at greater risk of passive fingerprinting.
When comparing race groups, Black users have slightly lower \% unique fingerprint values but we do not find a difference in their average anonymity set sizes. Income and age both show monotonic trends. When considering both \% unique and average anonymity set size, lower income groups are at greater fingerprinting risk, higher income groups are less at risk, and there is a monotonic decrease in fingerprinting risk as income increases. Similarly, there is a monotonic increase in \% unique fingerprints with age, where the oldest users are at greatest risk. The findings reported above are consistent across the metrics computed both with and without the sampling process.

\section{Risk of demographics inference}
\label{section:inferring demographics}

In this section we study and demonstrate how browser attributes present a risk beyond uniquely identifying users. This analysis addresses RQ3: Can sensitive demographic information be inferred from device fingerprinting signals? If users in large anonymity sets are considered protected by k-anonymity~\cite{kAnonymity1998}, this risk may be understood as a homogeneity attack: a k-anonymous group can leak demographic information about users due to lack of diversity~\cite{ldiversity}.

To make this risk more intuitive, we provide examples from our data: Hispanics represent only 11\% of our sample yet represent more than 45\% of the users with "es-US" in their Languages attribute. The Languages attribute may then be useful for inferring whether users are Hispanic. Similarly, the Device memory attribute may be useful for inferring user income, where users in lower or higher income groups may be more likely to buy devices with lower or higher memory. Users with household income under \$50k/year represent only 35\% of our sample, yet represent more than 60\% of users with ``Device memory"=2.0. Device memory is an informative example because this attribute's API was designed to curtail fingerprinting, rounding values to the nearest power of 2~\cite{w3cDeviceMemory, mozillaDeviceMemory}. Yet by helping group users by demographics, this may still put users at risk, where the risk is demographic inference rather than fingerprinting.
While these examples may be intuitive, there are less intuitive relationships between demographics and device attributes that take many possible values, and where the values are less human-interpretable, such as for the WebGL and User agent attributes.

Demographic inference poses potential risks to users as well as advertisers and ad serving platforms. Consider the case where a large portion of users sharing a fingerprint or attribute value are in a specific demographic group at risk of targeting (e.g. Hispanic users with es-US in their Languages). While an individual might not be identified, a site can better target the demographic, creating opportunities for price discrimination, targeted disinformation or other problematic and even inadvertent bias.
For example, in the 2016 US presidential race, Cambridge Analytica allegedly used Facebook's platform in a voter suppression campaign targeting a disproportionately Black group of users, potentially harming both Black voters and Facebook \cite{washpoCambridgeAnalytica2020, guardian-fb-fined}. 
In the US there are regulations that prohibit preferencing ads for housing (the Fair Housing Act)~\cite{fairhousingact24CFR} or employment~\cite{titleVII1964employment} based on demographics, such as gender or age, and in 2018 Facebook was sued for allegedly enabling advertisers to violate the Fair Housing Act~\cite{FBfairhousingactlawsuite}.
Targeting US employment and housing opportunity ads based on specific user demographics is now prohibited by the major ad serving platforms Facebook~\cite{FBAdPolicies} and Google~\cite{GoogleAdPolicies}.
Despite these policies, researchers have demonstrated how employment ads on platforms such as Facebook can still target US audiences based on these protected demographic categories, either intentionally or inadvertently, due to features not made clear in ad serving campaigns~\cite{kaplan2022}.
This presents an example of how problematic demographic biases can emerge in ad delivery systems, breaking the system's own policies.
Ad serving platforms often use machine learning models to optimize ad delivery, where these models leverage a variety of signals collected from users~\cite{FBmlAdDelivery}.
If these include signals predictive of demographics (e.g. browser attributes), ad platforms may unintentionally increase the risks targeted ads can pose.

In this section we first demonstrate how the web browser attributes commonly collected as browser fingerprinting signals can be used to predict demographics with machine learning.
Second, we measure information gain about demographics given the browser attributes. The machine learning analysis is used to demonstrate the risk. The second analysis is used to evaluate which browser attributes present the greatest risk, and for which demographics. Doing so can help inform privacy enhancing browser developments as well as improve the interpretability of results.

In the analysis below, we collapse age and income categories from 6 groups to 3 groups in order to provide larger class sizes for the machine learning models. Note we do not train models for the gender: “Other” and income: “Prefer not to say” groups because these groups are too small, but these users are still included in the data used to train models to predict other groups.

\subsection{Machine learning prediction}
\label{section:ml prediction}

\subsubsection{Methods}
The goal of this analysis is not to create an optimal model.  The goal is to assess whether a simple model, using our limited set of browser attributes, can predict user demographics. If this is possible, a company or platform that collects these signals along with more signals may more easily predict demographics.

\paragraph{Inputs:} The model inputs are the browser attributes, which are treated as categorical variables and one-hot-encoded. Many of the browser attributes have a large number of possible values. To handle this challenge we only encode values that appear at least k=100 times in the data, where k is treated as a hyperparameter (i.e. if a user has a browser attribute with a value occurring less than k times, this is represented in their feature vector as all zeros).

\paragraph{Models:} We develop a separate classifier model for each demographic group to formulate a binary prediction task. 
The models only differ in that they were trained with different labels.
For example, to evaluate the risk of inferring a user’s race is Asian, the model is trained to predict a boolean valued label: 1=Asian; 0=not Asian.
The classifiers have a consistent model architecture and set of hyperparameters: We use a fully connected neural network (multilayer perceptron classifier) with 3 hidden layers (sizes 32, 8, 2), a batch size of 1024, and the identity activation function. These hyperparameters were determined by using training data in a grid search over the parameter space with cross validation.

\paragraph{Training and testing:} We cut a 80/20 train/test sample and for each demographic group, $d$, we do the following process: Assign binary labels indicating if each user is in group $d$, train a classifier to predict $d$, and report results using test data.

\paragraph{Evaluation:} We report results using the area under the receiver operating characteristic curve (AUROC)~\cite{bradley1997}.  The metric is a standard way to evaluate machine learning classifiers over all classification thresholds and it offers the following interpretation: Consider sampling a random person A in demographic group $d$, and a random person B not in $d$. The probability that the model outputs a higher value on input A as compared to input B is the AUROC. Models with AUROC greater than 0.5 are more informative than random chance.

\subsubsection{Results}

\begin{table}
  \caption{Machine learning demographic prediction results.}
  \label{tab:ml_results}
\small
\begin{tabular}{l|lr}
\toprule
\multicolumn{2}{c}{\textbf{Demographic group}} & \textbf{AUROC} \\
\hline
\multirow{2}{*}{ \textbf{Gender} } & Male & 0.663 \\
& Female & 0.679 \\
\hline
\multirow{3}{*}{ \textbf{Age} } & 18 - 34 years & 0.639 \\
& 35 - 54 years & 0.547 \\
& 55 or older & 0.644 \\
\hline
\multirow{6}{*}{ \textbf{\makecell[l]{Ethnicity \\and race} }} & Hispanic & 0.6 \\
& Non-hispanic & 0.618 \\
& White & 0.572 \\
& Black & 0.677 \\
& Asian & 0.698 \\
& Other or mixed & 0.583 \\
\hline
\multirow{3}{*}{ \textbf{Income} } & Less than \$50,000 & 0.605 \\
& \$50,000 - \$99,999 & 0.516 \\
& \$100,000 or more & 0.617 \\
\bottomrule
\end{tabular}
\end{table}

Results are shown in Table~\ref{tab:ml_results}.
All models had AUROC scores above 0.5, indicating predictive power, but to varying degrees.
For inferring race, model results suggest risk is highest for Asian, then Black users, and lowest for White users. The “Other or mixed” race group had relatively lower AUROC scores versus the other minority groups, which may be expected considering this is a combination of groups who alone might have different attributes. For age, whether someone was in the oldest age group was relatively easier to infer. This greater risk for older users is consistent with results showing they are more likely to have unique fingerprints, underscoring the finding that they are also more likely to express concerns about fingerprinting. The model results also indicate that whether users are in lower or higher income groups can be inferred and overall, gender was an easier demographic to infer.

The models were designed to be simple, trained with only 13 browser attributes as inputs, yet still demonstrate predictive power for each demographic group. These results suggest how a more complex model using browser data with more types of signals may present risk for downstream users.

\subsection{Mutual information analysis}

Prior related work used entropy to quantify the level of identifying information in browser attributes, where higher entropy indicated more information, and hence more risk~\cite{eckersley2010, Laperdrix2016, GomezBoix2018, Laperdrix2020}.  We build on this prior work by using another well known measure from information theory to quantify the "information gain" about user demographics given browser attributes. The basic idea is that browser attributes and demographics can be considered random variables, which we show are not independent. We calculate the mutual information between these variables to compare the amount of information each browser attribute provides about demographics.

\subsubsection{Methods}
Mutual information measures the amount of information one random variable contains about another, and has been described as the reduction in the uncertainty of one random variable due to the knowledge of the other~\cite{cover1991}. Here we measure the reduction in uncertainty of demographics due to knowledge of browser attributes.

Consider the following as random variables:  $D$ as Demographic category (e.g., income) and $A$ as Browser attribute (e.g., Device memory). We calculate $I(A; D)$ as is the mutual information between $A$ and $D$:

\begin{displaymath}
    I(A; D) = \sum_A\sum_D{p(a,d) \log\left(\frac{p(a,d)}{p(a)p(d)}\right)}
\end{displaymath}

Mutual information is a non-negative metric, where a value of 0 indicates variables $A$ and $D$ are independent. Mutual information can also be expressed in terms of conditional entropy: 

\begin{displaymath}
    I(A; D) = H(D) - H(D|A)    
\end{displaymath}

Our analysis uses a normalized form of mutual information \cite{Vinh2010} in order to better handle the varying distributions for the different demographic categories. We calculate:

\begin{displaymath}
    \frac{I(A; D)}{H(D)} = 1 - \frac{H(D|A)}{H(D)}
\end{displaymath}

\noindent Where $H(D)$ is the marginal entropy of $D$, and $H(D|A)$ is the conditional entropy. $\frac{H(D|A)}{H(D)}$ represents the amount of uncertainty about $D$ after the attribute value, $A$, is known, normalized by the overall uncertainty of $D$. The normalized mutual information is therefore always between 0 and 1, with larger values indicating a stronger relationship between the demographic category and the browser attribute.

Given many of the browser attributes have infrequent values seen for only a few users, we define a parameter k and in the analysis for each browser attribute, we ignore user data corresponding to attribute values appearing fewer than k times.  We present and analyze results for k=100, consistent with the machine learning prediction task, which limits the one-hot-encoding of browser attributes to values with at least k=100. We also repeat this analysis for k=1 and k=50, with the full results and details in Appendix~\ref{SI:mutual_information}.

\subsubsection{Results}

\begin{figure}
    \centering
    \includegraphics[width=0.48\textwidth]{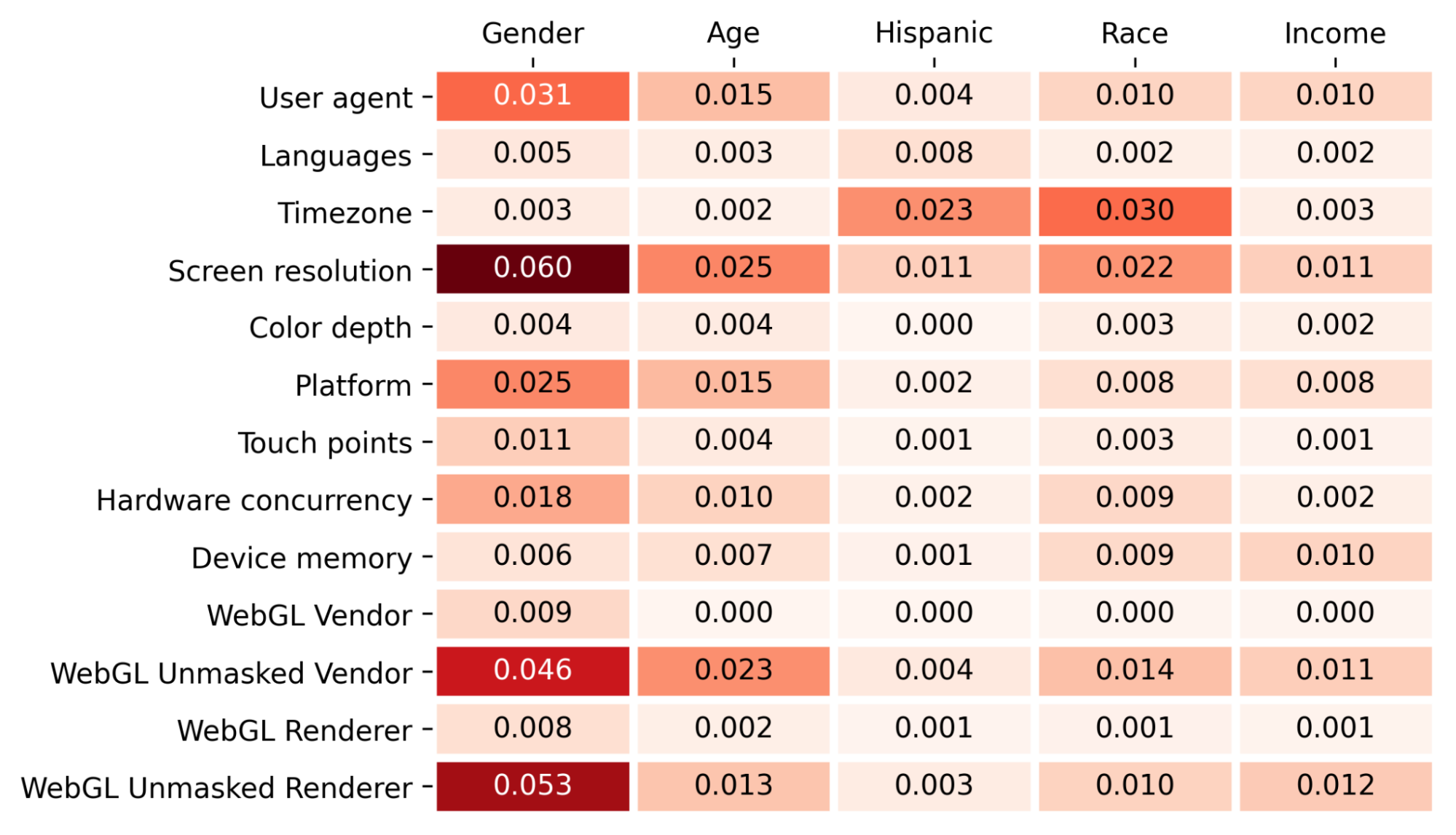}
    \caption{Mutual information analysis results.}
    \label{fig:mutual_info_results_k100}
\end{figure}

Results are shown in Figure~\ref{fig:mutual_info_results_k100}. Since the entropies for demographic categories differ, we do not make direct comparisons across demographics when interpreting these results. We instead compare values for browser attributes within each demographic category (along columns in Figure~\ref{fig:mutual_info_results_k100}), identifying which attributes are relatively more informative.

Overall the most informative attributes, for all demographic categories, are Screen resolution, User agent, and the WebGL unmasked attributes.  These attributes also had the smallest average anonymity set sizes, putting users at greater risk of both fingerprinting and demographic inference, compared to other attributes. Other browser attributes are more informative for specific demographic categories. For example, Device memory is relatively more informative for Income and Languages is relatively more informative for Hispanic, while this is not the case for other demographic categories. Note Languages was also relatively informative for race when the parameter k was smaller, partly because there are a small number of Languages values with a large fraction of Asian users (see Appendix~\ref{SI:mutual_information}). Other attributes add nuance to the findings regarding fingerprinting risk. For example, the Platform attribute has relatively higher mutual information for gender and age, yet presented minimal risk in the analysis of Section~\ref{section: fingerprinting risk}.

A potential limitation of these results is that they may be specific to the timing of data collection. For example, if short-term popularity of a device varied by demographic group, this may be reflected in shorter-term correlations between demographics and attributes specific to the device. Even so, these results highlight how relationships between attributes and demographics can present risks which browser developers should consider.

\section{Discussion}

\subsection{Survey data enhance analysis of device data}
In this work we collected a first-of-its-kind open dataset via a tool that included a survey. 
The resulting dataset includes previously studied browser attributes, along with users' demographics and other survey question responses. This combination of features allowed us to study novel research questions about relationships between demographics and browser attributes (RQ2 and RQ3), as well as participants' understanding and concerns regarding fingerprinting. Beyond asking new research questions, the demographic data also enable analyses that control for potential demographic biases.

In the process of data collection, we also conducted an experiment studying which demographic groups were more likely to share the browser data we requested, along with other factors impacting that decision. Results from this experiment, in combination with the analysis of browser attributes, cast important light on past work, and can inform future data collection efforts. In particular, researchers should be wary of whose data are not included in a dataset, and how resulting biases may impact research results. For example, our study found females were significantly less likely to share their data versus males. We also found important differences between genders with respect to fingerprinting risks.  While our study recruited participants from a balanced pool of male and female participants, some prior works recruited participants through potentially biased channels \cite{eckersley2010, Laperdrix2016}. These prior studies may then report results that have important gender biases, yet where bias is unknown given demographics were not integrated into the analyses. This concern might also apply to more recent work measuring entropy on the web using data from Chrome users who enabled the setting to "Share usage reports and crash analytics with Google"~\cite{Googlers2024}. If female users were less likely to opt in to such data sharing, consistent with our experiment, such results may also be biased.

Our survey also highlighted an opportunity to educate users. More than 70\% of participants said they were concerned about fingerprinting/tracking, yet only 43\% said they understand it. Our study also found specific demographic groups, namely females, were significantly less likely to say they understand fingerprinting. As developers continue efforts to reduce fingerprinting risks, our survey suggests they should also put more attention towards equitably educating users about fingerprinting.

In addition, our experiment analysis found that users who were shown their browser attributes were less likely to share the browser data, even when controlling for whether they said they understood or were concerned about browser fingerprinting. This may suggest that many users are not fully aware of the fingerprinting data collected from their devices as they browse the web, and would engage with the web differently if made more aware.  Future studies might further explore how better educating web users about fingerprinting, or otherwise increasing their awareness by showing what fingerprinting signals are collected, impacts users' privacy related behaviors.

\subsection{Risks differ by demographic group}
In this study we asked: (RQ2) Are some demographic groups more at risk than others?
Throughout the analyses we found this was the case. This included interesting consistent trends for age. Fingerprinting risks increased with age, and so did concern about fingerprinting; older study participants were both more concerned and more at risk. We also found monotonic trends regarding income, where fingerprinting risk increased as users' household income decreased, with the lowest income users at greatest risk.

Our analysis also demonstrated different types of risks for different demographic groups. For example, males had overall more unique fingerprints, yet females may be more at risk of passive fingerprinting, where data in web requests (e.g. User agent and Languages) can be used to fingerprint users without detection by browsers. Our results also show how specific attributes present different amounts of risk for different groups. For example, the Languages attribute presents higher risk for Hispanic (vs non-Hispanic) and Asian and "Other" race groups (vs White), by putting these users in smaller anonymity sets on average. Some browsers have already developed changes to reduce the fingerprinting risk from Languages~\cite{Brave2022}. As browser developers continue to find ways to improve user privacy, this study suggests they might further ask: "Privacy for whom?"

\subsection{Fingerprinting risks go beyond tracking}

In this study we showed that beyond tracking users, device attributes used as fingerprinting signals can also help infer user demographics (RQ3). This puts users at risk of targeted information campaigns, and puts advertisers and ad serving platforms at risk of violating regulatory policies. These results highlight an overlooked risk and also add nuance to (a) the evaluation of risk for demographic groups, and (b) how browser developers should prioritize changes that reduce fingerprinting risk.

For (a), consider that Black users in our analysis had slightly less unique fingerprints versus other race groups, suggesting lower risk. However, the machine learning analysis (Section~\ref{section:ml prediction}) showed Black users are at relatively higher risk of having their race inferred from browser attributes. This demonstrates the value of considering these different types of analyses together.   While some demographic groups (e.g. Black users) may be in larger anonymity sets on average, if these sets consist of more users from that same group (e.g. Black users), then while this might help protect these users from unique identification and tracking, it may put them at greater risk of demographic inference. Future work might more closely inspect the demographic distributions of anonymity sets.

For (b), the mutual information analysis helped identify browser attributes that present fingerprinting risks that the analysis in Section~\ref{section: fingerprinting risk} alone would have overlooked. For example, the Platform attribute had relatively low \% unique values, yet relatively high mutual information values for the age and gender demographics.  This analysis also demonstrated why previous approaches to reduce fingerprinting risk by reducing entropy, or the number of potential values a browser API can yield, are not enough.  The Device memory attribute provides an example, with an API designed to curtail fingerprinting by returning values rounded to the nearest power of 2~\cite{w3cDeviceMemory}. Yet our analysis found this attribute risks grouping users by income level, exposing them to demographic inference. 
Our analysis suggests future browser developments consider a balance between minimizing fingerprinting risk and the likelihood of grouping users by demographics.

\subsection{Limitations and future work}

While this work identifies risks, a limitation is that it does not explore risk mitigation strategies or solutions to address them. That is beyond the scope of this work. We hope by providing the dataset collected through this work, we can help enable future research that pursues such contributions.
Also, while our analyses identified differences in risks between demographic groups, these analyses were limited by how groups were defined and how they were represented in our data.  For example, we considered a limited definition of race, where even one race group, such as "Asian" or "Other or mixed" contains many subgroups. And while our analyses found that older users were more at risk, older users were underrepresented in our dataset. Furthermore, our analyses were limited to studying demographic groups in isolation (e.g. females versus males), while future work should also consider intersectionality (e.g. Black females versus White males).

There are further limitations due to how the survey data were collected. For example, it's possible that our analyses regarding participants' likelihood to share their browser data were impacted by an unmeasured response bias due to which participants chose to take the survey in the first place. Future work exploring this question (RQ1) might measure or mitigate this potential bias by using more generic survey descriptions when recruiting participants, or collecting demographics before providing an opportunity to leave the survey, or using alternative recruitment methods.
Another potential source of response bias is due to how our survey prompted participants to share their data before asking them (S2) about their concerns regarding fingerprinting/tracking. In the share prompt, the survey noted the resulting risks of tracking. This ethical disclosure may have primed participants to then be more wary of tracking when answering S2.
Also, our survey asked participants whether they think they understand fingerprinting (S1) without assessing whether they actually understand fingerprinting. Researchers who want to better assess differences between demographic groups' understanding of fingerprinting might use questions assessing that understanding and find different results than those here.

Our analyses were also limited by our dataset's scope and size. 
For example, there are more browser attributes and more active fingerprinting methods that our dataset does not include, and our analyses may overlook risks due to the most modern fingerprinting methods.
With respect to how the dataset's size limited analyses, an example is the analysis about  inferring demographics from browser attributes, where we collapsed some demographic categories in order to create larger groups for more robust analyses.  
Furthermore, our data are limited to the US and our results do not necessarily generalize to other geographies.
Even with these limitations, more research questions beyond the scope of this paper can be explored by leveraging our dataset.

\subsection{Summary of results and implications}

Here we highlight results that may be of particular interest to future researchers, developers, and policy makers. Browser fingerprinting is a commonly used mechanism to track users across the web~\cite{iqbal2021} and as browsers further transition users away from third-party cookies, the risks of fingerprinting may become more pertinent.

Our results show fingerprinting risks differ across demographic groups due to how device attributes and user configurations vary across these groups. This should be of importance to future researchers and developers. Researchers should be wary of which demographic groups are represented in their datasets, and how demographic biases in data can impact their results.  Developers should be wary of how their technologies, including those designed to mitigate fingerprinting risk, impact different demographics.

Our results also highlight an opportunity for developers and privacy advocates to more equitably educate users about fingerprinting risks. The majority of users in our study said they were concerned about fingerprinting/tracking, while less than half said they understand it, with female users significantly less likely to say they understand it.

We also find that device attributes can be used to help infer user demographics, a risk that goes beyond user tracking. This adds nuance to how browser developers should design solutions to mitigate fingerprinting risks. In particular, simply reducing entropy may not be enough. Our analysis suggests that when designing future changes to browser APIs, developers should balance entropy reduction with an effort to reduce the likelihood of grouping users by demographics.

\subsection{Future work is enabled by an open data collection framework}

In this work we demonstrated a consent-driven strategy to collect a dataset, enabling our research and future research by others. The resulting dataset contains browser attributes, including many studied in prior work, along with users' demographics.  Future research can use this dataset to revisit analyses described in prior works, which lacked open datasets or user demographics, as well as research potential strategies to mitigate the risks we studied.  If our dataset size is a limitation, our data collection method can be reused to ethically collect more data, further enabling future research.

\section*{Data and code availability}

The dataset is available via the following repository: \\
\url{https://github.com/aberke/fingerprinting-study}. \\
The repository also includes the data collection instrument, along with data preprocessing and analysis code.
The de-identified data are provided for research purposes only and should not be used in attempts to re-identify study participants. Access is granted to researchers who provide information about their research and accept the dataset terms of use.

\begin{acks}

The authors thank the crowdworkers who participated in this study. The Prolific study was funded by Google Privacy Sandbox.
In addition, we thank the following people who contributed to this work outside of authorship contributions: Mike Taylor, Michael Specter, Mihai Christodorescu

\end{acks}

\bibliographystyle{ACM-Reference-Format}
\bibliography{fingerprinting}


\begin{thebibliography}{68}


\ifx \showCODEN    \undefined \def \showCODEN     #1{\unskip}     \fi
\ifx \showDOI      \undefined \def \showDOI       #1{#1}\fi
\ifx \showISBNx    \undefined \def \showISBNx     #1{\unskip}     \fi
\ifx \showISBNxiii \undefined \def \showISBNxiii  #1{\unskip}     \fi
\ifx \showISSN     \undefined \def \showISSN      #1{\unskip}     \fi
\ifx \showLCCN     \undefined \def \showLCCN      #1{\unskip}     \fi
\ifx \shownote     \undefined \def \shownote      #1{#1}          \fi
\ifx \showarticletitle \undefined \def \showarticletitle #1{#1}   \fi
\ifx \showURL      \undefined \def \showURL       {\relax}        \fi
\providecommand\bibfield[2]{#2}
\providecommand\bibinfo[2]{#2}
\providecommand\natexlab[1]{#1}
\providecommand\showeprint[2][]{arXiv:#2}

\bibitem[tit(1964)]%
        {titleVII1964employment}
LII / Legal Information Institute \bibinfo{year}{1964}\natexlab{}.
\newblock \bibinfo{booktitle}{\emph{Title {{VII}} of the {{Civil Rights Act}} of 1964}}.
\newblock LII / Legal Information Institute.
\newblock
\urldef\tempurl%
\url{https://www.law.cornell.edu/wex/title_vii}
\showURL{%
\tempurl}


\bibitem[fai(nd)]%
        {fairhousingact24CFR}
LII / Legal Information Institute \bibinfo{year}{n.d}\natexlab{}.
\newblock \bibinfo{booktitle}{\emph{24 {{CFR}} § 100.75 - {{Discriminatory}} Advertisements, Statements and Notices.}}
\newblock LII / Legal Information Institute.
\newblock
\urldef\tempurl%
\url{https://www.law.cornell.edu/cfr/text/24/100.75}
\showURL{%
\tempurl}


\bibitem[Acar et~al\mbox{.}(2014)]%
        {acar2014}
\bibfield{author}{\bibinfo{person}{Gunes Acar}, \bibinfo{person}{Christian Eubank}, \bibinfo{person}{Steven Englehardt}, \bibinfo{person}{Marc Juarez}, \bibinfo{person}{Arvind Narayanan}, {and} \bibinfo{person}{Claudia Diaz}.} \bibinfo{year}{2014}\natexlab{}.
\newblock \showarticletitle{The web never forgets: Persistent tracking mechanisms in the wild}. In \bibinfo{booktitle}{\emph{Proceedings of the 2014 ACM SIGSAC conference on computer and communications security}}. \bibinfo{pages}{674--689}.
\newblock


\bibitem[Alliance(2018)]%
        {FBfairhousingactlawsuite}
\bibfield{author}{\bibinfo{person}{National Fair~Housing Alliance}.} \bibinfo{year}{2018}\natexlab{}.
\newblock \bibinfo{title}{Exhibit {{A}} – {{Programmatic Relief}}.}
\newblock
\newblock
\urldef\tempurl%
\url{https://nationalfairhousing.org/wp-content/uploads/2019/03/FINAL-Exhibit-A-3-18.pdf}
\showURL{%
\tempurl}


\bibitem[Alvestrand(2002)]%
        {alvestrand2002}
\bibfield{author}{\bibinfo{person}{Harald~T. Alvestrand}.} \bibinfo{year}{2002}\natexlab{}.
\newblock \bibinfo{booktitle}{\emph{Content {{Language Headers}}}}.
\newblock \bibinfo{type}{Request for {{Comments}}} RFC 3282. \bibinfo{institution}{Internet Engineering Task Force}.
\newblock
\urldef\tempurl%
\url{https://doi.org/10.17487/RFC3282}
\showDOI{\tempurl}


\bibitem[Andriamilanto et~al\mbox{.}(2021a)]%
        {Andriamilanto2021a}
\bibfield{author}{\bibinfo{person}{Nampoina Andriamilanto}, \bibinfo{person}{Tristan Allard}, {and} \bibinfo{person}{Gaëtan~Le Guelvouit}.} \bibinfo{year}{2021}\natexlab{a}.
\newblock \showarticletitle{“{{Guess Who}}?” {{Large-Scale Data-Centric Study}} of the {{Adequacy}} of {{Browser Fingerprints}} for {{Web Authentication}}}. In \bibinfo{booktitle}{\emph{Innovative {{Mobile}} and {{Internet Services}} in {{Ubiquitous Computing}}}} (Cham), \bibfield{editor}{\bibinfo{person}{Leonard Barolli}, \bibinfo{person}{Aneta Poniszewska-Maranda}, {and} \bibinfo{person}{Hyunhee Park}} (Eds.). \bibinfo{publisher}{Springer International Publishing}, \bibinfo{pages}{161--172}.
\newblock
\showISBNx{978-3-030-50399-4}
\urldef\tempurl%
\url{https://doi.org/10.1007/978-3-030-50399-4_16}
\showDOI{\tempurl}


\bibitem[Andriamilanto et~al\mbox{.}(2021b)]%
        {Andriamilanto2021b}
\bibfield{author}{\bibinfo{person}{Nampoina Andriamilanto}, \bibinfo{person}{Tristan Allard}, \bibinfo{person}{Gaëtan Le~Guelvouit}, {and} \bibinfo{person}{Alexandre Garel}.} \bibinfo{year}{2021}\natexlab{b}.
\newblock \showarticletitle{A {{Large-scale Empirical Analysis}} of {{Browser Fingerprints Properties}} for {{Web Authentication}}}.
\newblock \bibinfo{journal}{\emph{ACM Transactions on the Web}} \bibinfo{volume}{16}, \bibinfo{number}{1} (\bibinfo{year}{2021}), \bibinfo{pages}{4:1--4:62}.
\newblock
\showISSN{1559-1131}
\urldef\tempurl%
\url{https://doi.org/10.1145/3478026}
\showDOI{\tempurl}


\bibitem[Archive(2022)]%
        {webalmanac2021}
\bibfield{author}{\bibinfo{person}{HTTP Archive}.} \bibinfo{year}{2022}\natexlab{}.
\newblock \bibinfo{booktitle}{\emph{Privacy | {{The Web Almanac}}}}.
\newblock
\urldef\tempurl%
\url{https://almanac.httparchive.org/en/2021/privacy#fig-6}
\showURL{%
\tempurl}


\bibitem[Bacis et~al\mbox{.}(2024)]%
        {Googlers2024}
\bibfield{author}{\bibinfo{person}{Enrico Bacis}, \bibinfo{person}{Igor Bilogrevic}, \bibinfo{person}{Robert Busa-Fekete}, \bibinfo{person}{Asanka Herath}, \bibinfo{person}{Antonio Sartori}, {and} \bibinfo{person}{Umar Syed}.} \bibinfo{year}{2024}\natexlab{}.
\newblock \showarticletitle{Assessing Web Fingerprinting Risk}. In \bibinfo{booktitle}{\emph{Companion Proceedings of the ACM on Web Conference 2024}}. \bibinfo{pages}{245--254}.
\newblock


\bibitem[Berke et~al\mbox{.}(2024)]%
        {amazonCSCW2024}
\bibfield{author}{\bibinfo{person}{Alex Berke}, \bibinfo{person}{Robert Mahari}, \bibinfo{person}{Sandy Pentland}, \bibinfo{person}{Kent Larson}, {and} \bibinfo{person}{Dana Calacci}.} \bibinfo{year}{2024}\natexlab{}.
\newblock \showarticletitle{Insights from an experiment crowdsourcing data from thousands of US Amazon users: The importance of transparency, money, and data use}. In \bibinfo{booktitle}{\emph{Proc. ACM Hum.-Comput. Interact}}.
\newblock
\urldef\tempurl%
\url{https://doi.org/10.1145/3687005}
\showDOI{\tempurl}


\bibitem[Bradley(1997)]%
        {bradley1997}
\bibfield{author}{\bibinfo{person}{Andrew~P Bradley}.} \bibinfo{year}{1997}\natexlab{}.
\newblock \showarticletitle{The use of the area under the ROC curve in the evaluation of machine learning algorithms}.
\newblock \bibinfo{journal}{\emph{Pattern recognition}} \bibinfo{volume}{30}, \bibinfo{number}{7} (\bibinfo{year}{1997}), \bibinfo{pages}{1145--1159}.
\newblock


\bibitem[Brave(2020a)]%
        {BraveRandomPlugins}
\bibfield{author}{\bibinfo{person}{Brave}.} \bibinfo{year}{2020}\natexlab{a}.
\newblock \bibinfo{booktitle}{\emph{Fingerprinting 2.0: {{Plugins}} · {{Issue}} \#9435 · Brave/Brave-Browser}}.
\newblock GitHub.
\newblock
\urldef\tempurl%
\url{https://github.com/brave/brave-browser/issues/9435}
\showURL{%
\tempurl}


\bibitem[Brave(2020b)]%
        {brave2020}
\bibfield{author}{\bibinfo{person}{Brave}.} \bibinfo{year}{2020}\natexlab{b}.
\newblock \bibinfo{title}{Fingerprinting {{Protections}} v3 {$\cdot$} {{Issue}} \#11770 {$\cdot$} Brave/Brave-Browser}.
\newblock \bibinfo{howpublished}{https://github.com/brave/brave-browser/issues/11770}.
\newblock


\bibitem[Brave(2022)]%
        {Brave2022}
\bibfield{author}{\bibinfo{person}{Brave}.} \bibinfo{year}{2022}\natexlab{}.
\newblock \bibinfo{booktitle}{\emph{Protecting against Browser-Language Fingerprinting}}.
\newblock Brave.
\newblock
\urldef\tempurl%
\url{https://brave.com/privacy-updates/17-language-fingerprinting/}
\showURL{%
\tempurl}


\bibitem[Bureau(2023)]%
        {USCensusQuickFacts}
\bibfield{author}{\bibinfo{person}{U.S.~Census Bureau}.} \bibinfo{year}{2023}\natexlab{}.
\newblock \bibinfo{booktitle}{\emph{U.{{S}}. {{Census Bureau QuickFacts}}: {{United States}}}}.
\newblock
\urldef\tempurl%
\url{https://www.census.gov/quickfacts/fact/table/US/PST045223}
\showURL{%
\tempurl}


\bibitem[Cao et~al\mbox{.}(2017)]%
        {crossbrowserFP2017}
\bibfield{author}{\bibinfo{person}{Yinzhi Cao}, \bibinfo{person}{Song Li}, {and} \bibinfo{person}{Erik Wijmans}.} \bibinfo{year}{2017}\natexlab{}.
\newblock \showarticletitle{({{Cross-}}){{Browser Fingerprinting}} via {{OS}} and {{Hardware Level Features}}}. In \bibinfo{booktitle}{\emph{Proceedings 2017 {{Network}} and {{Distributed System Security Symposium}}}} (San Diego, CA). \bibinfo{publisher}{Internet Society}.
\newblock
\showISBNx{978-1-891562-46-4}
\urldef\tempurl%
\url{https://doi.org/10.14722/ndss.2017.23152}
\showDOI{\tempurl}


\bibitem[Chalise et~al\mbox{.}(2022)]%
        {Chalise2022}
\bibfield{author}{\bibinfo{person}{Shekhar Chalise}, \bibinfo{person}{Hoang~Dai Nguyen}, {and} \bibinfo{person}{Phani Vadrevu}.} \bibinfo{year}{2022}\natexlab{}.
\newblock \showarticletitle{Your Speaker or My Snooper?: Measuring the Effectiveness of Web Audio Browser Fingerprints}. In \bibinfo{booktitle}{\emph{Proceedings of the 22nd {{ACM Internet Measurement Conference}}}} (Nice France). \bibinfo{publisher}{ACM}, \bibinfo{pages}{349--357}.
\newblock
\showISBNx{978-1-4503-9259-4}
\urldef\tempurl%
\url{https://doi.org/10.1145/3517745.3561435}
\showDOI{\tempurl}


\bibitem[Chrome(2023)]%
        {chrome3PC2023}
\bibfield{author}{\bibinfo{person}{Chrome}.} \bibinfo{year}{2023}\natexlab{}.
\newblock \bibinfo{booktitle}{\emph{The next Step toward Phasing out Third-Party Cookies in {{Chrome}}}}.
\newblock Google.
\newblock
\urldef\tempurl%
\url{https://blog.google/products/chrome/privacy-sandbox-tracking-protection/}
\showURL{%
\tempurl}


\bibitem[Chromium(2021)]%
        {chromiumUserAgentReduction}
\bibfield{author}{\bibinfo{person}{Chromium}.} \bibinfo{year}{2021}\natexlab{}.
\newblock \bibinfo{booktitle}{\emph{User-{{Agent Reduction}}}}.
\newblock The Chromium Projects.
\newblock
\urldef\tempurl%
\url{https://www.chromium.org/updates/ua-reduction/#reduced-navigatorplatform-values-for-all-versions}
\showURL{%
\tempurl}


\bibitem[{chromium}(2021)]%
        {chromium2021}
\bibfield{author}{\bibinfo{person}{{chromium}}.} \bibinfo{year}{2021}\natexlab{}.
\newblock \bibinfo{title}{User-{{Agent Reduction}} {\textbar} {{The Chromium Projects}}}.
\newblock \bibinfo{howpublished}{https://www.chromium.org/updates/ua-reduction/\#reduced-navigatorplatform-values-for-all-versions}.
\newblock


\bibitem[Chromium(nd)]%
        {chromiumPlugins}
\bibfield{author}{\bibinfo{person}{Chromium}.} \bibinfo{year}{n.d}\natexlab{}.
\newblock \bibinfo{booktitle}{\emph{Dom\_plugin\_array.Cc - {{Chromium Code Search}}}}.
\newblock
\urldef\tempurl%
\url{https://source.chromium.org/chromium/chromium/src/+/main:third_party/blink/renderer/modules/plugins/dom_plugin_array.cc;l=187}
\showURL{%
\tempurl}


\bibitem[Coopamootoo et~al\mbox{.}(2022)]%
        {coopamootoo2022}
\bibfield{author}{\bibinfo{person}{Kovila P.~L. Coopamootoo}, \bibinfo{person}{Maryam Mehrnezhad}, {and} \bibinfo{person}{Ehsan Toreini}.} \bibinfo{year}{2022}\natexlab{}.
\newblock \showarticletitle{"{{I}} Feel Invaded, Annoyed, Anxious and {{I}} May Protect Myself": {{Individuals}}' {{Feelings}} about {{Online Tracking}} and Their {{Protective Behaviour}} across {{Gender}} and {{Country}}}. In \bibinfo{booktitle}{\emph{31st {{USENIX Security Symposium}} ({{USENIX Security}} 22)}}. \bibinfo{pages}{287--304}.
\newblock
\showISBNx{978-1-939133-31-1}


\bibitem[Coopamootoo and Ng(2023)]%
        {coopamootoo2023}
\bibfield{author}{\bibinfo{person}{Kovila P.~L. Coopamootoo} {and} \bibinfo{person}{Magdalene Ng}.} \bibinfo{year}{2023}\natexlab{}.
\newblock \showarticletitle{"\{\vphantom\}{{Un-Equal}}\vphantom\{\} {{Online Safety}}?" {{A Gender Analysis}} of {{Security}} and {{Privacy Protection Advice}} and {{Behaviour Patterns}}}. In \bibinfo{booktitle}{\emph{32nd {{USENIX Security Symposium}} ({{USENIX Security}} 23)}}. \bibinfo{pages}{5611--5628}.
\newblock
\showISBNx{978-1-939133-37-3}


\bibitem[Cooper et~al\mbox{.}(2013)]%
        {rfc6973}
\bibfield{author}{\bibinfo{person}{Alissa Cooper}, \bibinfo{person}{Hannes Tschofenig}, \bibinfo{person}{Bernard~D. Aboba}, \bibinfo{person}{Jon Peterson}, \bibinfo{person}{John Morris}, \bibinfo{person}{Marit Hansen}, {and} \bibinfo{person}{Rhys Smith}.} \bibinfo{year}{2013}\natexlab{}.
\newblock \bibinfo{title}{Privacy {{Considerations}} for {{Internet Protocols}}}.
\newblock
\newblock
\urldef\tempurl%
\url{https://doi.org/10.17487/RFC6973}
\showDOI{\tempurl}


\bibitem[Cover and Thomas(1991)]%
        {cover1991}
\bibfield{author}{\bibinfo{person}{T.~M. Cover} {and} \bibinfo{person}{Joy~A. Thomas}.} \bibinfo{year}{1991}\natexlab{}.
\newblock \bibinfo{booktitle}{\emph{Elements of Information Theory}}.
\newblock \bibinfo{publisher}{Wiley}.
\newblock
\showISBNx{978-0-471-06259-2}


\bibitem[Douglas et~al\mbox{.}(2023)]%
        {douglas2023}
\bibfield{author}{\bibinfo{person}{Benjamin~D. Douglas}, \bibinfo{person}{Patrick~J. Ewell}, {and} \bibinfo{person}{Markus Brauer}.} \bibinfo{year}{2023}\natexlab{}.
\newblock \showarticletitle{Data Quality in Online Human-Subjects Research: {{Comparisons}} between {{MTurk}}, {{Prolific}}, {{CloudResearch}}, {{Qualtrics}}, and {{SONA}}}.
\newblock \bibinfo{journal}{\emph{PLOS ONE}} \bibinfo{volume}{18}, \bibinfo{number}{3} (\bibinfo{year}{2023}), \bibinfo{pages}{e0279720}.
\newblock
\showISSN{1932-6203}
\urldef\tempurl%
\url{https://doi.org/10.1371/journal.pone.0279720}
\showDOI{\tempurl}


\bibitem[Eckersley(2010)]%
        {eckersley2010}
\bibfield{author}{\bibinfo{person}{Peter Eckersley}.} \bibinfo{year}{2010}\natexlab{}.
\newblock \showarticletitle{How {{Unique Is Your Web Browser}}?}. In \bibinfo{booktitle}{\emph{Privacy {{Enhancing Technologies}}}} (Berlin, Heidelberg) \emph{(\bibinfo{series}{Lecture {{Notes}} in {{Computer Science}}})}. \bibinfo{publisher}{Springer}, \bibinfo{pages}{1--18}.
\newblock
\showISBNx{978-3-642-14527-8}
\urldef\tempurl%
\url{https://doi.org/10.1007/978-3-642-14527-8_1}
\showDOI{\tempurl}


\bibitem[Fielding and Reschke(2014)]%
        {fielding2014}
\bibfield{author}{\bibinfo{person}{Roy~T. Fielding} {and} \bibinfo{person}{Julian Reschke}.} \bibinfo{year}{2014}\natexlab{}.
\newblock \bibinfo{booktitle}{\emph{Hypertext {{Transfer Protocol}} ({{HTTP}}/1.1): {{Semantics}} and {{Content}}}}.
\newblock \bibinfo{type}{Request for {{Comments}}} RFC 7231. \bibinfo{institution}{Internet Engineering Task Force}.
\newblock
\urldef\tempurl%
\url{https://doi.org/10.17487/RFC7231}
\showDOI{\tempurl}


\bibitem[FingerprintJS(2023a)]%
        {fingerprintJSAccuracy}
\bibfield{author}{\bibinfo{person}{Inc FingerprintJS}.} \bibinfo{year}{2023}\natexlab{a}.
\newblock \bibinfo{booktitle}{\emph{Fingerprint {{Pro}} vs. {{FingerprintJS}}}}.
\newblock Fingerprint.
\newblock
\urldef\tempurl%
\url{https://fingerprint.com/github/}
\showURL{%
\tempurl}


\bibitem[FingerprintJS(2023b)]%
        {fingerprintjsv3}
\bibfield{author}{\bibinfo{person}{Inc FingerprintJS}.} \bibinfo{year}{2023}\natexlab{b}.
\newblock \bibinfo{booktitle}{\emph{Fingerprintjs/Fingerprintjs at V3}}.
\newblock GitHub.
\newblock
\urldef\tempurl%
\url{https://github.com/fingerprintjs/fingerprintjs/tree/v3}
\showURL{%
\tempurl}


\bibitem[FingerprintJS(2024)]%
        {fingerprintjs2024}
\bibfield{author}{\bibinfo{person}{Inc FingerprintJS}.} \bibinfo{year}{2024}\natexlab{}.
\newblock \bibinfo{booktitle}{\emph{Fingerprintjs/Fingerprintjs}}.
\newblock Fingerprint ®.
\newblock
\urldef\tempurl%
\url{https://github.com/fingerprintjs/fingerprintjs}
\showURL{%
\tempurl}


\bibitem[Fowler(2019)]%
        {fowler2019}
\bibfield{author}{\bibinfo{person}{Geoffrey~A. Fowler}.} \bibinfo{year}{2019}\natexlab{}.
\newblock \showarticletitle{Perspective | {{Think}} You’re Anonymous Online? {{A}} Third of Popular Websites Are ‘Fingerprinting’ You.}
\newblock \bibinfo{journal}{\emph{Washington Post}} (\bibinfo{year}{2019}).
\newblock
\showISSN{0190-8286}
\urldef\tempurl%
\url{https://www.washingtonpost.com/technology/2019/10/31/think-youre-anonymous-online-third-popular-websites-are-fingerprinting-you/}
\showURL{%
\tempurl}


\bibitem[Gerber et~al\mbox{.}(2018)]%
        {gerber2018}
\bibfield{author}{\bibinfo{person}{Nina Gerber}, \bibinfo{person}{Paul Gerber}, {and} \bibinfo{person}{Melanie Volkamer}.} \bibinfo{year}{2018}\natexlab{}.
\newblock \showarticletitle{Explaining the Privacy Paradox: {{A}} Systematic Review of Literature Investigating Privacy Attitude and Behavior}.
\newblock \bibinfo{journal}{\emph{Computers \& Security}}  \bibinfo{volume}{77} (\bibinfo{year}{2018}), \bibinfo{pages}{226--261}.
\newblock
\showISSN{0167-4048}
\urldef\tempurl%
\url{https://doi.org/10.1016/j.cose.2018.04.002}
\showDOI{\tempurl}


\bibitem[Google(2024)]%
        {GoogleAdPolicies}
\bibfield{author}{\bibinfo{person}{Google}.} \bibinfo{year}{2024}\natexlab{}.
\newblock \bibinfo{booktitle}{\emph{Personalized Advertising - {{Advertising Policies Help}}}}.
\newblock
\urldef\tempurl%
\url{https://support.google.com/adspolicy/answer/143465?hl=en}
\showURL{%
\tempurl}


\bibitem[Gómez-Boix et~al\mbox{.}(2018)]%
        {GomezBoix2018}
\bibfield{author}{\bibinfo{person}{Alejandro Gómez-Boix}, \bibinfo{person}{Pierre Laperdrix}, {and} \bibinfo{person}{Benoit Baudry}.} \bibinfo{year}{2018}\natexlab{}.
\newblock \showarticletitle{Hiding in the {{Crowd}}: An {{Analysis}} of the {{Effectiveness}} of {{Browser Fingerprinting}} at {{Large Scale}}}. In \bibinfo{booktitle}{\emph{Proceedings of the 2018 {{World Wide Web Conference}}}} (Republic and Canton of Geneva, CHE) \emph{(\bibinfo{series}{{{WWW}} '18})}. \bibinfo{publisher}{International World Wide Web Conferences Steering Committee}, \bibinfo{pages}{309--318}.
\newblock
\showISBNx{978-1-4503-5639-8}
\urldef\tempurl%
\url{https://doi.org/10.1145/3178876.3186097}
\showDOI{\tempurl}


\bibitem[Iqbal et~al\mbox{.}(2021)]%
        {iqbal2021}
\bibfield{author}{\bibinfo{person}{Umar Iqbal}, \bibinfo{person}{Steven Englehardt}, {and} \bibinfo{person}{Zubair Shafiq}.} \bibinfo{year}{2021}\natexlab{}.
\newblock \showarticletitle{Fingerprinting the Fingerprinters: {{Learning}} to Detect Browser Fingerprinting Behaviors}. In \bibinfo{booktitle}{\emph{2021 {{IEEE Symposium}} on {{Security}} and {{Privacy}} ({{SP}})}}. \bibinfo{publisher}{IEEE}, \bibinfo{pages}{1143--1161}.
\newblock


\bibitem[Kaplan et~al\mbox{.}(2022)]%
        {kaplan2022}
\bibfield{author}{\bibinfo{person}{Levi Kaplan}, \bibinfo{person}{Nicole Gerzon}, \bibinfo{person}{Alan Mislove}, {and} \bibinfo{person}{Piotr Sapiezynski}.} \bibinfo{year}{2022}\natexlab{}.
\newblock \showarticletitle{Measurement and Analysis of Implied Identity in Ad Delivery Optimization}. In \bibinfo{booktitle}{\emph{Proceedings of the 22nd {{ACM Internet Measurement Conference}}}} (New York, NY, USA) \emph{(\bibinfo{series}{{{IMC}} '22})}. \bibinfo{publisher}{Association for Computing Machinery}, \bibinfo{pages}{195--209}.
\newblock
\showISBNx{978-1-4503-9259-4}
\urldef\tempurl%
\url{https://doi.org/10.1145/3517745.3561450}
\showDOI{\tempurl}


\bibitem[Laperdrix et~al\mbox{.}(2020)]%
        {Laperdrix2020}
\bibfield{author}{\bibinfo{person}{Pierre Laperdrix}, \bibinfo{person}{Nataliia Bielova}, \bibinfo{person}{Benoit Baudry}, {and} \bibinfo{person}{Gildas Avoine}.} \bibinfo{year}{2020}\natexlab{}.
\newblock \showarticletitle{Browser {{Fingerprinting}}: {{A Survey}}}.
\newblock \bibinfo{journal}{\emph{ACM Transactions on the Web}} \bibinfo{volume}{14}, \bibinfo{number}{2} (\bibinfo{year}{2020}), \bibinfo{pages}{8:1--8:33}.
\newblock
\showISSN{1559-1131}
\urldef\tempurl%
\url{https://doi.org/10.1145/3386040}
\showDOI{\tempurl}


\bibitem[Laperdrix et~al\mbox{.}(2016)]%
        {Laperdrix2016}
\bibfield{author}{\bibinfo{person}{Pierre Laperdrix}, \bibinfo{person}{Walter Rudametkin}, {and} \bibinfo{person}{Benoit Baudry}.} \bibinfo{year}{2016}\natexlab{}.
\newblock \showarticletitle{Beauty and the {{Beast}}: {{Diverting Modern Web Browsers}} to {{Build Unique Browser Fingerprints}}}. In \bibinfo{booktitle}{\emph{2016 {{IEEE Symposium}} on {{Security}} and {{Privacy}} ({{SP}})}} (San Jose, CA). \bibinfo{publisher}{IEEE}, \bibinfo{pages}{878--894}.
\newblock
\showISBNx{978-1-5090-0824-7}
\urldef\tempurl%
\url{https://doi.org/10.1109/SP.2016.57}
\showDOI{\tempurl}


\bibitem[Machanavajjhala et~al\mbox{.}(2007)]%
        {ldiversity}
\bibfield{author}{\bibinfo{person}{Ashwin Machanavajjhala}, \bibinfo{person}{Daniel Kifer}, \bibinfo{person}{Johannes Gehrke}, {and} \bibinfo{person}{Muthuramakrishnan Venkitasubramaniam}.} \bibinfo{year}{2007}\natexlab{}.
\newblock \showarticletitle{L-Diversity: {{Privacy}} beyond k-Anonymity}.
\newblock \bibinfo{journal}{\emph{Acm transactions on knowledge discovery from data (TKDD)}} \bibinfo{volume}{1}, \bibinfo{number}{1} (\bibinfo{year}{2007}), \bibinfo{pages}{3--es}.
\newblock


\bibitem[Mayer(2009)]%
        {mayer2009}
\bibfield{author}{\bibinfo{person}{Jonathan~R Mayer}.} \bibinfo{year}{2009}\natexlab{}.
\newblock \bibinfo{title}{“{{Any}} Person... a Pamphleteer:” {{Internet Anonymity}} in the {{Age}} of {{Web}} 2.0}.
\newblock
\newblock
\urldef\tempurl%
\url{https://jonathanmayer.org/publications/thesis09.pdf}
\showURL{%
\tempurl}


\bibitem[Meta(2020)]%
        {FBmlAdDelivery}
\bibfield{author}{\bibinfo{person}{Meta}.} \bibinfo{year}{2020}\natexlab{}.
\newblock \bibinfo{booktitle}{\emph{How {{Does Facebook Use Machine Learning}} to {{Deliver Ads}}?}}
\newblock Meta for Business.
\newblock
\urldef\tempurl%
\url{https://www.facebook.com/business/news/good-questions-real-answers-how-does-facebook-use-machine-learning-to-deliver-ads}
\showURL{%
\tempurl}


\bibitem[Meta(nd)]%
        {FBAdPolicies}
\bibfield{author}{\bibinfo{person}{Meta}.} \bibinfo{year}{n.d}\natexlab{}.
\newblock \bibinfo{booktitle}{\emph{About Audiences for Credit, Employment or Housing Campaigns}}.
\newblock Meta Business Help Center.
\newblock
\urldef\tempurl%
\url{https://www.facebook.com/business/help/2220749868045706}
\showURL{%
\tempurl}


\bibitem[Mishra et~al\mbox{.}(2020)]%
        {mishra2020}
\bibfield{author}{\bibinfo{person}{Vikas Mishra}, \bibinfo{person}{Pierre Laperdrix}, \bibinfo{person}{Antoine Vastel}, \bibinfo{person}{Walter Rudametkin}, \bibinfo{person}{Romain Rouvoy}, {and} \bibinfo{person}{Martin Lopatka}.} \bibinfo{year}{2020}\natexlab{}.
\newblock \showarticletitle{Don’t {{Count Me Out}}: {{On}} the {{Relevance}} of {{IP~Address}} in~the~{{Tracking~Ecosystem}}}. In \bibinfo{booktitle}{\emph{Proceedings of {{The Web Conference}} 2020}} (New York, NY, USA) \emph{(\bibinfo{series}{{{WWW}} '20})}. \bibinfo{publisher}{Association for Computing Machinery}, \bibinfo{pages}{808--815}.
\newblock
\showISBNx{978-1-4503-7023-3}
\urldef\tempurl%
\url{https://doi.org/10.1145/3366423.3380161}
\showDOI{\tempurl}


\bibitem[Mowery and Shacham(2012)]%
        {mowery2012}
\bibfield{author}{\bibinfo{person}{Keaton Mowery} {and} \bibinfo{person}{Hovav Shacham}.} \bibinfo{year}{2012}\natexlab{}.
\newblock \showarticletitle{Pixel perfect: Fingerprinting canvas in HTML5}.
\newblock \bibinfo{journal}{\emph{Proceedings of W2SP}}  \bibinfo{volume}{2012} (\bibinfo{year}{2012}).
\newblock


\bibitem[Mozilla(2023a)]%
        {firefox3PC2023}
\bibfield{author}{\bibinfo{person}{Mozilla}.} \bibinfo{year}{2023}\natexlab{a}.
\newblock \bibinfo{booktitle}{\emph{Firefox {{Rolls Out Total Cookie Protection By Default}} | {{The Mozilla Blog}}}}.
\newblock
\urldef\tempurl%
\url{https://blog.mozilla.org/en/mozilla/firefox-rolls-out-total-cookie-protection-by-default-to-all-users-worldwide/}
\showURL{%
\tempurl}


\bibitem[Mozilla(2023b)]%
        {mozillaDeviceMemory}
\bibfield{author}{\bibinfo{person}{Mozilla}.} \bibinfo{year}{2023}\natexlab{b}.
\newblock \bibinfo{booktitle}{\emph{Navigator: {{deviceMemory}} Property - {{Web APIs}} | {{MDN}}}}.
\newblock
\urldef\tempurl%
\url{https://developer.mozilla.org/en-US/docs/Web/API/Navigator/deviceMemory}
\showURL{%
\tempurl}


\bibitem[Munir et~al\mbox{.}(2023)]%
        {cookiegraph2023}
\bibfield{author}{\bibinfo{person}{Shaoor Munir}, \bibinfo{person}{Sandra Siby}, \bibinfo{person}{Umar Iqbal}, \bibinfo{person}{Steven Englehardt}, \bibinfo{person}{Zubair Shafiq}, {and} \bibinfo{person}{Carmela Troncoso}.} \bibinfo{year}{2023}\natexlab{}.
\newblock \showarticletitle{{{CookieGraph}}: {{Understanding}} and {{Detecting First-Party Tracking Cookies}}}. In \bibinfo{booktitle}{\emph{Proceedings of the 2023 {{ACM SIGSAC Conference}} on {{Computer}} and {{Communications Security}}}} (New York, NY, USA) \emph{(\bibinfo{series}{{{CCS}} '23})}. \bibinfo{publisher}{Association for Computing Machinery}, \bibinfo{pages}{3490--3504}.
\newblock
\showISBNx{9798400700507}
\urldef\tempurl%
\url{https://doi.org/10.1145/3576915.3616586}
\showDOI{\tempurl}


\bibitem[O’Neil(2001)]%
        {oneil2001}
\bibfield{author}{\bibinfo{person}{Dara O’Neil}.} \bibinfo{year}{2001}\natexlab{}.
\newblock \showarticletitle{Analysis of {{Internet Users}}’ {{Level}} of {{Online Privacy Concerns}}}.
\newblock \bibinfo{journal}{\emph{Social Science Computer Review}} \bibinfo{volume}{19}, \bibinfo{number}{1} (\bibinfo{year}{2001}), \bibinfo{pages}{17--31}.
\newblock
\showISSN{0894-4393}
\urldef\tempurl%
\url{https://doi.org/10.1177/089443930101900103}
\showDOI{\tempurl}


\bibitem[Pau et~al\mbox{.}(2023)]%
        {pau2023}
\bibfield{author}{\bibinfo{person}{Kiu~Nai Pau}, \bibinfo{person}{Vicki Wei~Qi Lee}, \bibinfo{person}{Shih~Yin Ooi}, {and} \bibinfo{person}{Ying~Han Pang}.} \bibinfo{year}{2023}\natexlab{}.
\newblock \showarticletitle{The {{Development}} of a {{Data Collection}} and {{Browser Fingerprinting System}}}.
\newblock \bibinfo{journal}{\emph{Sensors}} \bibinfo{volume}{23}, \bibinfo{number}{6} (\bibinfo{year}{2023}), \bibinfo{pages}{3087}.
\newblock
Issue 6.
\showISSN{1424-8220}
\urldef\tempurl%
\url{https://doi.org/10.3390/s23063087}
\showDOI{\tempurl}


\bibitem[Potoglou et~al\mbox{.}(2015)]%
        {Potoglou2015}
\bibfield{author}{\bibinfo{person}{Dimitris Potoglou}, \bibinfo{person}{Juan-Francisco Palacios}, {and} \bibinfo{person}{Claudio Feijóo}.} \bibinfo{year}{2015}\natexlab{}.
\newblock \showarticletitle{An Integrated Latent Variable and Choice Model to Explore the Role of Privacy Concern on Stated Behavioural Intentions in E-Commerce}.
\newblock \bibinfo{journal}{\emph{Journal of Choice Modelling}}  \bibinfo{volume}{17} (\bibinfo{year}{2015}), \bibinfo{pages}{10--27}.
\newblock
\showISSN{1755-5345}
\urldef\tempurl%
\url{https://doi.org/10.1016/j.jocm.2015.12.002}
\showDOI{\tempurl}


\bibitem[Pugliese et~al\mbox{.}(2020)]%
        {pugliese2020}
\bibfield{author}{\bibinfo{person}{Gaston Pugliese}, \bibinfo{person}{Christian Riess}, \bibinfo{person}{Freya Gassmann}, {and} \bibinfo{person}{Zinaida Benenson}.} \bibinfo{year}{2020}\natexlab{}.
\newblock \showarticletitle{Long-{{Term Observation}} on {{Browser Fingerprinting}}: {{Users}}’ {{Trackability}} and {{Perspective}}}.
\newblock \bibinfo{journal}{\emph{Proceedings on Privacy Enhancing Technologies}} \bibinfo{volume}{2020}, \bibinfo{number}{2} (\bibinfo{year}{2020}), \bibinfo{pages}{558--577}.
\newblock
\showISSN{2299-0984}
\urldef\tempurl%
\url{https://doi.org/10.2478/popets-2020-0041}
\showDOI{\tempurl}


\bibitem[Redmiles et~al\mbox{.}(2019)]%
        {redmiles2019}
\bibfield{author}{\bibinfo{person}{Elissa~M. Redmiles}, \bibinfo{person}{Sean Kross}, {and} \bibinfo{person}{Michelle~L. Mazurek}.} \bibinfo{year}{2019}\natexlab{}.
\newblock \showarticletitle{How {{Well Do My Results Generalize}}? {{Comparing Security}} and {{Privacy Survey Results}} from {{MTurk}}, {{Web}}, and {{Telephone Samples}}}. In \bibinfo{booktitle}{\emph{2019 {{IEEE Symposium}} on {{Security}} and {{Privacy}} ({{SP}})}}. \bibinfo{pages}{1326--1343}.
\newblock
\showISSN{2375-1207}
\urldef\tempurl%
\url{https://doi.org/10.1109/SP.2019.00014}
\showDOI{\tempurl}


\bibitem[Samarati and Sweeney(1998)]%
        {kAnonymity1998}
\bibfield{author}{\bibinfo{person}{Pierangela Samarati} {and} \bibinfo{person}{Latanya Sweeney}.} \bibinfo{year}{1998}\natexlab{}.
\newblock \showarticletitle{Protecting privacy when disclosing information: k-anonymity and its enforcement through generalization and suppression}.
\newblock  (\bibinfo{year}{1998}).
\newblock


\bibitem[Senol and Acar(2023)]%
        {senol2023}
\bibfield{author}{\bibinfo{person}{Asuman Senol} {and} \bibinfo{person}{Gunes Acar}.} \bibinfo{year}{2023}\natexlab{}.
\newblock \showarticletitle{Unveiling the {{Impact}} of {{User-Agent Reduction}} and {{Client Hints}}: {{A Measurement Study}}}. In \bibinfo{booktitle}{\emph{Proceedings of the 22nd {{Workshop}} on {{Privacy}} in the {{Electronic Society}}}} (Copenhagen Denmark). \bibinfo{publisher}{ACM}, \bibinfo{pages}{91--106}.
\newblock
\showISBNx{9798400702358}
\urldef\tempurl%
\url{https://doi.org/10.1145/3603216.3624965}
\showDOI{\tempurl}


\bibitem[StatCounter(2023)]%
        {statcounter2023}
\bibfield{author}{\bibinfo{person}{StatCounter}.} \bibinfo{year}{2023}\natexlab{}.
\newblock \bibinfo{title}{Operating {{System Market Share United States Of America}}}.
\newblock \bibinfo{howpublished}{https://gs.statcounter.com/os-market-share/all/united-states-of-america/}.
\newblock


\bibitem[Tang et~al\mbox{.}(2022)]%
        {tang2022}
\bibfield{author}{\bibinfo{person}{Jenny Tang}, \bibinfo{person}{Eleanor Birrell}, {and} \bibinfo{person}{Ada Lerner}.} \bibinfo{year}{2022}\natexlab{}.
\newblock \showarticletitle{Replication: {{How}} Well Do My Results Generalize Now? {{The}} External Validity of Online Privacy and Security Surveys}. In \bibinfo{booktitle}{\emph{Eighteenth Symposium on Usable Privacy and Security ({{SOUPS}} 2022)}} (Boston, MA). \bibinfo{publisher}{USENIX Association}, \bibinfo{pages}{367--385}.
\newblock
\showISBNx{978-1-939133-30-4}
\urldef\tempurl%
\url{https://www.usenix.org/system/files/soups2022-tang.pdf}
\showURL{%
\tempurl}


\bibitem[Team(2019)]%
        {microsoftEdge3PC}
\bibfield{author}{\bibinfo{person}{Microsoft~Edge Team}.} \bibinfo{year}{2019}\natexlab{}.
\newblock \bibinfo{booktitle}{\emph{Introducing Tracking Prevention, Now Available in {{Microsoft Edge}} Preview Builds}}.
\newblock Microsoft Edge Blog.
\newblock
\urldef\tempurl%
\url{https://blogs.windows.com/msedgedev/2019/06/27/tracking-prevention-microsoft-edge-preview/}
\showURL{%
\tempurl}


\bibitem[Timberg and Stanley-Becker(2020)]%
        {washpoCambridgeAnalytica2020}
\bibfield{author}{\bibinfo{person}{Craig Timberg} {and} \bibinfo{person}{Isaac Stanley-Becker}.} \bibinfo{year}{2020}\natexlab{}.
\newblock \showarticletitle{Cambridge {{Analytica}} Database Identified {{Black}} Voters as Ripe for ‘Deterrence,’ {{British}} Broadcaster Says}.
\newblock \bibinfo{journal}{\emph{Washington Post}} (\bibinfo{year}{2020}).
\newblock
\showISSN{0190-8286}
\urldef\tempurl%
\url{https://www.washingtonpost.com/technology/2020/09/28/trump-2016-cambridge-analytica-suppression/}
\showURL{%
\tempurl}


\bibitem[U.S. Census~Bureau(2023a)]%
        {USCensus-HINC-01}
\bibfield{author}{\bibinfo{person}{Current Population~Survey U.S. Census~Bureau}.} \bibinfo{year}{2023}\natexlab{a}.
\newblock \bibinfo{booktitle}{\emph{2023 {{Annual Social}} and {{Economic Supplement}} ({{CPS ASEC}}). {{Selected Characteristics}} of {{Households}} by {{Total Money Income}} in 2022. {{Table HINC-01}}.}}
\newblock


\bibitem[U.S. Census~Bureau(2023b)]%
        {NC-EST2022-ALLDATA}
\bibfield{author}{\bibinfo{person}{Population~Division U.S. Census~Bureau}.} \bibinfo{year}{2023}\natexlab{b}.
\newblock \bibinfo{booktitle}{\emph{Monthly {{Population Estimates}} by {{Age}}, {{Sex}}, {{Race}}, and {{Hispanic Origin}} for the {{United States}}: {{April}} 1, 2020 to {{July}} 1, 2022 ({{With}} Short-Term Projections to {{December}} 2023)}}.
\newblock
\urldef\tempurl%
\url{https://www2.census.gov/programs-surveys/popest/technical-documentation/file-layouts/2020-2022/NC-EST2022-ALLDATA.pdf}
\showURL{%
\tempurl}


\bibitem[{U.S. Census Bureau, Population Division}(2022)]%
        {census2022SCPRC-EST2022-18+POP}
\bibfield{author}{\bibinfo{person}{{U.S. Census Bureau, Population Division}}.} \bibinfo{year}{2022}\natexlab{}.
\newblock \bibinfo{title}{Estimates of the Total Resident Population and Resident Population Age 18 Years and Older for the United States, Regions, States, District of Columbia, and Puerto Rico: July 1, 2022}.
\newblock
\newblock
\urldef\tempurl%
\url{https://www.census.gov/data/tables/time-series/demo/popest/2020s-national-detail.html}
\showURL{%
\tempurl}
\newblock
\shownote{SCPRC-EST2022-18+POP}.


\bibitem[Vastel et~al\mbox{.}(2018)]%
        {vastel2018}
\bibfield{author}{\bibinfo{person}{Antoine Vastel}, \bibinfo{person}{Pierre Laperdrix}, \bibinfo{person}{Walter Rudametkin}, {and} \bibinfo{person}{Romain Rouvoy}.} \bibinfo{year}{2018}\natexlab{}.
\newblock \showarticletitle{{{FP-STALKER}}: {{Tracking Browser Fingerprint Evolutions}}}. In \bibinfo{booktitle}{\emph{2018 {{IEEE Symposium}} on {{Security}} and {{Privacy}} ({{SP}})}}. \bibinfo{pages}{728--741}.
\newblock
\urldef\tempurl%
\url{https://doi.org/10.1109/SP.2018.00008}
\showDOI{\tempurl}


\bibitem[Vinh et~al\mbox{.}(2010)]%
        {Vinh2010}
\bibfield{author}{\bibinfo{person}{Nguyen~Xuan Vinh}, \bibinfo{person}{Julien Epps}, {and} \bibinfo{person}{James Bailey}.} \bibinfo{year}{2010}\natexlab{}.
\newblock \showarticletitle{Information {{Theoretic Measures}} for {{Clusterings Comparison}}: {{Variants}}, {{Properties}}, {{Normalization}} and {{Correction}} for {{Chance}}}.
\newblock \bibinfo{journal}{\emph{Journal of Machine Learning Research}}  \bibinfo{volume}{11} (\bibinfo{year}{2010}).
\newblock
\urldef\tempurl%
\url{https://jmlr.csail.mit.edu/papers/volume11/vinh10a/vinh10a.pdf}
\showURL{%
\tempurl}


\bibitem[W3C(2019)]%
        {W3Cfingerprinting2019}
\bibfield{author}{\bibinfo{person}{W3C}.} \bibinfo{year}{2019}\natexlab{}.
\newblock \bibinfo{booktitle}{\emph{Mitigating {{Browser Fingerprinting}} in {{Web Specifications}}}}.
\newblock
\urldef\tempurl%
\url{https://www.w3.org/TR/fingerprinting-guidance/#types-of-fingerprinting}
\showURL{%
\tempurl}


\bibitem[W3C(2022)]%
        {w3cDeviceMemory}
\bibfield{author}{\bibinfo{person}{W3C}.} \bibinfo{year}{2022}\natexlab{}.
\newblock \bibinfo{booktitle}{\emph{Device {{Memory}} | {{W3C Working Draft}}, 22 {{July}} 2022}}.
\newblock
\urldef\tempurl%
\url{https://www.w3.org/TR/device-memory/}
\showURL{%
\tempurl}


\bibitem[Wilander(2020)]%
        {webkit3PC2020}
\bibfield{author}{\bibinfo{person}{John Wilander}.} \bibinfo{year}{2020}\natexlab{}.
\newblock \bibinfo{booktitle}{\emph{Full {{Third-Party Cookie Blocking}} and {{More}}}}.
\newblock WebKit.
\newblock
\urldef\tempurl%
\url{https://webkit.org/blog/10218/full-third-party-cookie-blocking-and-more/}
\showURL{%
\tempurl}


\bibitem[Wong(2019)]%
        {guardian-fb-fined}
\bibfield{author}{\bibinfo{person}{Julia~Carrie Wong}.} \bibinfo{year}{2019}\natexlab{}.
\newblock \showarticletitle{Facebook to Be Fined \$5bn for {{Cambridge Analytica}} Privacy Violations – Reports}.
\newblock \bibinfo{journal}{\emph{The Guardian}} (\bibinfo{year}{2019}).
\newblock
\showISSN{0261-3077}
\urldef\tempurl%
\url{https://www.theguardian.com/technology/2019/jul/12/facebook-fine-ftc-privacy-violations}
\showURL{%
\tempurl}


\end{thebibliography}

\appendix

\section{Data collection tool}
\label{SI:data collection tool}

We provide the data collection tool in the open repository\footnote{\url{https://github.com/aberke/fingerprinting-study/tree/master/instrument}}.
This includes the Qualtrics survey tool used by Prolific participants. 

A written copy of the survey content is included in this Appendix. It includes information about the survey shown to the participants, questions, answer options, and survey logic.

\section{Participants and data collection details}
\label{SI:participants and data collection details}

Data were collected in December of 2023. In order to better protect participant privacy, we avoid publishing a more precise date of collection\footnote{Vastel et al. (2018) find browser attributes change frequently, where 50\% of browser instances in their sample changed their fingerprints in less than 5 days, 80\% in less than 10 days~\cite{vastel2018}.}. 

Given that an important use of this dataset is to quantify the uniqueness of device attributes and users, there may be concern that devices contributed data multiple times, or that users submitted the survey more than once.
We note that this is also a risk for prior works which relied on cookies to deduplicate their data, and numerous other studies that leverage the Prolific platform to recruit participants.
Since our participants are recruited from Prolific, and our software is built on top of Qualtrics survey software, we have reason to believe this risk in our dataset is small. 
The Prolific system is set up so that each participant can have only one submission per study\footnote{\url{https://researcher-help.prolific.com/hc/en-gb/articles/360009220453-How-do-I-prevent-participants-from-taking-my-study-multiple-times}}.
Qualtrics provides a feature to prevent multiple submissions, which we utilized as well\footnote{\url{https://www.qualtrics.com/support/survey-platform/survey-module/survey-checker/fraud-detection/}}. 

\newpage
\clearpage



# Data collection tool

Data were collected via a Qualtrics survey with a custom software integration.
In addition to asking survey questions, this tool:
1. Collected participants' browser data, only after obtaining their informed consent
2. Embedded an experiment, where a random 50% of participants were shown the data we asked to collect.

This document contains a non-interactive copy of the survey questions and answer options.

# Participants and recruitment

Participants were limited to the US population and were required to be eligible English speakers and 18 years or older.

Below we show the details used to recruit participants on the Prolific platform.

---

**Estimated survey time:** 2 minutes
**Pay:** $0.60 ($18/hr)

**Title:**
Study about web user privacy

**Description:**

We are collecting data in order to study web user privacy and assess users' risks of being tracked across websites they visit.

Requirements:
- You are 18 years or older
- You are a U.S. resident

---

# Survey content

*Italicized text communicates information about the survey, not shown to participants. Horizontal lines indicate page breaks between questions shown to participants.*

---

**Introduction**

We are collecting data in order to study web user privacy and assess users' risks of being tracked across websites they visit.

This survey will ask you questions about your demographics.

This survey will also ask for your consent to collect information from your browser – the same kind of information that is used to track users across websites they visit. You will have the option to decline to share that data and your payment for participating in this survey will not depend on that choice.

The data you share through this survey will be made publicly available for research, but with your Prolific ID removed. Clicking to continue indicates that you meet the following requirements and agree to complete this survey voluntarily.

Requirements:
- You are 18 years or older
- You are a U.S. resident

☐ Continue
☐ Exit without payment

---

What is your Prolific ID?

[text field: *automatically fills with Prolific ID from URL parameter.*]

---

What is your age group?
- [ ] 18 - 24 years
- [ ] 25 - 34 years
- [ ] 35 - 44 years
- [ ] 45 - 54 years
- [ ] 55 - 64 years
- [ ] 65 or older

How do you describe yourself?
- [ ] Male
- [ ] Female
- [ ] Other

Are you of Spanish, Hispanic, or Latino origin?
- [ ] Yes
- [ ] No

Choose one or more races that you consider yourself to be
- [ ] White or Caucasian
- [ ] Black or African American
- [ ] American Indian/Native American or Alaska Native
- [ ] Asian
- [ ] Native Hawaiian or Other Pacific Islander
- [ ] Other

What was your total household income before taxes during the past 12 months?
- [ ] Less than $25,000
- [ ] $25,000 - $49,999
- [ ] $50,000 - $74,999
- [ ] $75,000 - $99,999
- [ ] $100,000 - $149,999
- [ ] $150,000 or more
- [ ] Prefer not to say

Select no options and leave this empty. This is an attention check.
- [ ] 1
- [ ] 2
- [ ] 3
- [ ] 4+
- [ ] Prefer not to say

Which U.S. State do you live in?

[drop down select]
*Options include 50 states, "District of Columbia", and "I did not reside in the United States"*

*Participants who fail the attention check are automatically redirected out of the survey.*

We are asking to collect attributes from your web browser in order to study how devices may be tracked across websites. The data you share will be published for research. It is possible that if another website collects these attributes, they can be used to identify your device in the published dataset. If you click "Decline", then we will not collect your browser attributes.

*A random subset of participants (with the "showdata=true" experiment flag) see the below table in a scrollable interface; other participants do not see the table. Below shows example values, where long values have been replaced with "..." for the purpose of this document. Browser data are only collected after the participant's explicit consent.*

Here are the attributes we'd like to collect:

| Attribute | Value |
| --- | --- |
| User agent | Mozilla/5.0 (Macintosh; Intel Mac OS X 10_15_7) AppleWebKit/537.36... |
| Languages | en-US,en,ca |
| Fonts | ["Arial Unicode MS","Gill Sans","Helvetica Neue","Menlo"] |
| Plugins | ["PDF Viewer","Chrome PDF Viewer","Chromium PDF Viewer", ...] |
| Local storage | true |
| Timezone | America/New_York |
| Screen resolution | [900,1440] |
| Color depth | 30 |
| Platform | MacIntel |
| Touch points | 0 |
| Cookies enabled | true |
| WebGL Vendor | WebKit |
| WebGL Renderer | WebKit WebGL |
| WebGL Unmasked Vendor | Google Inc. (Apple) |
| WebGL Unmasked Renderer | ANGLE (Apple, ANGLE Metal Renderer: Apple M2, Unspecified Version) |
| Hardware concurrency | 8 |
| Device memory | 8 |
| UA high entropy values | {"architecture":"arm","brands":[{"brand":"Not A(Brand)","version":"99"},...]} |
| Canvas text | data:image/png;base64,iVBORw0KGgoAAAANSUhEUgAAAPA… |
| Canvas geometry | data:image/png;base64,iVBORw0KGgoAAAANSUhEUgAAAHg… |
| Canvas text hashed | [1159232804,-959516503,1697727883,147613198,1522606278] |
| Canvas geometry hashed | [1369032737,1864480827,-2065969454,-955405403] |

22 rows (scroll)

    [ X ] Consent to share (*checked by default*)
    [   ] Decline

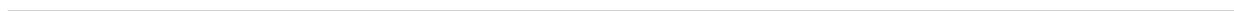

To what extent do you agree with the following statements?

"I think that I understand how browser fingerprinting works."
- ☐ Strongly disagree
- ☐ Somewhat disagree
- ☐ Neutral
- ☐ Somewhat agree
- ☐ Strongly agree

"I am concerned that websites and companies try to fingerprint/track my browser."
- ☐ Strongly disagree
- ☐ Somewhat disagree
- ☐ Neutral
- ☐ Somewhat agree
- ☐ Strongly agree

---

Thank you for participating.
Please provide any comments (optional)
[ text field ]

---

*Participants are automatically redirected back to Prolific.*



\clearpage

\section{Browser attributes collected}
\label{SI:browser attributes collected}

Table \ref{tab:browser_attributes_all} lists the browser attributes collected and provided in the dataset. It includes the number of distinct values in our dataset for each attribute, and the most frequently occurring value is provided as an example.

We collected all browser attributes via client-side JavaScript that we included in a webpage that participants loaded into their browser when taking the survey, similarly to common fingerprinting scripts.  All data were collected with participants' informed consent.

The following attributes were collected via the open license\footnote{https://github.com/fingerprintjs/fingerprintjs/blob/v3/LICENSE} version of the FingerprintJS (v3) library:
Timezone, Screen resolution, Color depth, Platform, Touch points, Fonts, Plugins, Local storage, Cookies enabled.

The other attributes were collected via additional scripting.

\begin{table*}
  \caption{Full set of browser attributes collected for the dataset.}
  \label{tab:browser_attributes_all}
\footnotesize
\begin{tabular}{llll}
\toprule
Attributes & \makecell[l]{Distinct\\values} & \makecell[l]{\% \\Unique} & Example (most frequent) value \\
\hline
User agent & 434 & 2.8 & Mozilla/5.0 (Windows NT 10.0; Win64; x64) AppleWebKit/537.36 (KHTML, like Gecko) Chrome/120.0.0.0 Safari/537.36 \\
Languages & 264 & 2.4 & en-US,en \\
Timezone & 49 & 0.2 & America/New\_York \\
Screen resolution & 572 & 4.5 & [1920,1080] \\
Color depth & 3 & 0 & 24 \\
Platform & 12 & 0 & Win32 \\
Touch points & 11 & 0 & 0 \\
Hardware concurrency & 24 & 0.1 & 4 \\
Device memory & 7 & 0 & 8 \\
WebGL Vendor & 3 & 0 & WebKit \\
WebGL Unmasked Vendor & 36 & 0.1 & Google Inc. (Intel) \\
WebGL Renderer & 36 & 0.1 & WebKit WebGL \\
WebGL Unmasked Renderer & 654 & 3.2 & Apple GPU \\
Fonts & 555 & 4.7 & ["Gill Sans","Helvetica Neue","Menlo"] \\
Plugins & 214 & 2.4 & ["PDF Viewer","Chrome PDF Viewer","Chromium PDF Viewer","Microsoft Edge PDF Viewer","WebKit built-in PDF"] \\
Local storage & 1 & 0 & true \\
Cookies enabled & 2 & 0 & true \\
UA high entropy values & 903 & 6.4 & \\
Canvas text & 951 & 7.9 & data:image/png;base64,iVBORw0KGgoAAAANSUhEUgAAAPAAAAA8CAYAAABYfzddAAAAAXNSR0IArs4c6... \\
Canvas geometry & 522 & 4.9 & data:image/png;base64,iVBORw0KGgoAAAANSUhEUgAAAHgAAAB4CAYAAAA5ZDbSAAAAAXNSR0IArs4c6... \\
Canvas text hashed & 951 & 7.9 & [922093378,-1490499838,-1506162041,-986675839,651032246] \\
Canvas geometry hashed & 522 & 4.9 & [-1339238269,-218892239,25977246,-975268297] \\
\bottomrule
\end{tabular}
\end{table*}

\paragraph{\textbf{User agent:}}
Same as the User-Agent HTTP header; it helps servers better identify the application, operating system, vendor, and/or version to better serve the user content. 

Collected via $window.navigator.userAgent$.

\paragraph{\textbf{Languages:}}
Same as the Accept-Language HTTP header; represents the user's preferred languages.

Collected via $navigator.languages$.

\paragraph{\textbf{Timezone:}}
Represents the user's current timezone, and is subject to change with the user's location. 

Collected via FingerprintJS.

\paragraph{\textbf{Screen resolution:}}
Indicates screen width, height pair, accessed via the screen property.

Collected via FingerprintJS.

\paragraph{\textbf{Color depth:}}
The depth of the device screen's color palette in bits per pixel (1, 4, 8, 15, 16, 24, 32, or 48).

Accessed via $window.screen.colorDepth$.

Collected via FingerprintJS.

\paragraph{\textbf{Platform:}}
Identifies the platform on which the user's browser is running.

Collected via FingerprintJS.

\paragraph{\textbf{Touch points:}}
The maximum number of simultaneous touch contact points supported by the device. 

Accessed via $navigator.maxTouchPoints$ and set to 0 when unavailable.

Collected via FingerprintJS

\paragraph{\textbf{Hardware concurrency:}}
Represents the number of logical processors available to run threads on the user's computer. Not available in later versions of Safari.

Collected via $navigator.hardwareConcurrency$.

\paragraph{\textbf{Device memory:}}
To curtail fingerprinting this value is rounded to a nearest value among 0.25, 0.5, 1, 2, 4, 8.

Collected via $navigator.deviceMemory$.

\paragraph{\textbf{WebGL:}}
WebGL (Web Graphics Library) is a JavaScript API that draws interactive 2D and 3D graphics. We access the WebGL attributes via the webgl context after creating a canvas element.

We access the WebGL "Unmasked" Vendor and Renderer attributes by further querying the $WEBGL\_debug\_renderer\_info$ extension.
The availability of these WebGL Unmasked attributes is dependent on the privacy settings in the browser.

\paragraph{\textbf{Fonts:}}
This is a list of fonts representing a subset of fonts available to the browser, as only the common fonts are tested for.

Collected via FingerprintJS.

\paragraph{\textbf{Plugins:}}
Accessible via $navigator.plugins$.

Collected via FingerprintJS.

\paragraph{\textbf{Local storage:}}
Boolean indicating whether the window localStorage property is accessible.

Collected via FingerprintJS.

\paragraph{\textbf{Cookies enabled:}}
Boolean indicating whether cookies can be set.

Collected via FingerprintJS.

\paragraph{\textbf{UA high entropy values:}}
Part of the recently developed User-Agent Client Hints API, which was designed to reduce passive fingerprinting risks.
At the time of data collection, this API was not yet a Web standard.

Collected via the $navigator.userAgentData.getHighEntropyValues$ API called with a list of the following parameters: `architecture', `model', `platform', `platformVersion'.

\paragraph{\textbf{Canvas elements:}}
Canvas fingerprinting was described in 2012 by Mowery and Shacham~\cite{mowery2012} and studied on the Web at scale in 2014 by Acar et al~\cite{acar2014}.
It is achieved by performing drawing operations in an HTML5 Canvas element and then reading and hashing the resulting data. Since drawing operations can render differently depending on device software and hardware characteristics, this method helps produce a fingerprint. 

The open source FingerprintJS library draws two different canvases, referred to as the "Canvas text" and "Canvas geometry" and collects the fingerprintable data from the canvas element via $toDataURL()$.

We use the same drawing operations to render these same canvases and similarly read the data via 
$toDataURL()$.

The resulting images can be viewed by inserting the returned values directly into the browser. We display examples below.

\begin{figure}[h]
    \centering
    \includegraphics[width=2cm]{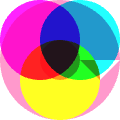}
    \caption{Canvas Geometry}
\end{figure}

\begin{figure}[h]
    \centering
    \includegraphics[width=5cm]{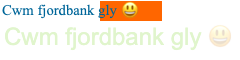}
    \caption{Canvas Text}
\end{figure}

In order to better study canvas fingerprinting, we also broke the drawing operations for each canvas into a list of steps, and then read and hashed the dataURL after each step.
The resulting list of hashed values are the "Canvas geometry hashed" and "Canvas text hashed" values collected.

\newpage

\section{Survey participant geographic distribution}

\begin{table}
  \caption{Participation by US state/territory compared to 2023 population estimates from the US Census Bureau.}
  \label{tab:participant_states}
  \small
\begin{tabular}{r|rr|rr|rr}
\toprule
& \multicolumn{2}{c}{US Census} & \multicolumn{4}{c}{Survey participants} \\
& \multicolumn{2}{c}{} & \multicolumn{2}{c}{All} & \multicolumn{2}{c}{Shared data} \\
US state/territory & n & \% & n & \% & n & \% \\
\midrule
Alabama & 3977628 & 1.5 & 176 & 1.4 & 118 & 1.4 \\
Alaska & 557899 & 0.2 & 11 & 0.1 & 9 & 0.1 \\
Arizona & 5848310 & 2.2 & 254 & 2 & 174 & 2.1 \\
Arkansas & 2362124 & 0.9 & 99 & 0.8 & 65 & 0.8 \\
California & 30519524 & 11.5 & 1361 & 10.9 & 929 & 11.1 \\
Colorado & 4662926 & 1.8 & 180 & 1.4 & 125 & 1.5 \\
Connecticut & 2894190 & 1.1 & 124 & 1 & 88 & 1 \\
Delaware & 819952 & 0.3 & 42 & 0.3 & 30 & 0.4 \\
District of Columbia & 552380 & 0.2 & 31 & 0.2 & 22 & 0.3 \\
Florida & 18229883 & 6.9 & 916 & 7.4 & 634 & 7.5 \\
Georgia & 8490546 & 3.2 & 475 & 3.8 & 327 & 3.9 \\
Hawaii & 1141525 & 0.4 & 35 & 0.3 & 15 & 0.2 \\
Idaho & 1497384 & 0.6 & 48 & 0.4 & 27 & 0.3 \\
Illinois & 9844167 & 3.7 & 489 & 3.9 & 330 & 3.9 \\
Indiana & 5274945 & 2.0 & 242 & 1.9 & 169 & 2 \\
Iowa & 2476882 & 0.9 & 111 & 0.9 & 75 & 0.9 \\
Kansas & 2246209 & 0.8 & 109 & 0.9 & 77 & 0.9 \\
Kentucky & 3509259 & 1.3 & 190 & 1.5 & 135 & 1.6 \\
Louisiana & 3506600 & 1.3 & 132 & 1.1 & 90 & 1.1 \\
Maine & 1146670 & 0.4 & 44 & 0.4 & 31 & 0.4 \\
Maryland & 4818337 & 1.8 & 258 & 2.1 & 167 & 2 \\
Massachusetts & 5659598 & 2.1 & 266 & 2.1 & 182 & 2.2 \\
Michigan & 7925350 & 3.0 & 381 & 3.1 & 246 & 2.9 \\
Minnesota & 4436981 & 1.7 & 190 & 1.5 & 123 & 1.5 \\
Mississippi & 2259864 & 0.9 & 87 & 0.7 & 62 & 0.7 \\
Missouri & 4821686 & 1.8 & 227 & 1.8 & 157 & 1.9 \\
Montana & 897161 & 0.3 & 23 & 0.2 & 15 & 0.2 \\
Nebraska & 1497381 & 0.6 & 65 & 0.5 & 45 & 0.5 \\
Nevada & 2508220 & 0.9 & 134 & 1.1 & 85 & 1 \\
New Hampshire & 1150004 & 0.4 & 50 & 0.4 & 26 & 0.3 \\
New Jersey & 7280551 & 2.7 & 318 & 2.6 & 211 & 2.5 \\
New Mexico & 1663024 & 0.6 & 57 & 0.5 & 41 & 0.5 \\
New York & 15611308 & 5.9 & 780 & 6.3 & 487 & 5.8 \\
North Carolina & 8498868 & 3.2 & 439 & 3.5 & 302 & 3.6 \\
North Dakota & 599192 & 0.2 & 22 & 0.2 & 18 & 0.2 \\
Ohio & 9207681 & 3.5 & 490 & 3.9 & 331 & 3.9 \\
Oklahoma & 3087217 & 1.2 & 136 & 1.1 & 95 & 1.1 \\
Oregon & 3401528 & 1.3 & 206 & 1.7 & 138 & 1.6 \\
Pennsylvania & 10332678 & 3.9 & 631 & 5.1 & 403 & 4.8 \\
Rhode Island & 892124 & 0.3 & 48 & 0.4 & 32 & 0.4 \\
South Carolina & 4229354 & 1.6 & 183 & 1.5 & 123 & 1.5 \\
South Dakota & 697420 & 0.3 & 18 & 0.1 & 9 & 0.1 \\
Tennessee & 5555761 & 2.1 & 259 & 2.1 & 171 & 2 \\
Texas & 22942176 & 8.7 & 969 & 7.8 & 679 & 8.1 \\
Utah & 2484582 & 0.9 & 103 & 0.8 & 72 & 0.9 \\
Vermont & 532828 & 0.2 & 30 & 0.2 & 18 & 0.2 \\
Virginia & 6834154 & 2.6 & 371 & 3 & 258 & 3.1 \\
Washington & 6164810 & 2.3 & 321 & 2.6 & 202 & 2.4 \\
West Virginia & 1417859 & 0.5 & 73 & 0.6 & 52 & 0.6 \\
Wisconsin & 4661826 & 1.8 & 241 & 1.9 & 170 & 2 \\
Wyoming & 454508 & 0.2 & 15 & 0.1 & 9 & 0.1 \\
Puerto Rico & 2707012 & 1.0 & 1 & 0 & 1 & 0 \\
\bottomrule
\end{tabular}
\end{table}

Since our survey participants were required to be US residents and 18 years or older, when comparing our participants’ geographic distribution to US Census Bureau estimates, we use census data limited to the 18+ population~\cite{census2022SCPRC-EST2022-18+POP}.

Our participants are overall representative of the US population with respect to state populations, with a Pearson correlation of r=0.988 (p<0.001). See Table~\ref{tab:participant_states}.

\newpage

\section{Survey regression analysis expanded results}
\label{SI: survey regression analysis}

This section expands the description of the models in Section~\ref{section:survey regression analysis}.

After testing the impact of demographic variables, we made the following simplifications to the regression analyses, which were applied consistently across the models.
We excluded participants who identified their gender as "Other" due to the small number of responses. We aggregated race variables to the categories: "White", "Black", "Asian", "Other or mixed", where "Other or mixed" includes participants who selected "Other", "American Indian/Alaska Native", or multiple races.
We did not include "Hispanic" or income group variables (income variables had no statistical significance in the S1 or S2 models).

Below we show the expanded version of the model regression results, with the log odds coefficients, odds ratios (OR) and 95\% CIs.

As robustness checks, we repeat the regression analyses using all of the demographic attributes as they were collected.
We confirm that the additional variables did not impact the significance of variables in the main regressions.

For S1 and S2 we use 1=Agree; 0=Otherwise.
An alternative approach would be to code the variables for S1 and S2 as 3 or 5 integers, given there was a range of agreement options, but we determined the resulting interpretation could not be as well justified. For example, if 2=Strongly agree and 1=Somewhat agree, 1.5 does not necessarily correspond to a real response exactly in their middle.

\begin{table}[hbt!]
  \caption{S1 model regression results.}
  \label{tab:S1_regression}
  \small
        \begin{tabular}{r|lll}
        \toprule
        \multicolumn{4}{l}{S1: “I think that I understand how browser fingerprinting works.”} \\
        \midrule
        Predictor & B (log odds) & OR & OR 95\% CI \\
        \midrule
        Intercept & 0.121** & 1.129 & [1.041, 1.225] \\
        Gender (Reference: Male) & & & \\
        Female & -0.664*** & 0.515 & [0.478, 0.554] \\
        Age (Reference: 35 - 44 years) & & & \\
        18 - 24 years & -0.327*** & 0.721 & [0.636, 0.818] \\
        25 - 34 years & -0.153** & 0.858 & [0.779, 0.945] \\
        45 - 54 years & 0.073 & 1.076 & [0.955, 1.212] \\
        55 - 64 years & 0.037 & 1.038 & [0.903, 1.192] \\
        65 and older & -0.256** & 0.774 & [0.644, 0.930] \\
        Race (Reference: White) & & & \\
        Asian & -0.23** & 0.794 & [0.697, 0.906] \\
        Black & 0.281*** & 1.324 & [1.179, 1.487] \\
        Other or mixed & 0.086 & 1.089 & [0.952, 1.247] \\
        \midrule
        N=12210 & & & \\
        pseudo R-squared = 0.024 & & & \\
        \midrule
        \multicolumn{4}{l}{Note: *p<0.05; **p<0.01; ***p<0.001} \\
        \bottomrule
        \end{tabular}
\end{table}

\begin{table}[hbt!]
  \caption{S2 model regression results.}
  \label{tab:S2_regression}
  \small
    \begin{tabular}{r|lll}
    \toprule
    \multicolumn{4}{l}{\makecell[l]{S2: “I am concerned that websites and companies try to \\fingerprint/track my browser.”}} \\
    \midrule
    Predictor & B (log odds) & OR & OR 95\% CI \\
    \midrule
    Intercept & 1.288*** & 3.626 & [3.185, 4.128] \\
    Gender (Reference: Male) & & & \\
    Female & 0.012 & 1.012 & [0.932, 1.098] \\
    Age (Reference: 35 - 44 years) & & & \\
    18 - 24 years & -0.22** & 0.803 & [0.704, 0.916] \\
    25 - 34 years & 0.004 & 1.004 & [0.903, 1.116] \\
    45 - 54 years & 0.24** & 1.272 & [1.111, 1.456] \\
    55 - 64 years & 0.458*** & 1.582 & [1.342, 1.864] \\
    65 and older & 0.527*** & 1.694 & [1.361, 2.108] \\
    Race (Reference: White) & & & \\
    Asian & 0.35*** & 1.419 & [1.229, 1.639] \\
    Black & -0.025 & 0.975 & [0.858, 1.109] \\
    Other or mixed & 0.212** & 1.236 & [1.062, 1.439] \\
    \midrule
    S1 agreement & 0.626*** & 1.871 & [1.719, 2.036] \\
    share & -1.106*** & 0.331 & [0.300, 0.365] \\
    showdata & -0.017 & 0.983 & [0.907, 1.065] \\
    \midrule
    N=12210 & & & \\
    pseudo R-squared = 0.059 & & & \\
    \bottomrule
    \end{tabular}
\end{table}

\begin{table}[hbt!]
  \caption{Share model regression results.}
  \label{tab:share_regression}
  \small
    \begin{tabular}{r|lll}
    \toprule
    Predictor & B (log odds) & OR & OR 95\% CI \\
    \midrule
    Intercept & 1.289*** & 3.629 & [3.174, 4.150] \\
    Gender (Reference: Male) & & & \\
    Female & -0.095* & 0.909 & [0.840, 0.984] \\
    Age (Reference: 35 - 44 years) & & & \\
    18 - 24 years & 0.417*** & 1.517 & [1.319, 1.746] \\
    25 - 34 years & 0.13* & 1.139 & [1.026, 1.264] \\
    45 - 54 years & -0.041 & 0.959 & [0.845, 1.089] \\
    55 - 64 years & -0.119 & 0.888 & [0.768, 1.026] \\
    65 and older & -0.229* & 0.795 & [0.660, 0.958] \\
    Race (Reference: White) & & & \\
    Asian & 0.142* & 1.152 & [1.001, 1.327] \\
    Black & -0.042 & 0.959 & [0.846, 1.087] \\
    Other or mixed & -0.05 & 0.951 & [0.824, 1.099] \\
    \midrule
    showdata & -0.138** & 0.871 & [0.806, 0.942] \\
    S1 agreement & 0.974*** & 2.648 & [2.142, 3.272] \\
    S2 agreement & -0.891*** & 0.41 & [0.366, 0.460] \\
    S1 x S2 agreement & -0.75*** & 0.472 & [0.376, 0.594] \\
    \midrule
    N=12210 & & & \\
    pseudo R-squared = 0.050 & & & \\
    \bottomrule
    \end{tabular}
\end{table}

\begin{table}[hbt!]
  \caption{S1 model regression robustness check results.}
  \label{tab:S1_regression_robustness}
  \small
  \begin{tabular}{r|lll}
    \toprule
    \multicolumn{4}{c}{S1: “I think that I understand how browser fingerprinting works.”} \\
    \midrule
    Predictor & B (log odds) & OR & OR 95\% CI \\
    \midrule
    Intercept & 0.133* & 1.143 & [1.025, 1.274] \\
    Gender (Ref: Male) & & & \\
    Female & -0.662*** & 0.516 & [0.479, 0.555] \\
    Other & -0.395** & 0.674 & [0.519, 0.875] \\
    Age (Ref: 35 - 44 years) & & & \\
    18 - 24 years & -0.324*** & 0.723 & [0.638, 0.820] \\
    25 - 34 years & -0.146** & 0.864 & [0.785, 0.952] \\
    45 - 54 years & 0.082 & 1.085 & [0.963, 1.222] \\
    55 - 64 years & 0.05 & 1.051 & [0.915, 1.208] \\
    65 and older & -0.241* & 0.786 & [0.654, 0.945] \\
    Race (Ref: White) & & & \\
    Asian & -0.229** & 0.796 & [0.698, 0.907] \\
    Black & 0.277*** & 1.32 & [1.176, 1.482] \\
    American Indian/Alaska Native & 0.204 & 1.226 & [0.770, 1.953] \\
    Other or mixed & 0.092 & 1.097 & [0.952, 1.262] \\
    Hispanic & -0.004 & 0.996 & [0.881, 1.126] \\
    Income (Ref: \$50,000 - \$74,999) & & & \\
    Less than \$25,000 & -0.051 & 0.95 & [0.835, 1.082] \\
    \$25,000 - \$49,999 & -0.028 & 0.972 & [0.870, 1.086] \\
    \$75,000 - \$99,999 & -0.03 & 0.971 & [0.860, 1.096] \\
    \$100,000 - \$149,999 & 0.019 & 1.019 & [0.905, 1.148] \\
    \$150,000 or more & -0.016 & 0.984 & [0.860, 1.127] \\
    Prefer not to say & -0.168 & 0.845 & [0.648, 1.103] \\
    \midrule
    N=12461 & & & \\
    pseudo R-squared = 0.024 & & & \\
    \bottomrule
    \end{tabular}
\end{table}

\begin{table}[hbt!]
  \caption{S2 model regression robustness check results.}
  \label{tab:S2_regression_robustness}
  \small
  \begin{tabular}{r|lll}
    \toprule
    \multicolumn{4}{l}{\makecell[l]{S2: “I am concerned that websites and companies try to \\fingerprint/track my browser.”}} \\
    \midrule
    Predictor & B (log odds) & OR & OR 95\% CI \\
    \midrule
    Intercept & 1.296*** & 3.656 & [3.143, 4.253] \\
    Gender (Ref: Male) & & & \\
    Female & 0.013 & 1.013 & [0.933, 1.100] \\
    Other & 0.275 & 1.317 & [0.983, 1.765] \\
    Age (Ref: 35 - 44 years) & & & \\
    18 - 24 years & -0.234** & 0.792 & [0.694, 0.904] \\
    25 - 34 years & -0.002 & 0.998 & [0.898, 1.109] \\
    45 - 54 years & 0.243*** & 1.275 & [1.114, 1.459] \\
    55 - 64 years & 0.461*** & 1.586 & [1.346, 1.868] \\
    65 and older & 0.523*** & 1.687 & [1.356, 2.099] \\
    Race (Ref: White) & & & \\
    Asian & 0.373*** & 1.452 & [1.256, 1.677] \\
    Black & -0.023 & 0.977 & [0.860, 1.110] \\
    American Indian/Alaska Native & -0.086 & 0.918 & [0.556, 1.516] \\
    Other or mixed & 0.249** & 1.283 & [1.094, 1.505] \\
    Hispanic & 0.053 & 1.054 & [0.921, 1.206] \\
    Income (Ref: \$50,000 - \$74,999) & & & \\
    Less than \$25,000 & 0.028 & 1.029 & [0.892, 1.187] \\
    \$25,000 - \$49,999 & -0.038 & 0.963 & [0.852, 1.088] \\
    \$75,000 - \$99,999 & -0.096 & 0.908 & [0.794, 1.038] \\
    \$100,000 - \$149,999 & 0.027 & 1.028 & [0.900, 1.174] \\
    \$150,000 or more & -0.061 & 0.941 & [0.810, 1.093] \\
    Prefer not to say & 0.17 & 1.185 & [0.873, 1.608] \\
    \midrule
    S1 agreement & 0.633*** & 1.883 & [1.732, 2.048] \\
    share & -1.1*** & 0.333 & [0.302, 0.366] \\
    showdata & -0.021 & 0.98 & [0.904, 1.061] \\
    \midrule
    N=12461 & & & \\
    pseudo R-squared = 0.060 & & & \\
    \bottomrule
    \end{tabular}

\end{table}

\begin{table}[hbt!]
  \caption{Share model regression robustness check results.}
  \label{tab:share_regression_robustness}
  \small
  \begin{tabular}{r|lll}
    \toprule
    Predictor & B (log odds) & OR & OR 95\% CI \\
    \midrule
    Intercept & 1.216*** & 3.373 & [2.891, 3.934] \\
    Gender (Ref: Male) & & & \\
    Female & -0.089* & 0.915 & [0.845, 0.991] \\
    Other & 0.171 & 1.186 & [0.885, 1.590] \\
    Age (Ref: 35 - 44 years) & & & \\
    18 - 24 years & 0.43*** & 1.537 & [1.336, 1.768] \\
    25 - 34 years & 0.131* & 1.14 & [1.028, 1.266] \\
    45 - 54 years & -0.045 & 0.956 & [0.843, 1.085] \\
    55 - 64 years & -0.129 & 0.879 & [0.761, 1.016] \\
    65 and older & -0.238* & 0.788 & [0.654, 0.950] \\
    Race (Ref: White) & & & \\
    Asian & 0.174* & 1.19 & [1.033, 1.372] \\
    Black & -0.027 & 0.973 & [0.859, 1.102] \\
    American Indian/Alaska Native & 0.02 & 1.02 & [0.608, 1.709] \\
    Other or mixed & -0.015 & 0.986 & [0.847, 1.147] \\
    Hispanic & -0.002 & 0.998 & [0.874, 1.141] \\
    Income (Ref: \$50,000 - \$74,999) & & & \\
    Less than \$25,000 & 0.107 & 1.113 & [0.969, 1.279] \\
    \$25,000 - \$49,999 & 0.067 & 1.069 & [0.949, 1.204] \\
    \$75,000 - \$99,999 & -0.024 & 0.976 & [0.858, 1.111] \\
    \$100,000 - \$149,999 & 0.15* & 1.161 & [1.021, 1.321] \\
    \$150,000 or more & 0.163* & 1.178 & [1.017, 1.364] \\
    Prefer not to say & -0.641*** & 0.527 & [0.404, 0.686] \\
    \midrule
    showdata & -0.134** & 0.875 & [0.809, 0.945] \\
    S1 agreement & 0.996*** & 2.708 & [2.192, 3.346] \\
    S2 agreement & -0.877*** & 0.416 & [0.372, 0.465] \\
    S1 x S2 agreement & -0.781*** & 0.458 & [0.365, 0.576] \\
    \midrule
    N=12461 & & & \\
    pseudo R-squared = 0.053 & & & \\
    \bottomrule
    \end{tabular}
\end{table}

Note there are some differences by income group in the share model. As to be expected, participants who answered "Prefer not to say" were less likely to share. Higher income participants were more likely to share, though this may be confounded with what motivates higher income people to participate in research on Prolific.

\clearpage

\section{Participants shown their data had longer survey durations on average}
\label{SI:survey_durations}

In addition to the main survey analysis, we did a diagnostic check to answer the following question: Did participants inspect their data when it was shown to them? 

We cannot know, but we might infer the answer is yes because we check whether users in the showdata=true experiment arm had longer survey durations on average (we automatically collected durations via the Qualtrics survey software).

The mean duration for showdata=true is 1.18x than for showdata=false (95.27 versus 80.63 seconds, respectively) and median duration for showdata=true is 1.20x than for showdata=false (72 versus 60 seconds, respectively).

\begin{figure}[hbt!]
    \centering
    \includegraphics[width=0.36\textwidth]{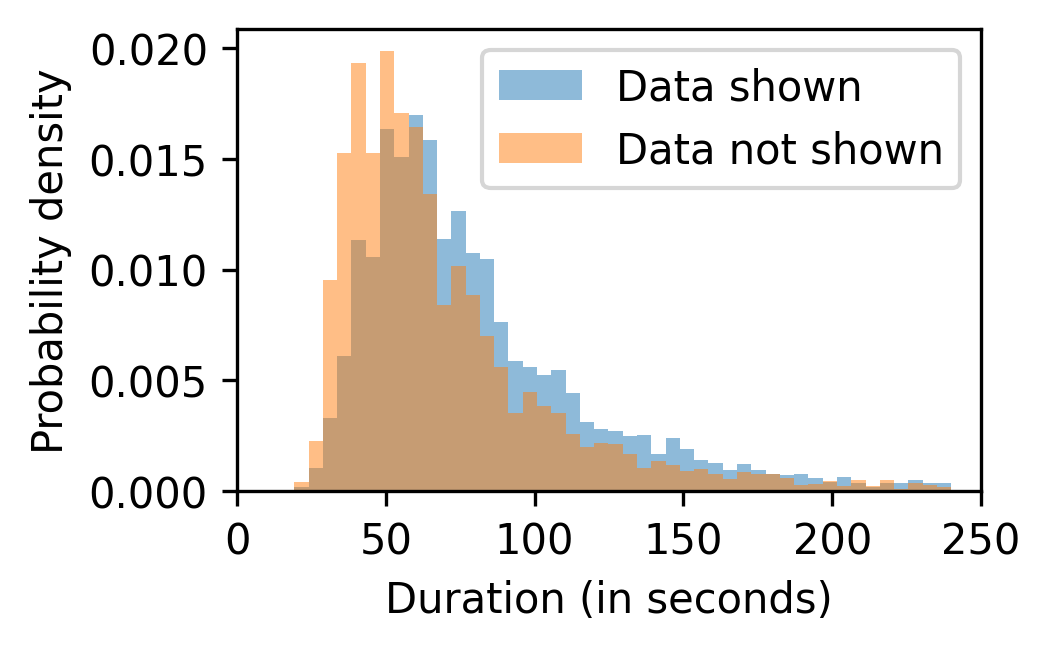}
    \caption{Survey duration times comparing participants who were shown their data versus not shown their data.}
    \label{fig:survey_durations}
\end{figure}

A one-sided t-test finds the difference in the mean durations to be statistically significant (p<0.001).
We also use a (non-parametric) Mann-Whitney U test because this data distribution is not normal. Again, the test finds the difference in the mean durations to be statistically significant (p<0.001). See Figure~\ref{fig:survey_durations}.

\section{Entropy changes with gender}
\label{SI:entropy changes with gender}

We demonstrate the impact of gender on entropy measurements in our dataset with the following methods. For a varying fraction of male users, ranging from 0 to 1 in increments of 0.1, we draw a random subsample (N=4000) from our data, without replacement, where the subsample has the given fraction of male users. From this subsample we then calculate normalized entropy values for each of the browser attributes. We do this sampling process 1000 times to estimate means with 95\% confidence intervals. Note this analysis does not include a significance test and is not meant to communicate a statistical result. Instead it is meant to motivate further analysis.

\begin{figure}
    \centering
    \includegraphics[width=0.4\textwidth]{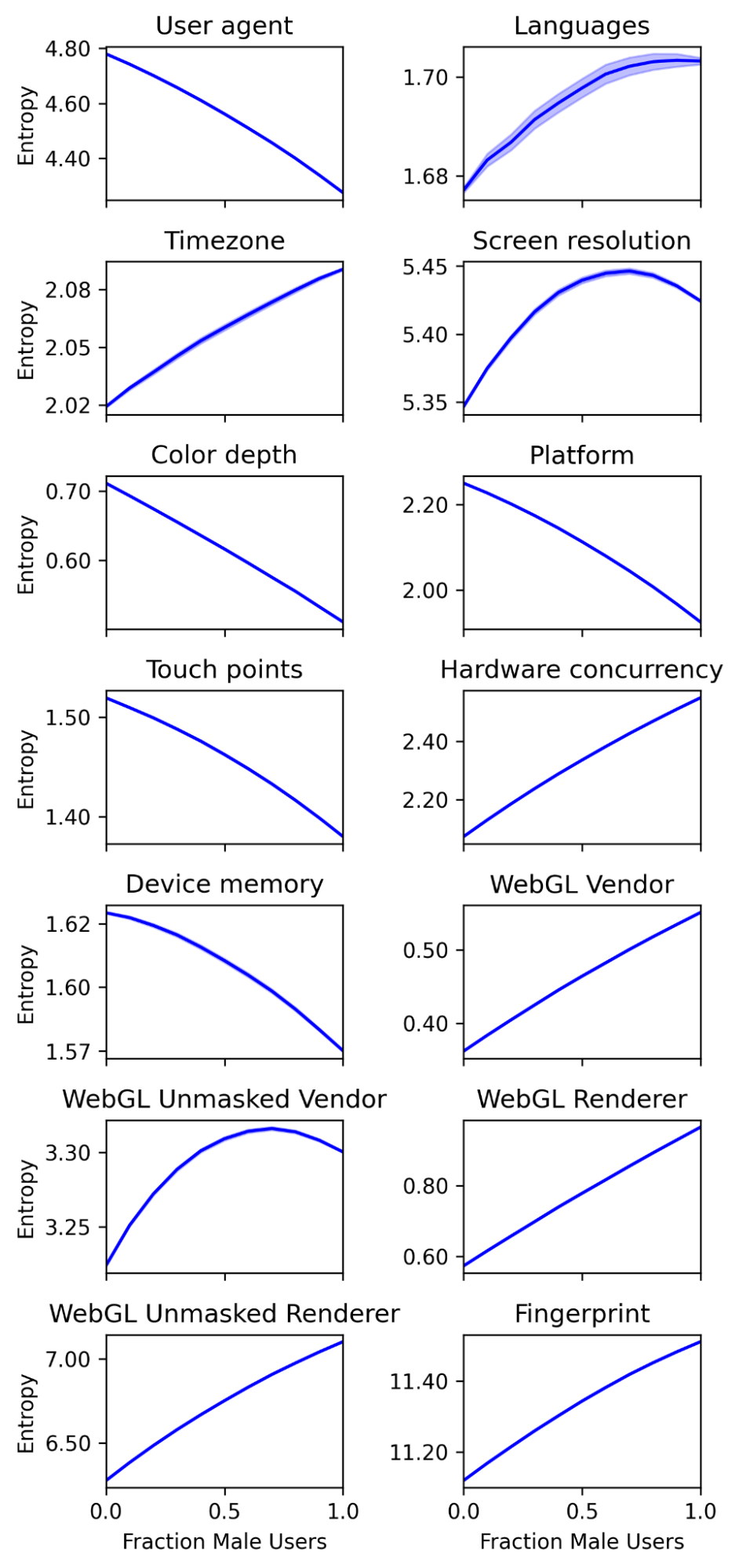}
    \caption{Changes in Shannon entropy estimates as male/non-male gender ratio changes.}
    \label{fig:entropy_changes_with_gender}
\end{figure}

Results are in Table~\ref{tab:entropy_changes_with_gender} and plotted in Figure~\ref{fig:entropy_changes_with_gender}.

\begin{table*}
  \caption{Changes in Shannon entropy estimates as male/non-male gender ratio changes.}
  \label{tab:entropy_changes_with_gender}
  \small
  \begin{tabular}{r|lllllllllll}
  \toprule
\multirow{2}{*}{ Attribute } & \multicolumn{11}{c}{Fraction male users and resulting mean entropy measurement} \\
& 0 & 0.1 & 0.2 & 0.3 & 0.4 & 0.5 & 0.6 & 0.7 & 0.8 & 0.9 & 1 \\
\midrule
User agent & 4.78 & 4.743 & 4.701 & 4.657 & 4.611 & 4.561 & 4.51 & 4.456 & 4.399 & 4.338 & 4.274 \\
Languages & 1.677 & 1.683 & 1.687 & 1.691 & 1.695 & 1.698 & 1.701 & 1.702 & 1.703 & 1.703 & 1.703 \\
Timezone & 2.024 & 2.032 & 2.039 & 2.046 & 2.053 & 2.059 & 2.064 & 2.07 & 2.075 & 2.08 & 2.084 \\
Screen resolution & 5.347 & 5.375 & 5.397 & 5.416 & 5.43 & 5.439 & 5.445 & 5.446 & 5.443 & 5.435 & 5.424 \\
Color depth & 0.711 & 0.693 & 0.674 & 0.655 & 0.636 & 0.617 & 0.597 & 0.576 & 0.556 & 0.534 & 0.512 \\
Platform & 2.249 & 2.227 & 2.201 & 2.174 & 2.145 & 2.113 & 2.08 & 2.045 & 2.008 & 1.968 & 1.926 \\
Touch points & 1.52 & 1.51 & 1.499 & 1.488 & 1.476 & 1.463 & 1.448 & 1.433 & 1.416 & 1.399 & 1.38 \\
Hardware concurrency & 2.074 & 2.131 & 2.186 & 2.238 & 2.289 & 2.336 & 2.382 & 2.427 & 2.469 & 2.51 & 2.55 \\
Device memory & 1.629 & 1.627 & 1.624 & 1.621 & 1.616 & 1.61 & 1.605 & 1.598 & 1.591 & 1.583 & 1.575 \\
WebGL Vendor & 0.362 & 0.384 & 0.405 & 0.425 & 0.445 & 0.464 & 0.483 & 0.501 & 0.518 & 0.535 & 0.551 \\
WebGL Unmasked Vendor & 3.224 & 3.251 & 3.272 & 3.288 & 3.301 & 3.309 & 3.314 & 3.316 & 3.314 & 3.308 & 3.3 \\
WebGL Renderer & 0.573 & 0.616 & 0.657 & 0.698 & 0.739 & 0.778 & 0.816 & 0.854 & 0.892 & 0.928 & 0.964 \\
WebGL Unmasked Renderer & 6.277 & 6.385 & 6.486 & 6.58 & 6.669 & 6.752 & 6.833 & 6.908 & 6.977 & 7.042 & 7.103 \\
Fingerprint & 11.119 & 11.168 & 11.214 & 11.259 & 11.302 & 11.343 & 11.382 & 11.419 & 11.452 & 11.483 & 11.512 \\
\bottomrule
\end{tabular}
\end{table*}

\clearpage

\section{Anonymity set sizes for attributes}
\label{SI:anonymity_set_sizes}

\begin{figure}[hbt!]
    \centering
    \includegraphics[width=0.65\textwidth]{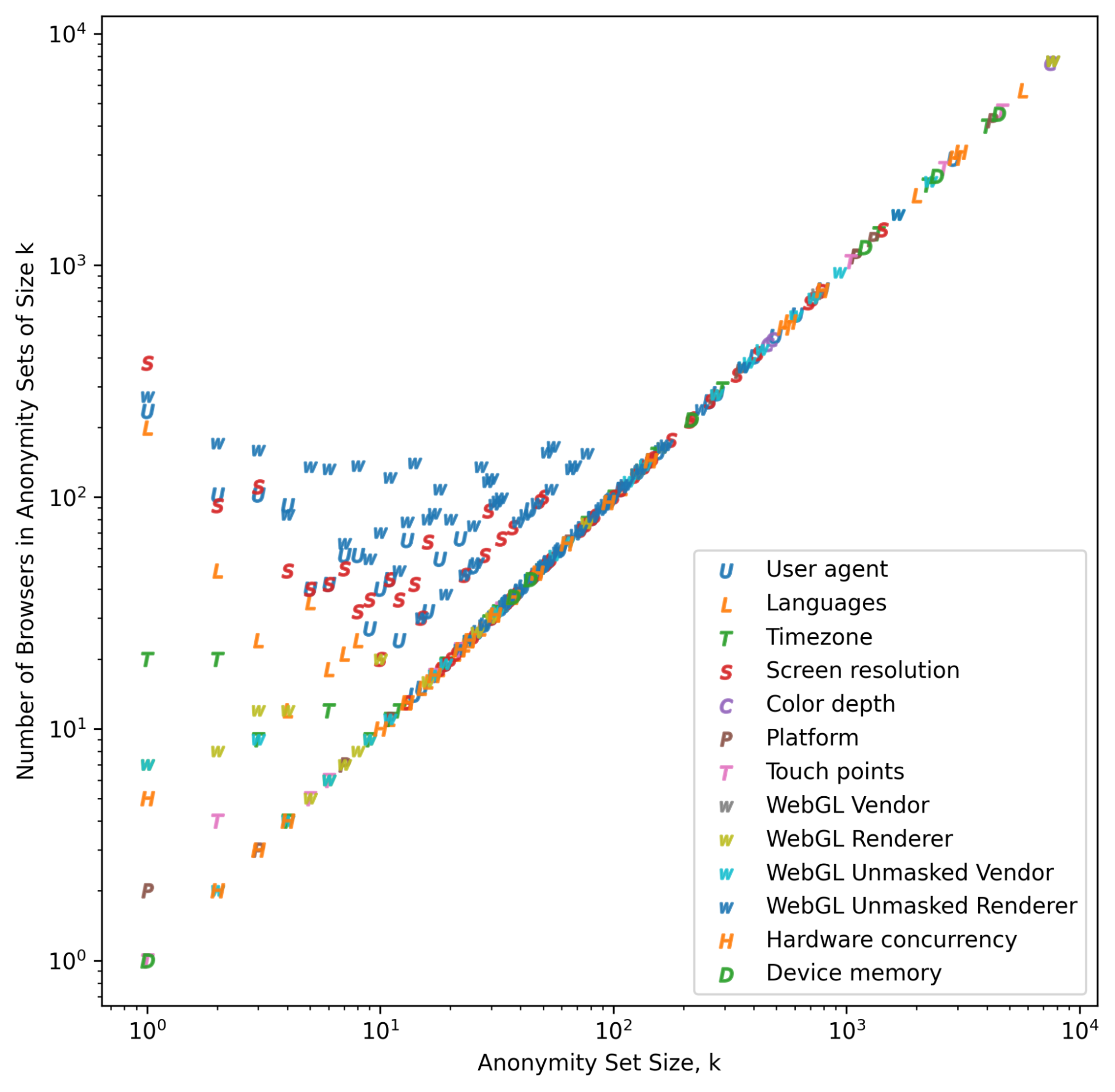}
    \caption{Number of users in anonymity sets of different sizes, considering each variable separately.}
    \label{fig:anonymity_set_sizes}
\end{figure}

In Figure~\ref{fig:anonymity_set_sizes} we recreate Fig. 3 from Eckersley’s ``How Unique is Your Web Browser" analysis~\cite{eckersley2010} using the browser attributes in this paper.
Points in the upper left are of greater concern for fingerprinting risk.

\clearpage

\section{Changes in fingerprinting metrics vs sample size}
\label{SI:fingerprinting metrics scaling}

To help understand the relationship between sample size and our fingerprinting metrics, we use repeated random sampling to create a series of subsamples of incremental sizes, n, and compute the percent unique and average anonymity set size for each such n.

\begin{figure}[hbt!]
    \centering
    \includegraphics[width=0.4\textwidth]{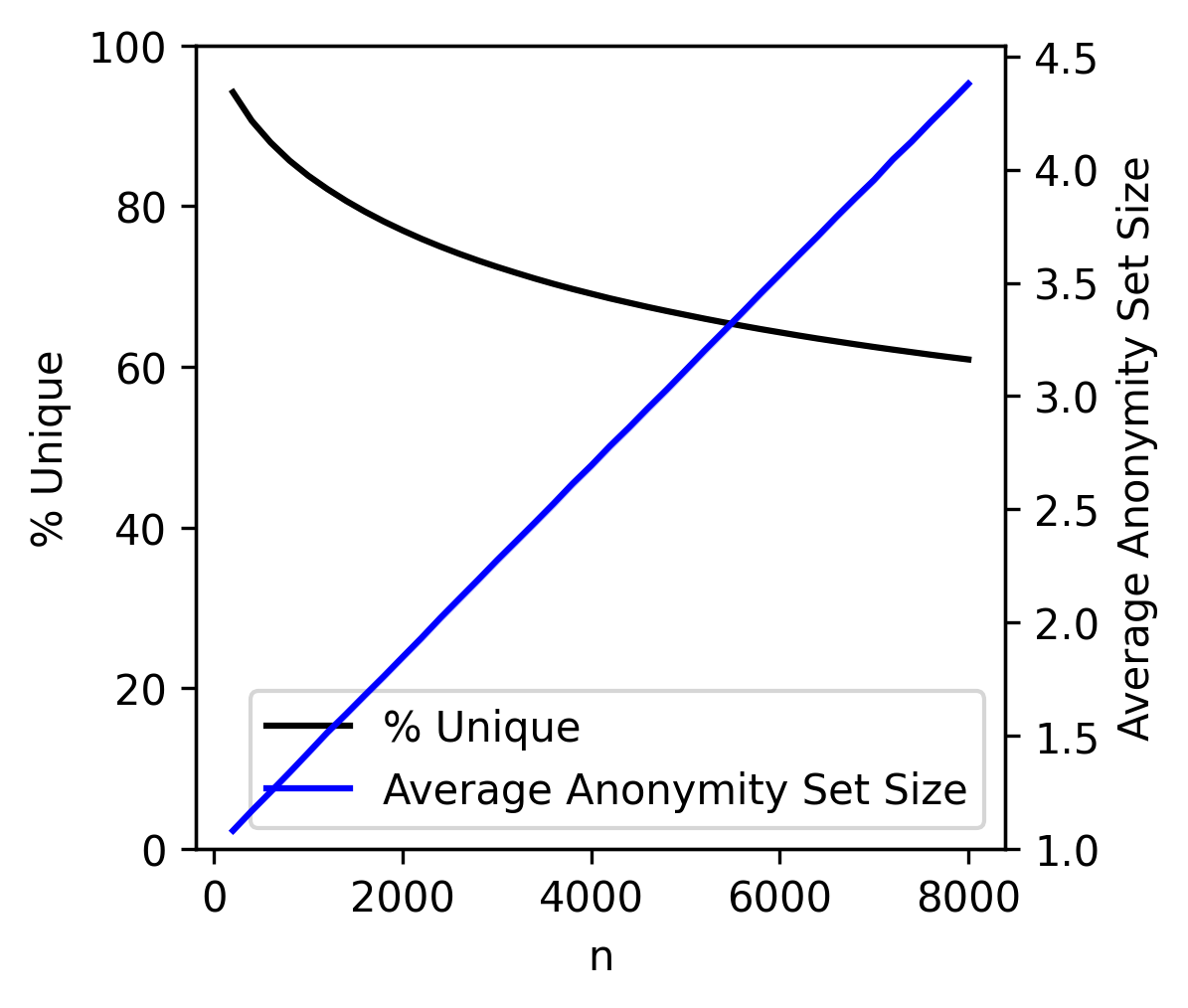}
    \caption{Estimated \% Unique and Average anonymity set size metrics for varying sample size (n).}
    \label{fig:metrics_scaling}
\end{figure}

We create subsamples for n ranging from 200 to 8000, in increments of 200.
For each n, we use random sampling, without replacement, to create the subsample and then compute the metrics. We repeat this process 1000 times for each n in order to compute the mean values with upper and lower 95\% confidence intervals.
Results are shown in Figure~\ref{fig:metrics_scaling}.
Note the CIs are plotted yet cannot be seen because they are so close to the means. 

The plot demonstrates how the average anonymity set size metric grows linearly with n. 
It also shows that the \% unique metric is not scale invariant. While it may approach a limit (as suggested in 2021 work by Andriamilanto et al~\cite{Andriamilanto2021b}), our sample is not large enough to expose this limit.

\section{Tables for fingerprinting metrics and demographic groups}
\label{SI:fingerprinting_metrics}

This section of the Appendix includes tables for the percent unique and average anonymity set size fingerprinting metrics, both with and without the subsampling process.

\begin{table*}[hbt!]
  \caption{Percent unique by gender, with equal sized demographic groups, showing means and 95\% confidence intervals.}
  \label{tab:pct_unique_gender_subsampled}
  \small
  \begin{tabular}{r|ll}
  \toprule
    Attribute & Male & Female \\
    \midrule
    n & 900 & 900 \\
    \midrule
    User agent & 5.5 [5.5, 5.6] & 6.3 [6.2, 6.3] \\
    Languages & 3.9 [3.8, 3.9] & 2.9 [2.9, 3.0] \\
    Timezone & 0.6 [0.6, 0.6] & 0.6 [0.6, 0.6] \\
    Screen resolution & 8.7 [8.7, 8.8] & 5.4 [5.4, 5.5] \\
    Color depth & 0.0 [0.0, 0.0] & 0.0 [0.0, 0.0] \\
    Platform & 0.1 [0.1, 0.1] & 0.1 [0.1, 0.1] \\
    Touch points & 0.1 [0.1, 0.1] & 0.1 [0.1, 0.1] \\
    Hardware concurrency & 0.3 [0.3, 0.3] & 0.0 [0.0, 0.0] \\
    Device memory & 0.0 [0.0, 0.0] & 0.0 [0.0, 0.0] \\
    WebGL Vendor & 0.0 [0.0, 0.0] & 0.0 [0.0, 0.0] \\
    WebGL Unmasked Vendor & 0.3 [0.3, 0.3] & 0.1 [0.1, 0.1] \\
    WebGL Renderer & 0.7 [0.7, 0.7] & 0.2 [0.1, 0.2] \\
    WebGL Unmasked Renderer & 10.7 [10.7, 10.8] & 6.5 [6.5, 6.6] \\
    Fingerprint & 82.1 [82.0, 82.2] & 73.5 [73.4, 73.6] \\
    \bottomrule
    \end{tabular}
\end{table*}

\begin{table*}
  \caption{Percent unique by age group, with equal sized demographic groups, showing means and 95\% confidence intervals.}
  \label{tab:pct_unique_age_subsampled}
  \small
  \begin{tabular}{r|llllll}
  \toprule
Attribute & 18 - 24 years & 25 - 34 years & 35 - 44 years & 45 - 54 years & 55 - 64 years & 65 or older \\
\midrule
n & 300 & 300 & 300 & 300 & 300 & 300 \\
\midrule
User agent & 6.2 [6.2, 6.3] & 6.0 [5.9, 6.1] & 5.8 [5.7, 5.9] & 5.2 [5.1, 5.3] & 6.1 [6.0, 6.1] & 6.3 [6.3, 6.4] \\
Languages & 3.0 [3.0, 3.1] & 3.5 [3.4, 3.5] & 3.7 [3.6, 3.7] & 3.5 [3.5, 3.6] & 3.5 [3.5, 3.6] & 3.2 [3.1, 3.2] \\
Timezone & 0.8 [0.8, 0.9] & 0.6 [0.6, 0.6] & 0.6 [0.5, 0.6] & 0.8 [0.8, 0.8] & 0.5 [0.4, 0.5] & 0.4 [0.4, 0.5] \\
Screen resolution & 4.0 [3.9, 4.0] & 7.2 [7.1, 7.3] & 7.4 [7.3, 7.5] & 7.2 [7.1, 7.3] & 9.1 [9.0, 9.1] & 9.8 [9.7, 9.8] \\
Color depth & 0.0 [0.0, 0.0] & 0.0 [0.0, 0.0] & 0.0 [0.0, 0.0] & 0.0 [0.0, 0.0] & 0.0 [0.0, 0.0] & 0.0 [0.0, 0.0] \\
Platform & 0.0 [0.0, 0.0] & 0.1 [0.1, 0.1] & 0.0 [0.0, 0.0] & 0.1 [0.1, 0.1] & 0.1 [0.1, 0.1] & 0.3 [0.3, 0.4] \\
Touch points & 0.1 [0.1, 0.1] & 0.1 [0.1, 0.1] & 0.2 [0.1, 0.2] & 0.1 [0.1, 0.1] & 0.1 [0.1, 0.1] & 0.0 [0.0, 0.0] \\
Hardware concurrency & 0.2 [0.2, 0.3] & 0.2 [0.2, 0.2] & 0.1 [0.1, 0.2] & 0.1 [0.0, 0.1] & 0.1 [0.1, 0.1] & 0.3 [0.3, 0.3] \\
Device memory & 0.0 [-0.0, 0.0] & 0.0 [-0.0, 0.0] & 0.0 [0.0, 0.0] & 0.0 [0.0, 0.0] & 0.0 [-0.0, 0.0] & 0.0 [0.0, 0.0] \\
WebGL Vendor & 0.0 [0.0, 0.0] & 0.0 [0.0, 0.0] & 0.0 [0.0, 0.0] & 0.0 [0.0, 0.0] & 0.0 [0.0, 0.0] & 0.0 [-0.0, 0.0] \\
WebGL Unmasked Vendor & 0.0 [0.0, 0.0] & 0.2 [0.2, 0.2] & 0.3 [0.3, 0.4] & 0.2 [0.1, 0.2] & 0.3 [0.3, 0.3] & 0.5 [0.5, 0.5] \\
WebGL Renderer & 0.1 [0.1, 0.1] & 0.5 [0.4, 0.5] & 0.7 [0.6, 0.7] & 0.3 [0.3, 0.3] & 0.5 [0.5, 0.6] & 0.5 [0.5, 0.5] \\
WebGL Unmasked Renderer & 7.2 [7.1, 7.2] & 8.7 [8.6, 8.8] & 9.6 [9.5, 9.7] & 8.4 [8.3, 8.5] & 10.5 [10.4, 10.6] & 10.9 [10.8, 10.9] \\
Fingerprint & 75.1 [74.9, 75.2] & 77.4 [77.2, 77.5] & 79.5 [79.4, 79.7] & 80.2 [80.0, 80.3] & 81.1 [80.9, 81.2] & 84.5 [84.4, 84.5] \\
\bottomrule
\end{tabular}
\end{table*}

\begin{table*}
  \caption{Percent unique by income group, with equal sized demographic groups, showing means and 95\% confidence intervals.}
  \label{tab:pct_unique_income_subsampled}
  \small
  \begin{tabular}{r|llllll}
  \toprule
Attribute & Less than \$25,000 & \$25,000 - \$49,999 & \$50,000 - \$74,999 & \$75,000 - \$99,999 & \$100,000 - \$149,999 & \$150,000 or more \\
\midrule
n & 300 & 300 & 300 & 300 & 300 & 300 \\
\midrule
User agent & 8.6 [8.5, 8.7] & 6.4 [6.4, 6.5] & 5.7 [5.6, 5.7] & 5.0 [4.9, 5.0] & 4.8 [4.7, 4.9] & 4.7 [4.7, 4.8] \\
Languages & 3.3 [3.3, 3.4] & 3.4 [3.3, 3.5] & 3.3 [3.3, 3.4] & 3.7 [3.6, 3.7] & 3.2 [3.2, 3.3] & 3.6 [3.6, 3.7] \\
Timezone & 0.6 [0.6, 0.6] & 0.3 [0.2, 0.3] & 0.9 [0.8, 0.9] & 0.6 [0.6, 0.7] & 0.5 [0.5, 0.5] & 0.8 [0.8, 0.8] \\
Screen resolution & 9.1 [9.0, 9.2] & 7.2 [7.1, 7.3] & 7.6 [7.5, 7.7] & 6.8 [6.8, 6.9] & 6.7 [6.6, 6.8] & 6.2 [6.1, 6.2] \\
Color depth & 0.0 [0.0, 0.0] & 0.0 [0.0, 0.0] & 0.0 [0.0, 0.0] & 0.0 [0.0, 0.0] & 0.0 [0.0, 0.0] & 0.0 [0.0, 0.0] \\
Platform & 0.1 [0.1, 0.1] & 0.1 [0.1, 0.1] & 0.1 [0.1, 0.1] & 0.0 [0.0, 0.0] & 0.1 [0.1, 0.1] & 0.0 [0.0, 0.0] \\
Touch points & 0.0 [0.0, 0.0] & 0.1 [0.1, 0.1] & 0.1 [0.1, 0.1] & 0.1 [0.1, 0.1] & 0.1 [0.1, 0.1] & 0.2 [0.2, 0.2] \\
Hardware concurrency & 0.1 [0.1, 0.1] & 0.1 [0.1, 0.1] & 0.3 [0.3, 0.3] & 0.2 [0.2, 0.2] & 0.1 [0.1, 0.1] & 0.1 [0.1, 0.1] \\
Device memory & 0.0 [-0.0, 0.0] & 0.0 [0.0, 0.0] & 0.1 [0.0, 0.1] & 0.0 [-0.0, 0.0] & 0.0 [-0.0, 0.0] & 0.0 [0.0, 0.0] \\
WebGL Vendor & 0.0 [0.0, 0.0] & 0.0 [-0.0, 0.0] & 0.0 [-0.0, 0.0] & 0.0 [0.0, 0.0] & 0.0 [0.0, 0.0] & 0.0 [-0.0, 0.0] \\
WebGL Unmasked Vendor & 0.4 [0.4, 0.4] & 0.2 [0.1, 0.2] & 0.2 [0.2, 0.2] & 0.3 [0.3, 0.3] & 0.2 [0.2, 0.2] & 0.2 [0.2, 0.2] \\
WebGL Renderer & 0.9 [0.9, 1.0] & 0.4 [0.4, 0.4] & 0.2 [0.2, 0.2] & 0.5 [0.5, 0.6] & 0.3 [0.3, 0.3] & 0.3 [0.3, 0.4] \\
WebGL Unmasked Renderer & 10.7 [10.6, 10.8] & 10.0 [10.0, 10.1] & 8.7 [8.6, 8.8] & 8.4 [8.3, 8.5] & 7.5 [7.4, 7.6] & 6.4 [6.4, 6.5] \\
Fingerprint & 82.6 [82.5, 82.8] & 79.9 [79.7, 80.0] & 78.9 [78.7, 79.0] & 76.8 [76.6, 76.9] & 75.9 [75.8, 76.1] & 72.6 [72.4, 72.7] \\
\bottomrule
\end{tabular}
\end{table*}

\begin{table*}
  \caption{Percent unique by Hispanic vs Non-hispanic ethnicity, with equal sized demographic groups, showing means and 95\% confidence intervals.}
  \label{tab:pct_unique_hisp_subsampled}
  \small
  
\begin{tabular}{r|ll}
\toprule
Attribute & Hispanic & Non-hispanic \\
\midrule
n & 900 & 900 \\
\midrule
User agent & 5.4 [5.4, 5.4] & 6.1 [6.0, 6.1] \\
Languages & 4.0 [4.0, 4.0] & 3.3 [3.3, 3.4] \\
Timezone & 0.5 [0.5, 0.5] & 0.6 [0.6, 0.6] \\
Screen resolution & 6.7 [6.7, 6.8] & 7.5 [7.4, 7.5] \\
Color depth & 0.0 [0.0, 0.0] & 0.0 [0.0, 0.0] \\
Platform & 0.0 [0.0, 0.0] & 0.0 [0.0, 0.1] \\
Touch points & 0.0 [0.0, 0.0] & 0.1 [0.1, 0.1] \\
Hardware concurrency & 0.1 [0.1, 0.1] & 0.1 [0.1, 0.1] \\
Device memory & 0.0 [0.0, 0.0] & 0.0 [0.0, 0.0] \\
WebGL Vendor & 0.0 [-0.0, 0.0] & 0.0 [0.0, 0.0] \\
WebGL Unmasked Vendor & 0.2 [0.2, 0.2] & 0.3 [0.2, 0.3] \\
WebGL Renderer & 0.4 [0.4, 0.4] & 0.5 [0.5, 0.5] \\
WebGL Unmasked Renderer & 7.1 [7.1, 7.1] & 9.1 [9.0, 9.1] \\
Fingerprint & 79.0 [79.0, 79.1] & 79.0 [78.9, 79.1] \\
\bottomrule
\end{tabular}
\end{table*}

\begin{table*}
  \caption{Percent unique by race, with equal sized demographic groups, showing means and 95\% confidence intervals.}
  \label{tab:pct_unique_race_subsampled}
  \small
  \begin{tabular}{r|llll}
  \toprule
Attribute & White & Black & Asian & Other or mixed \\
\midrule
n & 450 & 450 & 450 & 450 \\
\midrule
User agent & 5.8 [5.7, 5.8] & 7.2 [7.1, 7.2] & 4.7 [4.6, 4.7] & 7.6 [7.6, 7.7] \\
Languages & 3.0 [2.9, 3.0] & 3.4 [3.3, 3.4] & 4.9 [4.9, 5.0] & 4.0 [4.0, 4.1] \\
Timezone & 0.7 [0.7, 0.7] & 0.6 [0.6, 0.6] & 0.2 [0.2, 0.2] & 0.3 [0.3, 0.3] \\
Screen resolution & 7.3 [7.2, 7.4] & 6.7 [6.7, 6.8] & 7.6 [7.5, 7.6] & 8.2 [8.2, 8.3] \\
Color depth & 0.0 [0.0, 0.0] & 0.0 [0.0, 0.0] & 0.0 [0.0, 0.0] & 0.0 [0.0, 0.0] \\
Platform & 0.1 [0.1, 0.1] & 0.1 [0.1, 0.1] & 0.0 [0.0, 0.0] & 0.0 [0.0, 0.0] \\
Touch points & 0.1 [0.1, 0.1] & 0.0 [0.0, 0.0] & 0.3 [0.3, 0.3] & 0.1 [0.0, 0.1] \\
Hardware concurrency & 0.1 [0.1, 0.1] & 0.1 [0.1, 0.1] & 0.5 [0.4, 0.5] & 0.2 [0.1, 0.2] \\
Device memory & 0.0 [0.0, 0.0] & 0.0 [0.0, 0.0] & 0.0 [0.0, 0.0] & 0.0 [0.0, 0.0] \\
WebGL Vendor & 0.0 [0.0, 0.0] & 0.0 [0.0, 0.0] & 0.0 [0.0, 0.0] & 0.0 [0.0, 0.0] \\
WebGL Unmasked Vendor & 0.3 [0.3, 0.3] & 0.2 [0.2, 0.2] & 0.1 [0.1, 0.1] & 0.2 [0.2, 0.2] \\
WebGL Renderer & 0.6 [0.6, 0.6] & 0.5 [0.5, 0.5] & 0.2 [0.2, 0.2] & 0.1 [0.1, 0.1] \\
WebGL Unmasked Renderer & 9.0 [8.9, 9.1] & 8.2 [8.2, 8.3] & 7.0 [6.9, 7.0] & 8.1 [8.0, 8.1] \\
Fingerprint & 78.5 [78.3, 78.6] & 76.0 [75.9, 76.1] & 77.9 [77.7, 78.0] & 79.1 [79.0, 79.2] \\
\bottomrule
\end{tabular}
\end{table*}

\begin{table*}
  \caption{Average anonymity set size by gender, with equal sized demographic groups, showing means and 95\% confidence intervals.}
  \label{tab:avg_k_gender_subsampled}
  \small
  \begin{tabular}{r|ll}
  \toprule
Attribute & Male & Female \\
n & 900 & 900 \\
\midrule
User agent & 280.7 [279.9, 281.5] & 226.4 [225.7, 227.1] \\
Languages & 947.3 [946.0, 948.6] & 908.3 [907.1, 909.5] \\
Timezone & 580.7 [579.9, 581.6] & 595.0 [594.2, 595.9] \\
Screen resolution & 118.0 [117.6, 118.4] & 99.2 [98.9, 99.4] \\
Color depth & 1468.7 [1467.4, 1470.0] & 1393.1 [1391.6, 1394.5] \\
Platform & 613.9 [612.8, 615.0] & 500.6 [499.8, 501.5] \\
Touch points & 785.8 [784.8, 786.7] & 724.7 [724.0, 725.5] \\
Hardware concurrency & 455.7 [455.0, 456.4] & 533.5 [532.7, 534.2] \\
Device memory & 722.1 [721.2, 723.0] & 677.7 [676.8, 678.5] \\
WebGL Vendor & 1459.4 [1458.1, 1460.7] & 1551.6 [1550.5, 1552.7] \\
WebGL Unmasked Vendor & 258.7 [258.2, 259.2] & 272.9 [272.3, 273.4] \\
WebGL Renderer & 1443.4 [1442.0, 1444.9] & 1544.0 [1542.9, 1545.2] \\
WebGL Unmasked Renderer & 65.5 [65.2, 65.8] & 104.2 [103.7, 104.6] \\
Fingerprint & 1.5 [1.5, 1.5] & 2.0 [2.0, 2.1] \\
\bottomrule
\end{tabular}
\end{table*}

\begin{table*}
  \caption{Average anonymity set size by age group, with equal sized demographic groups, showing means and 95\% confidence intervals.}
  \label{tab:avg_k_age_subsampled}
  \footnotesize
  \begin{tabular}{r|llllll}
  \toprule
Attribute & 18 - 24 years & 25 - 34 years & 35 - 44 years & 45 - 54 years & 55 - 64 years & 65 or older \\
n & 300 & 300 & 300 & 300 & 300 & 300 \\
\midrule
User agent & 244.4 [243.4, 245.4] & 256.3 [255.2, 257.5] & 268.5 [267.3, 269.6] & 263.1 [262.0, 264.2] & 274.5 [273.5, 275.4] & 318.2 [317.6, 318.8] \\
Languages & 914.5 [912.8, 916.2] & 923.6 [921.7, 925.4] & 942.7 [940.9, 944.5] & 972.3 [970.6, 974.1] & 988.4 [986.9, 989.9] & 989.6 [988.6, 990.6] \\
Timezone & 575.1 [574.0, 576.1] & 582.7 [581.5, 583.9] & 593.5 [592.4, 594.6] & 594.9 [593.8, 595.9] & 617.3 [616.3, 618.3] & 587.3 [586.7, 587.8] \\
Screen resolution & 115.1 [114.7, 115.5] & 107.5 [107.0, 107.9] & 108.2 [107.7, 108.6] & 106.2 [105.8, 106.6] & 108.5 [108.1, 108.8] & 110.1 [109.8, 110.3] \\
Color depth & 1358.8 [1356.7, 1361.0] & 1425.3 [1423.3, 1427.3] & 1468.9 [1467.1, 1470.7] & 1493.2 [1491.6, 1494.8] & 1496.9 [1495.4, 1498.3] & 1484.9 [1484.1, 1485.8] \\
Platform & 549.2 [547.9, 550.5] & 566.7 [565.2, 568.2] & 570.1 [568.6, 571.6] & 580.0 [578.5, 581.4] & 596.1 [594.7, 597.4] & 644.7 [643.8, 645.5] \\
Touch points & 768.0 [766.8, 769.1] & 761.4 [760.2, 762.7] & 740.8 [739.7, 741.9] & 735.7 [734.6, 736.8] & 735.9 [734.9, 737.0] & 756.2 [755.5, 756.9] \\
Hardware concurrency & 483.7 [482.8, 484.6] & 486.6 [485.6, 487.6] & 493.5 [492.5, 494.5] & 515.7 [514.8, 516.6] & 534.7 [533.9, 535.5] & 510.2 [509.7, 510.6] \\
Device memory & 740.0 [738.8, 741.1] & 709.9 [708.7, 711.1] & 685.8 [684.6, 687.0] & 657.8 [656.7, 658.9] & 671.0 [670.0, 672.0] & 702.4 [701.8, 703.0] \\
WebGL Vendor & 1521.1 [1519.5, 1522.7] & 1493.2 [1491.4, 1495.1] & 1491.2 [1489.4, 1493.0] & 1494.0 [1492.4, 1495.7] & 1506.0 [1504.5, 1507.5] & 1492.9 [1492.1, 1493.7] \\
WebGL Unmasked Vendor & 254.0 [253.4, 254.7] & 257.8 [257.1, 258.6] & 274.8 [274.1, 275.6] & 299.2 [298.4, 300.0] & 317.7 [316.9, 318.5] & 337.8 [337.3, 338.3] \\
WebGL Renderer & 1511.4 [1509.7, 1513.1] & 1480.3 [1478.3, 1482.3] & 1477.9 [1475.9, 1479.8] & 1481.1 [1479.3, 1483.0] & 1494.0 [1492.4, 1495.6] & 1480.4 [1479.5, 1481.3] \\
WebGL Unmasked Renderer & 93.8 [93.3, 94.4] & 83.5 [82.9, 84.0] & 75.0 [74.5, 75.4] & 70.9 [70.4, 71.3] & 71.5 [71.1, 71.9] & 59.0 [58.8, 59.2] \\
Fingerprint & 1.9 [1.9, 1.9] & 1.8 [1.8, 1.8] & 1.6 [1.6, 1.6] & 1.5 [1.5, 1.5] & 1.7 [1.7, 1.7] & 1.3 [1.3, 1.3] \\
\bottomrule
\end{tabular}
\end{table*}

\begin{table*}
  \caption{Average anonymity set size by income group, with equal sized demographic groups, showing means and 95\% confidence intervals.}
  \label{tab:avg_k_income_subsampled}
  \footnotesize
  \begin{tabular}{r|llllll}
  \toprule
Attribute & Less than \$25,000 & \$25,000 - \$49,999 & \$50,000 - \$74,999 & \$75,000 - \$99,999 & \$100,000 - \$149,999 & \$150,000 or more \\
n & 300 & 300 & 300 & 300 & 300 & 300 \\
\midrule
User agent & 249.2 [248.2, 250.1] & 251.6 [250.5, 252.6] & 255.8 [254.7, 256.8] & 262.4 [261.4, 263.4] & 261.0 [259.9, 262.0] & 234.3 [233.4, 235.1] \\
Languages & 955.1 [953.5, 956.8] & 948.4 [946.6, 950.1] & 921.3 [919.5, 923.0] & 913.6 [912.0, 915.3] & 925.0 [923.3, 926.6] & 871.6 [870.1, 873.2] \\
Timezone & 565.5 [564.5, 566.5] & 590.2 [589.0, 591.3] & 581.8 [580.7, 582.9] & 604.7 [603.5, 605.8] & 582.4 [581.3, 583.5] & 586.5 [585.5, 587.6] \\
Screen resolution & 107.5 [107.1, 107.9] & 112.3 [111.9, 112.8] & 107.2 [106.8, 107.6] & 109.9 [109.5, 110.3] & 107.0 [106.6, 107.4] & 104.7 [104.3, 105.0] \\
Color depth & 1452.0 [1450.3, 1453.7] & 1459.6 [1457.8, 1461.3] & 1428.6 [1426.7, 1430.5] & 1413.2 [1411.3, 1415.1] & 1422.7 [1420.9, 1424.6] & 1394.8 [1393.0, 1396.5] \\
Platform & 566.2 [564.9, 567.6] & 567.5 [566.1, 569.0] & 560.9 [559.6, 562.3] & 568.6 [567.3, 569.9] & 563.9 [562.6, 565.2] & 511.6 [510.4, 512.7] \\
Touch points & 760.7 [759.5, 761.8] & 761.5 [760.4, 762.7] & 750.6 [749.4, 751.7] & 763.5 [762.4, 764.6] & 739.4 [738.3, 740.5] & 738.5 [737.4, 739.5] \\
Hardware concurrency & 488.6 [487.7, 489.6] & 486.0 [485.0, 487.0] & 497.3 [496.4, 498.3] & 483.2 [482.3, 484.1] & 504.3 [503.4, 505.2] & 496.2 [495.3, 497.1] \\
Device memory & 655.3 [654.1, 656.5] & 666.8 [665.6, 668.0] & 698.6 [697.5, 699.8] & 730.1 [728.9, 731.2] & 739.4 [738.3, 740.5] & 748.8 [747.8, 749.8] \\
WebGL Vendor & 1474.1 [1472.4, 1475.8] & 1480.4 [1478.6, 1482.2] & 1506.9 [1505.4, 1508.5] & 1508.1 [1506.5, 1509.7] & 1501.6 [1500.0, 1503.2] & 1519.4 [1517.9, 1520.8] \\
WebGL Unmasked Vendor & 242.6 [242.0, 243.3] & 256.8 [256.1, 257.5] & 265.6 [265.0, 266.3] & 269.4 [268.8, 270.1] & 282.9 [282.2, 283.6] & 280.2 [279.6, 280.8] \\
WebGL Renderer & 1459.2 [1457.4, 1461.0] & 1466.6 [1464.6, 1468.5] & 1495.4 [1493.6, 1497.1] & 1496.8 [1495.1, 1498.5] & 1489.3 [1487.5, 1491.1] & 1508.5 [1506.9, 1510.1] \\
WebGL Unmasked Renderer & 64.9 [64.5, 65.4] & 76.7 [76.1, 77.2] & 85.7 [85.1, 86.2] & 88.2 [87.6, 88.7] & 87.3 [86.8, 87.9] & 105.9 [105.3, 106.5] \\
Fingerprint & 1.6 [1.5, 1.6] & 1.6 [1.6, 1.6] & 1.7 [1.7, 1.8] & 1.9 [1.9, 1.9] & 1.9 [1.9, 1.9] & 2.3 [2.3, 2.3] \\
\bottomrule
\end{tabular}

\end{table*}

\begin{table*}
  \caption{Average anonymity set size by Hispanic vs non-Hispanic origin, with equal sized demographic groups, showing means and 95\% confidence intervals.}
  \label{tab:avg_k_hisp_subsampled}
  \small
    \begin{tabular}{r|ll}
  \toprule
Attribute & Hispanic & Non-hispanic \\
n & 900 & 900 \\
\midrule
User agent & 235.6 [235.3, 235.8] & 250.5 [249.7, 251.3] \\
Languages & 828.3 [827.9, 828.7] & 905.0 [903.8, 906.2] \\
Timezone & 506.6 [506.4, 506.7] & 561.0 [560.3, 561.7] \\
Screen resolution & 110.3 [110.1, 110.4] & 109.1 [108.8, 109.5] \\
Color depth & 1420.7 [1420.2, 1421.2] & 1430.8 [1429.5, 1432.1] \\
Platform & 544.4 [544.1, 544.8] & 557.9 [556.9, 558.9] \\
Touch points & 758.9 [758.6, 759.2] & 756.1 [755.3, 756.9] \\
Hardware concurrency & 474.8 [474.6, 475.1] & 487.3 [486.7, 488.0] \\
Device memory & 709.1 [708.8, 709.5] & 703.0 [702.1, 703.9] \\
WebGL Vendor & 1509.0 [1508.6, 1509.5] & 1498.9 [1497.7, 1500.1] \\
WebGL Unmasked Vendor & 254.7 [254.5, 254.9] & 261.5 [261.0, 262.0] \\
WebGL Renderer & 1497.7 [1497.2, 1498.2] & 1486.5 [1485.2, 1487.9] \\
WebGL Unmasked Renderer & 100.5 [100.3, 100.7] & 86.9 [86.5, 87.3] \\
Fingerprint & 1.9 [1.9, 1.9] & 1.8 [1.8, 1.8] \\
\bottomrule
\end{tabular}
\end{table*}

\begin{table*}
  \caption{Average anonymity set size by race, with equal sized demographic groups, showing means and 95\% confidence intervals.}
  \label{tab:avg_k_race_subsampled}
  \small
\begin{tabular}{r|llll}
\toprule
Attribute & White & Black & Asian & Other or mixed \\
n & 450 & 450 & 450 & 450 \\
\midrule
User agent & 255.4 [254.5, 256.4] & 263.8 [263.0, 264.6] & 294.4 [293.6, 295.2] & 237.4 [236.8, 238.0] \\
Languages & 919.1 [917.6, 920.6] & 923.4 [922.2, 924.6] & 822.3 [821.2, 823.5] & 849.3 [848.3, 850.4] \\
Timezone & 575.0 [574.0, 576.0] & 606.5 [605.7, 607.4] & 543.0 [542.3, 543.6] & 514.2 [513.7, 514.8] \\
Screen resolution & 109.1 [108.7, 109.5] & 105.8 [105.5, 106.0] & 116.1 [115.8, 116.4] & 106.5 [106.2, 106.7] \\
Color depth & 1425.9 [1424.2, 1427.6] & 1460.9 [1459.8, 1462.1] & 1341.7 [1340.3, 1343.1] & 1386.9 [1385.7, 1388.1] \\
Platform & 562.3 [561.0, 563.5] & 592.4 [591.4, 593.4] & 589.9 [589.0, 590.9] & 540.4 [539.6, 541.2] \\
Touch points & 755.4 [754.3, 756.5] & 719.4 [718.6, 720.2] & 790.6 [789.7, 791.5] & 750.4 [749.7, 751.1] \\
Hardware concurrency & 489.8 [488.9, 490.7] & 522.6 [521.8, 523.3] & 451.1 [450.4, 451.8] & 485.3 [484.7, 485.9] \\
Device memory & 705.4 [704.3, 706.4] & 655.2 [654.3, 656.0] & 789.5 [788.7, 790.3] & 705.0 [704.3, 705.7] \\
WebGL Vendor & 1500.9 [1499.3, 1502.4] & 1512.4 [1511.2, 1513.6] & 1543.1 [1542.1, 1544.2] & 1517.2 [1516.2, 1518.3] \\
WebGL Unmasked Vendor & 263.8 [263.2, 264.4] & 288.6 [288.1, 289.2] & 260.4 [260.0, 260.8] & 249.1 [248.8, 249.5] \\
WebGL Renderer & 1488.7 [1487.1, 1490.4] & 1501.3 [1500.0, 1502.6] & 1534.5 [1533.4, 1535.6] & 1506.6 [1505.5, 1507.7] \\
WebGL Unmasked Renderer & 84.0 [83.5, 84.5] & 78.4 [78.0, 78.8] & 89.2 [88.8, 89.6] & 98.9 [98.5, 99.3] \\
Fingerprint & 1.8 [1.7, 1.8] & 1.8 [1.8, 1.8] & 1.8 [1.7, 1.8] & 1.8 [1.8, 1.8] \\
\bottomrule
\end{tabular}
\end{table*}

\begin{table*}
  \caption{Fingerprinting metrics for the overall sample, computed without resampling.}
  \label{tab:overall_fingerprinting_metrics_notsubsampled}
  \small
  
\begin{tabular}{r|lll}
\toprule
& \multicolumn{3}{c}{\multirow{2}{*}{ Overall (N=8400) }} \\
& \multicolumn{3}{c}{} \\
\midrule
Attribute & Distinct values & \% Unique & Avg k \\
\midrule
User agent & 434 & 2.8 & 1185.2 \\
Languages & 264 & 2.4 & 4333 \\
Timezone & 49 & 0.2 & 2735.2 \\
Screen resolution & 572 & 4.5 & 505.3 \\
Color depth & 3 & 0 & 6684.5 \\
Platform & 12 & 0 & 2618.4 \\
Touch points & 11 & 0 & 3520.5 \\
Hardware concurrency & 24 & 0.1 & 2290.3 \\
Device memory & 7 & 0 & 3272 \\
WebGL Vendor & 3 & 0 & 6983.5 \\
WebGL Unmasked Vendor & 36 & 0.1 & 1232.6 \\
WebGL Renderer & 36 & 0.1 & 6924.7 \\
WebGL Unmasked Renderer & 654 & 3.2 & 385.3 \\
Fingerprint & 5973 & 60.2 & 4.6 \\
\bottomrule
\end{tabular}
\end{table*}

\begin{table*}
  \caption{Percent unique by gender, computed without resampling.}
  \label{tab:pct_unique_gender_notsubsampled}
  \small
  \begin{tabular}{r|lll}
  \toprule
Attribute & Male & Female & Other \\
n & 4227 & 3990 & 183 \\
\midrule
User agent & 2.7 & 2.8 & 3.8 \\
Languages & 2.6 & 2 & 4.4 \\
Timezone & 0.2 & 0.3 & 0 \\
Screen resolution & 5.8 & 2.9 & 8.2 \\
Color depth & 0 & 0 & 0 \\
Platform & 0 & 0.1 & 0 \\
Touch points & 0 & 0 & 0 \\
Hardware concurrency & 0.1 & 0 & 0 \\
Device memory & 0 & 0 & 0 \\
WebGL Vendor & 0 & 0 & 0 \\
WebGL Unmasked Vendor & 0.1 & 0 & 0.5 \\
WebGL Renderer & 0.1 & 0 & 1.1 \\
WebGL Unmasked Renderer & 4.1 & 2.3 & 4.9 \\
Fingerprint & 65.2 & 54.6 & 69.4 \\
\bottomrule
\end{tabular}
\end{table*}

\begin{table*}
  \caption{Percent unique by age group, computed without resampling.}
  \label{tab:pct_unique_age_notsubsampled}
  \small
  \begin{tabular}{r|llllll}
  \toprule
Attribute & 18 - 24 years & 25 - 34 years & 35 - 44 years & 45 - 54 years & 55 - 64 years & 65 or older \\
n & 1302 & 2859 & 2024 & 1158 & 712 & 345 \\
\midrule
User agent & 2.9 & 2.8 & 2.8 & 2.3 & 2.9 & 3.2 \\
Languages & 2 & 2.4 & 2.3 & 2.8 & 2.5 & 1.7 \\
Timezone & 0.5 & 0.1 & 0.1 & 0.4 & 0.3 & 0 \\
Screen resolution & 2.5 & 4.8 & 4.9 & 4 & 5.9 & 6.7 \\
Color depth & 0 & 0 & 0 & 0 & 0 & 0 \\
Platform & 0 & 0 & 0 & 0 & 0 & 0.3 \\
Touch points & 0 & 0 & 0 & 0 & 0.1 & 0 \\
Hardware concurrency & 0.1 & 0.1 & 0 & 0 & 0 & 0.3 \\
Device memory & 0 & 0 & 0 & 0 & 0 & 0 \\
WebGL Vendor & 0 & 0 & 0 & 0 & 0 & 0 \\
WebGL Unmasked Vendor & 0 & 0.1 & 0.1 & 0.1 & 0.1 & 0 \\
WebGL Renderer & 0 & 0.1 & 0.1 & 0.1 & 0 & 0.3 \\
WebGL Unmasked Renderer & 1.9 & 2.8 & 3.7 & 3.5 & 4.4 & 5.8 \\
Fingerprint & 55.2 & 58.5 & 61.6 & 62.6 & 64.6 & 69 \\
\bottomrule
\end{tabular}
\end{table*}

\begin{table*}
  \caption{Percent unique by income group, computed without resampling.}
  \label{tab:pct_unique_income_notsubsampled}
  \small
  \begin{tabular}{r|llllll}
  \toprule
Attribute & Less than \$25,000 & \$25,000 - \$49,999 & \$50,000 - \$74,999 & \$75,000 - \$99,999 & \$100,000 - \$149,999 & \$150,000 or more \\
n & 1097 & 1842 & 1692 & 1267 & 1411 & 951 \\
\midrule
User agent & 4.5 & 3.4 & 2.7 & 2 & 2.3 & 1.6 \\
Languages & 2.3 & 2.6 & 2.2 & 2.7 & 2 & 2.4 \\
Timezone & 0.3 & 0.1 & 0.3 & 0.2 & 0.2 & 0.4 \\
Screen resolution & 5.8 & 4.3 & 4.7 & 4.7 & 3.8 & 3.7 \\
Color depth & 0 & 0 & 0 & 0 & 0 & 0 \\
Platform & 0 & 0.1 & 0.1 & 0 & 0 & 0 \\
Touch points & 0 & 0 & 0.1 & 0 & 0 & 0 \\
Hardware concurrency & 0 & 0.1 & 0.1 & 0.1 & 0 & 0 \\
Device memory & 0 & 0 & 0.1 & 0 & 0 & 0 \\
WebGL Vendor & 0 & 0 & 0 & 0 & 0 & 0 \\
WebGL Unmasked Vendor & 0.2 & 0 & 0.1 & 0.2 & 0.1 & 0 \\
WebGL Renderer & 0.4 & 0 & 0 & 0.2 & 0.1 & 0 \\
WebGL Unmasked Renderer & 4 & 3.4 & 3.4 & 3.2 & 3 & 2.5 \\
Fingerprint & 67.5 & 61.9 & 60.9 & 58 & 56.9 & 55.3 \\
\midrule
\end{tabular}
\end{table*}

\begin{table*}
  \caption{Percent unique by Hispanic vs non-Hispanic origin, computed without resampling.}
  \label{tab:pct_unique_hisp_notsubsampled}
  \small
  \begin{tabular}{r|ll}
  \toprule
Attribute & Hispanic & Non-hispanic \\
n & 923 & 7477 \\
\midrule
User agent & 3 & 2.8 \\
Languages & 2.9 & 2.3 \\
Timezone & 0.2 & 0.2 \\
Screen resolution & 4.6 & 4.5 \\
Color depth & 0 & 0 \\
Platform & 0 & 0 \\
Touch points & 0 & 0 \\
Hardware concurrency & 0 & 0.1 \\
Device memory & 0 & 0 \\
WebGL Vendor & 0 & 0 \\
WebGL Unmasked Vendor & 0 & 0.1 \\
WebGL Renderer & 0.1 & 0.1 \\
WebGL Unmasked Renderer & 2.4 & 3.3 \\
Fingerprint & 62.9 & 59.9 \\
\bottomrule
\end{tabular}
\end{table*}

\begin{table*}
  \caption{Percent unique by race, computed without resampling.}
  \label{tab:pct_unique_race_notsubsampled}
  \small
  \begin{tabular}{r|llll}
  \toprule
Attribute & White & Black & Asian & Other or mixed \\
n & 5911 & 938 & 842 & 709 \\
\midrule
User agent & 2.5 & 3.6 & 2.1 & 4.7 \\
Languages & 2 & 2.2 & 4.3 & 3.1 \\
Timezone & 0.3 & 0.2 & 0 & 0 \\
Screen resolution & 4.4 & 4.2 & 5.1 & 4.8 \\
Color depth & 0 & 0 & 0 & 0 \\
Platform & 0 & 0 & 0 & 0 \\
Touch points & 0 & 0 & 0.1 & 0 \\
Hardware concurrency & 0 & 0 & 0.2 & 0.1 \\
Device memory & 0 & 0 & 0 & 0 \\
WebGL Vendor & 0 & 0 & 0 & 0 \\
WebGL Unmasked Vendor & 0.1 & 0 & 0 & 0.1 \\
WebGL Renderer & 0.1 & 0 & 0 & 0 \\
WebGL Unmasked Renderer & 3.5 & 2 & 2.5 & 3.4 \\
Fingerprint & 60.8 & 54.2 & 62.1 & 61.4 \\
\bottomrule
\end{tabular}
\end{table*}

\begin{table*}
  \caption{Average anonymity set size by gender, computed without resampling.}
  \label{tab:avg_k_gender_notsubsampled}
  \small
  \begin{tabular}{r|lll}
  \toprule
Attribute & Male & Female & Other \\
n & 4227 & 3990 & 183 \\
\midrule
User agent & 1309.9 & 1055.7 & 1128.3 \\
Languages & 4422.8 & 4237.2 & 4346.5 \\
Timezone & 2706.8 & 2772.7 & 2573.9 \\
Screen resolution & 549.2 & 459.7 & 484.2 \\
Color depth & 6856.7 & 6503.1 & 6661.1 \\
Platform & 2875.5 & 2340.2 & 2747 \\
Touch points & 3664.7 & 3378.1 & 3293.6 \\
Hardware concurrency & 2119.1 & 2476.1 & 2192.1 \\
Device memory & 3372 & 3164.7 & 3303.9 \\
WebGL Vendor & 6789.4 & 7218.5 & 6341.6 \\
WebGL Unmasked Vendor & 1206.2 & 1263.3 & 1173.7 \\
WebGL Renderer & 6712.3 & 7181.9 & 6223.2 \\
WebGL Unmasked Renderer & 300.4 & 477.3 & 341.8 \\
Fingerprint & 3.4 & 5.8 & 3.8 \\
\bottomrule
\end{tabular}
\end{table*}

\begin{table*}
  \caption{Average anonymity set size by age group, computed without resampling.}
  \label{tab:avg_k_age_notsubsampled}
  \small
  \begin{tabular}{r|llllll}
  \toprule
Attribute & 18 - 24 years & 25 - 34 years & 35 - 44 years & 45 - 54 years & 55 - 64 years & 65 or older \\
n & 1302 & 2859 & 2024 & 1158 & 712 & 345 \\
\midrule
User agent & 1103.5 & 1159.1 & 1216 & 1184.5 & 1236.3 & 1425.4 \\
Languages & 4215.9 & 4251.5 & 4339.7 & 4465.5 & 4540.3 & 4538.3 \\
Timezone & 2672.2 & 2706.9 & 2756.2 & 2764.3 & 2862.8 & 2724.1 \\
Screen resolution & 542.3 & 501.9 & 502.4 & 485.1 & 496.1 & 496.7 \\
Color depth & 6287.5 & 6602.7 & 6803.8 & 6914.2 & 6933.3 & 6876 \\
Platform & 2525.1 & 2595.5 & 2615.4 & 2640.9 & 2711.8 & 2910.4 \\
Touch points & 3591.1 & 3574.1 & 3477.9 & 3438.4 & 3439.1 & 3502.7 \\
Hardware concurrency & 2240.9 & 2249.6 & 2278.9 & 2360.7 & 2439.8 & 2336 \\
Device memory & 3480.6 & 3332.1 & 3210.7 & 3075.9 & 3138.8 & 3280.8 \\
WebGL Vendor & 7094.6 & 6961.6 & 6945.5 & 6961.6 & 7022.8 & 6960.7 \\
WebGL Unmasked Vendor & 1164.9 & 1175.9 & 1227.6 & 1306.7 & 1375.8 & 1442.6 \\
WebGL Renderer & 7049.6 & 6900.9 & 6881.9 & 6900 & 6965.8 & 6900.7 \\
WebGL Unmasked Renderer & 459.3 & 408 & 363.8 & 340.8 & 343.1 & 281.9 \\
Fingerprint & 5.5 & 5 & 4.3 & 3.6 & 4.3 & 2.4 \\
\bottomrule
\end{tabular}
\end{table*}

\begin{table*}
  \caption{Average anonymity set size by income group, computed without resampling.}
  \label{tab:avg_k_income_notsubsampled}
  \small
  \begin{tabular}{r|llllll}
  \toprule
Attribute & Less than \$25,000 & \$25,000 - \$49,999 & \$50,000 - \$74,999 & \$75,000 - \$99,999 & \$100,000 - \$149,999 & \$150,000 or more \\
n & 1097 & 1842 & 1692 & 1267 & 1411 & 951 \\
\midrule
User agent & 1168.9 & 1173 & 1200.6 & 1231 & 1220.2 & 1090.1 \\
Languages & 4469.5 & 4440.6 & 4313.8 & 4276.4 & 4328.7 & 4072.7 \\
Timezone & 2637.8 & 2758.1 & 2714.3 & 2823.5 & 2719.1 & 2736.1 \\
Screen resolution & 501.2 & 523.7 & 500.1 & 512.2 & 498.2 & 484.8 \\
Color depth & 6787.9 & 6822 & 6674.2 & 6602.7 & 6651 & 6515.6 \\
Platform & 2654.3 & 2658.5 & 2631.7 & 2666.6 & 2640.3 & 2389.1 \\
Touch points & 3553.6 & 3556.3 & 3510.6 & 3568.5 & 3451.8 & 3450.3 \\
Hardware concurrency & 2276.3 & 2262.1 & 2312 & 2248.7 & 2348 & 2309.8 \\
Device memory & 3046 & 3100.9 & 3250.4 & 3391.9 & 3438.7 & 3476.7 \\
WebGL Vendor & 6874.8 & 6903.1 & 7027.1 & 7033 & 7007 & 7085.9 \\
WebGL Unmasked Vendor & 1128.1 & 1192.9 & 1231.8 & 1252.4 & 1312.1 & 1297.3 \\
WebGL Renderer & 6805 & 6838 & 6972.7 & 6979.9 & 6949.6 & 7035.1 \\
WebGL Unmasked Renderer & 297.4 & 351.3 & 389 & 404.9 & 400.6 & 486.2 \\
Fingerprint & 3.5 & 3.7 & 4.3 & 4.9 & 5.2 & 6.8 \\
\bottomrule
\end{tabular}
\end{table*}

\begin{table*}
  \caption{Average anonymity set size by Hispanic vs non-Hispanic origin, computed without resampling.}
  \label{tab:avg_k_hisp_notsubsampled}
  \small
  \begin{tabular}{r|ll}
  \toprule
Attribute & Hispanic & Non-hispanic \\
n & 923 & 7477 \\
\midrule
User agent & 1121.3 & 1193.1 \\
Languages & 3986.8 & 4375.7 \\
Timezone & 2400.8 & 2776.5 \\
Screen resolution & 505.9 & 505.2 \\
Color depth & 6645.5 & 6689.3 \\
Platform & 2558.9 & 2625.8 \\
Touch points & 3531.8 & 3519.1 \\
Hardware concurrency & 2236.2 & 2297 \\
Device memory & 3294.7 & 3269.2 \\
WebGL Vendor & 7025 & 6978.3 \\
WebGL Unmasked Vendor & 1190.6 & 1237.8 \\
WebGL Renderer & 6970.3 & 6919.1 \\
WebGL Unmasked Renderer & 438.6 & 378.8 \\
Fingerprint & 4.9 & 4.5 \\
\bottomrule
\end{tabular}
\end{table*}

\begin{table*}
  \caption{Average anonymity set size by race, computed without resampling.}
  \label{tab:avg_k_race_notsubsampled}
  \small
\begin{tabular}{r|llll}
\toprule
Attribute & White & Black & Asian & Other or mixed \\
n & 5911 & 938 & 842 & 709 \\
\midrule
User agent & 1170.8 & 1211 & 1339.7 & 1087.8 \\
Languages & 4406 & 4431.7 & 3937.4 & 4063.4 \\
Timezone & 2774.7 & 2953.1 & 2481.2 & 2419.3 \\
Screen resolution & 505 & 487.3 & 537.8 & 492.7 \\
Color depth & 6721.9 & 6889.6 & 6317.9 & 6536.5 \\
Platform & 2599.2 & 2743.8 & 2717.1 & 2495.5 \\
Touch points & 3524.6 & 3357.6 & 3685.2 & 3506.1 \\
Hardware concurrency & 2293.7 & 2436.9 & 2119.7 & 2270.3 \\
Device memory & 3258.4 & 3035.5 & 3641 & 3260.3 \\
WebGL Vendor & 6951.3 & 7004.5 & 7147.9 & 7028.9 \\
WebGL Unmasked Vendor & 1227.4 & 1351.2 & 1202.4 & 1155.2 \\
WebGL Renderer & 6889.5 & 6947.7 & 7104.2 & 6974.8 \\
WebGL Unmasked Renderer & 380.5 & 353.1 & 402.3 & 447.9 \\
Fingerprint & 4.6 & 4.8 & 4.3 & 4.5 \\
\bottomrule
\end{tabular}
\end{table*}

\clearpage

\section{Additional details for mutual information analysis}
\label{SI:mutual_information}

We computed $ \frac{I(A_k; D)}{H(D)}$, representing the normalized mutual information between a given browser attribute, $A$, and demographic category, $D$, for multiple values of the parameter $k$: $k=1$, $k=50$, $k=100$.
Note that many browser attributes have a longtail distribution, and higher values of k cut off the long tail, limiting analysis to attribute values shared by at least k users.
The choice of parameter k impacts the number of distinct attribute values for each attribute, the number of users in the computation, and therefore the resulting value for normalized mutual information.

In Tables \ref{tab:mutual_info_k50} and \ref{tab:mutual_info_k100}, N shows the number of users in the analysis for each browser attribute, after dropping data for attribute values with fewer than k users. Metrics were computed separately for each attribute, A, resulting in different values of N when k>1.

\begin{table*}
  \caption{Normalized mutual information between browser attributes and demographic categories (k=1; N=8400; i.e. no restriction on attribute value anonymity set size).}
  \label{tab:mutual_info_k1}
  \small
  \begin{tabular}{rr|lllll}
  \toprule
Attribute & Distinct values & Gender & Age & Hispanic & Race & Income \\
\midrule
User agent & 434 & 0.092 & 0.077 & 0.063 & 0.077 & 0.068 \\
Languages & 264 & 0.04 & 0.038 & 0.053 & 0.053 & 0.035 \\
Timezone & 49 & 0.01 & 0.008 & 0.031 & 0.046 & 0.011 \\
Screen resolution & 572 & 0.123 & 0.104 & 0.082 & 0.103 & 0.086 \\
Color depth & 3 & 0.004 & 0.004 & 0 & 0.003 & 0.002 \\
Platform & 12 & 0.026 & 0.016 & 0.002 & 0.009 & 0.009 \\
Touch points & 11 & 0.014 & 0.006 & 0.002 & 0.005 & 0.004 \\
Hardware concurrency & 24 & 0.027 & 0.013 & 0.005 & 0.015 & 0.006 \\
Device memory & 7 & 0.009 & 0.008 & 0.002 & 0.009 & 0.01 \\
WebGL Vendor & 3 & 0.009 & 0 & 0 & 0.001 & 0.001 \\
WebGL Unmasked Vendor & 36 & 0.05 & 0.027 & 0.008 & 0.019 & 0.016 \\
WebGL Renderer & 36 & 0.016 & 0.008 & 0.006 & 0.006 & 0.007 \\
WebGL Unmasked Renderer & 654 & 0.146 & 0.122 & 0.103 & 0.123 & 0.107 \\
\bottomrule
\end{tabular}
\end{table*}

\begin{table*}
  \caption{Normalized mutual information between browser attributes and demographic categories (k=50).}
  \label{tab:mutual_info_k50}
  \small
  \begin{tabular}{rrr|lllll}
  \toprule
Attribute & Distinct values & N & Gender & Age & Hispanic & Race & Income \\
\midrule
User agent & 17 & 6741 & 0.034 & 0.02 & 0.006 & 0.011 & 0.013 \\
Languages & 3 & 7798 & 0.005 & 0.003 & 0.008 & 0.002 & 0.002 \\
Timezone & 7 & 8223 & 0.004 & 0.002 & 0.023 & 0.031 & 0.003 \\
Screen resolution & 25 & 6542 & 0.058 & 0.027 & 0.015 & 0.025 & 0.013 \\
Color depth & 3 & 8400 & 0.004 & 0.004 & 0 & 0.003 & 0.002 \\
Platform & 7 & 8377 & 0.025 & 0.015 & 0.002 & 0.008 & 0.008 \\
Touch points & 3 & 8312 & 0.011 & 0.004 & 0.001 & 0.003 & 0.001 \\
Hardware concurrency & 8 & 8185 & 0.02 & 0.011 & 0.002 & 0.011 & 0.004 \\
Device memory & 4 & 8318 & 0.006 & 0.007 & 0.001 & 0.009 & 0.01 \\
WebGL Vendor & 2 & 8374 & 0.009 & 0 & 0 & 0 & 0 \\
WebGL Unmasked Vendor & 15 & 8181 & 0.046 & 0.024 & 0.005 & 0.016 & 0.013 \\
WebGL Renderer & 5 & 8120 & 0.009 & 0.002 & 0.001 & 0.002 & 0.001 \\
WebGL Unmasked Renderer & 36 & 4880 & 0.053 & 0.027 & 0.012 & 0.029 & 0.027 \\
\bottomrule
\end{tabular}
\end{table*}

\begin{table*}
  \caption{Normalized mutual information between browser attributes and demographic categories (k=100).}
  \label{tab:mutual_info_k100}
  \small
  \begin{tabular}{rrr|lllll}
  \toprule
Attribute & Distinct values & N & Gender & Age & Hispanic & Race & Income \\
\midrule
User agent & 12 & 6380 & 0.031 & 0.015 & 0.004 & 0.01 & 0.01 \\
Languages & 3 & 7798 & 0.005 & 0.003 & 0.008 & 0.002 & 0.002 \\
Timezone & 6 & 8146 & 0.003 & 0.002 & 0.023 & 0.03 & 0.003 \\
Screen resolution & 16 & 5972 & 0.06 & 0.025 & 0.011 & 0.022 & 0.011 \\
Color depth & 3 & 8400 & 0.004 & 0.004 & 0 & 0.003 & 0.002 \\
Platform & 6 & 8313 & 0.025 & 0.015 & 0.002 & 0.008 & 0.008 \\
Touch points & 3 & 8312 & 0.011 & 0.004 & 0.001 & 0.003 & 0.001 \\
Hardware concurrency & 6 & 8027 & 0.018 & 0.01 & 0.002 & 0.009 & 0.002 \\
Device memory & 4 & 8318 & 0.006 & 0.007 & 0.001 & 0.009 & 0.01 \\
WebGL Vendor & 2 & 8374 & 0.009 & 0 & 0 & 0 & 0 \\
WebGL Unmasked Vendor & 11 & 7915 & 0.046 & 0.023 & 0.004 & 0.014 & 0.011 \\
WebGL Renderer & 4 & 8043 & 0.008 & 0.002 & 0.001 & 0.001 & 0.001 \\
WebGL Unmasked Renderer & 9 & 3068 & 0.053 & 0.013 & 0.003 & 0.01 & 0.012 \\
\bottomrule
\end{tabular}
\end{table*}


\end{document}